\documentclass[onecolumn]{aastex63}
\usepackage[utf8]{inputenc}
\usepackage{graphicx}
\usepackage{xcolor}
\usepackage{amsmath,amssymb}
\usepackage{gensymb}
\usepackage{times}
\usepackage{nicefrac}
\usepackage{amsmath}
\usepackage{xcolor}
\usepackage[normalem]{ulem} 

\usepackage{savesym}
\savesymbol{tablenum}
\usepackage{siunitx}
\usepackage{xspace}
\restoresymbol{SIX}{tablenum}
\DeclareSIUnit\year{yr}
\DeclareSIUnit\parsec{pc}
\DeclareSIUnit\msun{M_\odot}
\DeclareSIUnit\Rsun{R_\odot}

\DeclareSIUnit\mstar{M_\star}

\newcommand{\given}{\ensuremath{ \,|\, }}

\newcommand{\bs}[1]{\ensuremath{\boldsymbol{#1}}}
\newcommand{\pdf}{PDF\xspace}
\newcommand{\pdfs}{PDFs\xspace}

\sloppypar

\begin{document}
\title{\textsl{Lux}: \\
A generative, multi-output, latent-variable model for astronomical data with noisy labels
}

\newcommand{\affcca}{
    Center for Computational Astrophysics, Flatiron Institute,
    162 Fifth Ave, New York, NY 10010, USA
}

\newcommand{\affcolumbia}{
    Department of Astronomy, Columbia University,
    550 West 120th Street, New York, NY 10027, USA
}

\newcommand{\affanu}{
Research School of Astronomy $\&$ Astrophysics, Australian National University, Canberra, ACT 2611, Australia
}

\newcommand{\affuw}{
    Department of Astronomy, Box 351580, University of Washington, Seattle, WA 98195
}

\newcommand{\affccpp}{
    Center for Cosmology and Particle Physics, Department of Physics, New York University,
    726 Broadway, New York, NY 10003, USA
}
\newcommand{\affmpia}{
    Max-Planck-Institut f\"ur Astronomie,
    K\"onigstuhl 17, D-69117 Heidelberg, Germany
}

\newcommand{\affmonash}{
    School of Physics $\&$ Astronomy, Monash University, Clayton 3800, Victoria, Australia
}

\newcommand{\affmon}{
Faculty of Information Technology, Monash University, Clayton 3800, Victoria, Australia
}

\author[0000-0003-1856-2151]{Danny Horta}
\affiliation{\affcca}

\author[0000-0003-0872-7098]{Adrian~M.~Price-Whelan}
\affiliation{\affcca}

\author[0000-0003-2866-9403]{David~W.~Hogg}
\affiliation{\affcca}
\affiliation{\affmpia}
\affiliation{\affccpp}

\author[0000-0001-5082-6693]{Melissa~K.~Ness}
\affiliation{\affcca}
\affiliation{\affcolumbia}
\affiliation{\affanu}

\author[0000-0003-0174-0564]{Andrew~R.~Casey}
\affiliation{\affcca}
\affiliation{\affmonash}
\affiliation{\affmon}

\correspondingauthor{Danny Horta}
\email{dhortadarrington@gmail.com}

\begin{abstract}\noindent
The large volume of spectroscopic data available now and from near-future surveys will enable high-dimensional measurements of stellar parameters and properties. Current methods for determining stellar labels from spectra use physics-driven models, which are computationally expensive and have limitations in their accuracy due to simplifications. While machine learning methods provide efficient paths toward emulating physics-based pipelines, they often do not properly account for uncertainties and have complex model structure, both of which can lead to biases and inaccurate label inference.
Here we present \textsl{Lux}: a data-driven framework for modeling stellar spectra and labels that addresses prior limitations.
\textsl{Lux} is a generative, multi-output, latent variable model framework built on \texttt{JAX} for computational efficiency and flexibility.
As a generative model, \textsl{Lux} properly accounts for uncertainties and missing data in the input stellar labels and spectral data and can either be used in probabilistic or discriminative settings.
Here, we present several examples of how \textsl{Lux} can successfully emulate methods for precise stellar label determinations for stars ranging in stellar type and signal-to-noise from the \textsl{APOGEE} survey.
We also show how a simple \textsl{Lux} model is successful at performing label transfer between the \textsl{APOGEE} and \textsl{GALAH} surveys.
\textsl{Lux} is a powerful new framework for the analysis of large-scale spectroscopic survey data.
Its ability to handle uncertainties while maintaining high precision makes it particularly valuable for stellar survey label inference and cross-survey analysis, and the flexible model structure allows for easy extension to other data types.
\end{abstract}

\keywords{methods: data analysis — methods: statistical — techniques: spectroscopic}

\section{Introduction}
The vast amounts of high-quality spectroscopic data the astronomy community is collecting with ground and space-based telescopes is unprecedented.
It both provides an opportunity, and generates a need, for the development of novel statistical and machine-learning models.
Specifically, large-scale spectroscopic surveys of the Galaxy (e.g., \textsl{APOGEE}: \citealp{Majewski2017}, \textsl{GALAH}: \citealp{Freeman2012}, \textsl{LAMOST}: \citealp{Zhao2012}, among others, including now \textsl{Gaia}: \citealp[][]{Gaia2023}), are providing multi-band, multi-resolution data sets for Galaxy science.
From these stellar spectra it is possible to determine the intrinsic properties of stars, such as stellar parameters and detailed element abundances (i.e., stellar labels).
It is also possible to obtain precise radial velocities, that can be combined with celestial positions, distances, and proper motions to deliver full 6D phase-space information, and thus kinematics or orbits.

Traditionally, stellar labels are determined from comparison of a spectrum with a grid of synthetic stellar model spectra (\citealp[e.g.,][]{Steinmetz2006,Yanny2009,Gilmore2012,Zhao2012,Perez2015,Martell2017}). However, the stellar photosphere models that are used have physical ingredients that are incomplete or simplified. For example, 1D stellar photosphere models are almost always used, assumed to be in local thermal equilibrium, for large surveys (for computational feasibility). Moreover, it is often typical to apply a post-calibration step to ensure that stellar labels derived using some minimization technique on model and observed spectra match some external higher-fidelity information, like benchmark stars in globular clusters (\citealp[e.g.,][]{Kordopatis2013,Meszaros2013,Jofre2014, Cunha2017}). With the advent of large-scale stellar surveys that deliver spectra for millions of stars in the Milky Way, these requirements become very computationally expensive.

In an attempt to circumvent these requirements, in the last decade there has been a push to use data-driven methods to determine stellar parameters and element abundances of stars using linear regression (e.g., \citealp{Ness2015, Casey2016}) or machine-learning methods (e.g., \citealp{Ting2019, Ciuca2022, Andrae2023, Buck2024, rozansnki2024}), that are more suited to deal with high-volume data. These methods are considered latent variable models, that aim to infer a mapping between observables and measurables via some latent space (e.g., \citealp[][]{Nolan2006, Hall2019, Eilers2019}), and can fall under the umbrella of ``\textit{emulator}'' or ``\textit{label transfer}'' approaches depending on the training and testing data employed. In essence, they are used to: 1) train a model on a set of (trustworthy) input stellar spectra and stellar labels; 2) optimize a set of (latent) model parameters; 3) use the trained model to predict stellar labels for some catalog data. While the functional form of the model may vary between these approaches (e.g., quadratic, neural-network, etc...), the process is still the same in practice. These models have proven extremely successful in delivering accurate and precise stellar labels in a fast and cost-effective manner \citep[e.g.,][]{Ness2016_cannon, Ho2017,Ho2017b,Xiang2017,Xiang2019,Rene2023, Li2024, Guiglion2024,Ness2024}. However, they also come with some limitations. For example, these models assume the input training stellar labels are ground-truth (i.e., no uncertainties are taken into account), and are not able to train a model with missing label data (for example, stars that have $T_{\mathrm{eff}}$ information but no [Fe/H]), although see \citet[][]{Leung_2018_astroNN} or \citet[][]{Nolan2024} for counter examples. As a result, many good data are not used in the training step, which hinders the stellar label regime that can be probed. This restriction also limits the ability to perform a two-way label transfer between two spectroscopic data sets, as typically stars will have a set of labels from one survey but not the other.

In this paper, we present \textsl{Lux}\footnote{\textsl{Lux} is the Latin word for light.}, a new multi-output generative latent variable model that is able to circumvent many of these limitations to infer stellar labels from stellar spectra. Unlike many past data-driven approaches, \textsl{Lux} is a generative model of stellar spectra and stellar labels. \textsl{Lux} can: 1) account for input stellar label uncertainties; 2) train a model using partial missing label data; 3) use simple model forms to capture and model a wide stellar label space; 4) estimate stellar labels in a fast and cost-effective manner, using \texttt{JAX} \citep{jax2018github}.

In Section~\ref{sec_data} we introduce the data and samples used. In Section~\ref{sec_model} we describe the framework of the \textsl{Lux} model, introduce the likelihood function, explain the routine for employing the model, and highlight the novel aspects of \textsl{Lux}. In Section~\ref{sec_model_res} we present a range of results that illustrate how \textsl{Lux} can precisely infer stellar labels from the \textsl{APOGEE} data. We also show results of the \textsl{Lux} mode using synthetic stellar spectra generated with \textsl{Korg} \citep[][]{Wheeler2023} in Appendix~\ref{app_korg}. In Section~\ref{sec_galah} we illustrate how \textsl{Lux} can effectively perform multi-survey translation between the \textsl{APOGEE-GALAH} surveys. We finish by discussing the idea of \textsl{Lux} in the wider context of stellar label determination and machine-learning in Section~\ref{sec_discussion}, and close by providing our summary and concluding remarks in Section~\ref{sec_conclusions}.

\section{Data}
\label{sec_data}

\begin{figure*}
    \centering
    \includegraphics[width=1\textwidth]{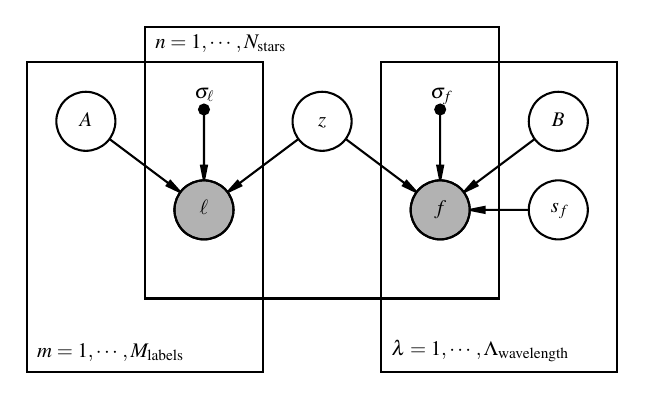}
    \caption{Graphical model of \textsl{Lux}. Here, $\boldsymbol{\ell}$ represents labels, $\boldsymbol{f}$ represents flux, $\boldsymbol{z}$ are the latent variables, $\boldsymbol{A}$ and $\boldsymbol{B}$ represent the matrices that project the latent variables into stellar labels and stellar fluxes, respectively, $\boldsymbol{s}$ is the vector of scatter terms at every pixel to account for underestimated uncertainties in the flux measurements, and $\boldsymbol{\sigma}_{\ell}$ and $\boldsymbol{\sigma}_{f}$ represent the uncertainties in the labels and flux, respectively. See Section~\ref{sec_model_struc} for details.}
    \label{fig:gpm}
\end{figure*}

We use data from the latest spectroscopic data-release from the \textsl{APOGEE} survey \citep[DR17;][]{Majewski2017, SDSSDR17}. The \textsl{APOGEE} data are based on observations collected by two high-resolution, multi-fibre spectrographs \citep{Wilson2019} attached to the 2.5m Sloan telescope at Apache Point Observatory \citep{Gunn2006} and the du Pont 2.5~m telescope at Las Campanas Observatory \citep{BowenVaughan1973}, respectively. Element abundances and stellar parameters are derived using the \textsl{ASPCAP} pipeline \citep{Perez2015} based on the \textsl{FERRE} code \citep[][]{Prieto2006} and the line lists from \citet{Cunha2017} and \citet[][]{Smith2021}. The spectra themselves were
reduced by a customized pipeline \citep{Nidever2015}. For details on
target selection criteria, see \citet{Zasowski2013} for \textsl{APOGEE}, \citet{Zasowski2017} for \textsl{APOGEE}-2, \citet[][]{Beaton2021} for \textsl{APOGEE} north, and \citet[][]{Santana2021} for \textsl{APOGEE} south.

We also make use of the second version of the \textsl{GALAH} DR3 data (\citealp{Martell2017,Buder2020}), a high-resolution ($R \approx 28,000$) optical survey that uses the \textsl{HERMES} spectrograph \citep{Sheinis2015} with 2dF fiber positioning system \citep[][]{Lewis2002} mounted on the 3.9-meter Anglo-Australian Telescope at Siding Spring Observatory, Australia. All data from \textsl{HERMES} were reduced with the \texttt{iraf} pipeline, and is analyzed with the Spectroscopy Made Easy (\texttt{SME}) software \citep[][]{Piskunov2016}, using the \texttt{MARCS} theoretical 1D hydrostatic models \citep[][]{Gustafsson2008}.
\vspace{1cm}
\subsection{Cleaning and preparing the data}

\textsl{Lux} can be executed on either continuum-normalised or flux-normalised spectra. For this paper, we work with continuum-normalised spectra. Before running \textsl{Lux}, we prepare the spectral data in the following way: we replace any bad flux measurements (i.e., flux values that are zero or inverse variances are smaller than 0.1) with a value equal to the median of flux for that star across all wavelengths (or for continuum-normalized spectra, like in the one used in this work, we set the flux to unity and the flux error to a large value --- namely, $9999$). For the labels, as \textsl{Lux} is able to include the uncertainty on the measurement in the training step of our model, we input the value and corresponding uncertainty for every label of every star. However, for stars with no stellar label measurement determined (i.e., label measurements that are missing or \texttt{NaN}), we set the value of the measurement for that star as the median of the distribution in the training sample, and then inflate its error to a very high value (namely, $\sigma_{\ell, n} = 9999$); during training, due to the large label uncertainty value for these stars will effectively be ignored by the likelihood function (we do not set this value to infinity because that leads to improper gradients of the likelihood).

\subsection{Train and test samples}
\label{sec_train_test_samples}

As we aim to assess how well the model performs across different regimes of the data, we divide our parent data set into multiple sub-samples that will either be used for training or testing. All the sub-samples we use are listed as follows:

\begin{description}
    \item[A. High-SNR field RGB-train] \hfill\\ 5,000 high signal-to-noise ($>100$ SNR) field red giant branch stars ($3,500 < T_{\mathrm{eff}} < 5,500$~K and $0 < \log~g <3.5$).
    \item[B. High-SNR field RGB-test] \hfill\\ 10,000 high signal-to-noise ($>100$ SNR) field red giant branch stars ($3,500 < T_{\mathrm{eff}} < 5,500$~K and $0 < \log~g <3.5$).
    \item[C. Low-SNR field RGB-test] \hfill\\ 5,000 low signal-to-noise ($30<$ SNR $<60$) field red giant branch stars ($3,500 < T_{\mathrm{eff}} < 5,500$~K and $0 < \log~g <3.5$).
    \item[D. High-SNR OC RGB-test] \hfill\\ 790 high signal-to-noise ($>100$ SNR) red giant branch stars ($3,500 < T_{\mathrm{eff}} < 5,500$~K and $0 < \log~g <3.5$) in nine open clusters, taken from the value added catalog from \citep[][]{Myers2022}. The open clusters these stars are associated with are: ASCC\,11, Berkeley\,66, Collinder\,34, FSR\,0496, FSR\,0542, IC\,166, NGC\,188, NGC\,752, and NGC\,1857.
    \item[E. High-SNR field all-train] \hfill\\ 4,000 high signal-to-noise ($>100$ SNR) red giant branch, main-sequence, and dwarf stars ($3,000 < T_{\mathrm{eff}} < 6,500$~K and $0 < \log~g <6$).
    \item[F. High-SNR field all-test] \hfill\\ 1,000 high signal-to-noise ($>100$ SNR) red giant branch, main-sequence, and dwarf stars ($3,000 < T_{\mathrm{eff}} < 6,500$~K and $0 < \log~g <6$).
    \item [G. \textsl{GALAH-APOGEE} field giants-train] \hfill\\ 4,000 medium signal-to-noise ($>50$ SNR$_{APOGEE}$) red giant branch stars ($3,800 < T_{\mathrm{eff}} < 6,000$~K and $0 < \log~g <3.5$) taken from a cross match between the \textsl{APOGEE} DR17 and \textsl{GALAH} DR3 surveys.
    \item [H. \textsl{GALAH-APOGEE} field giants-test] \hfill\\ 1,000 medium signal-to-noise ($>50$ SNR$_{APOGEE}$) red giant branch stars ($3,800 < T_{\mathrm{eff}} < 6,000$~K and $0 < \log~g <3.5$) taken from a cross match between the \textsl{APOGEE} DR17 and \textsl{GALAH} DR3 surveys.
\end{description}
Samples A--F all contain data solely from \textsl{APOGEE}, whereas samples G and H are comprised of overlapping stars between the \textsl{APOGEE} and \textsl{GALAH} surveys, and use spectral fluxes from \textsl{APOGEE} and stellar labels from \textsl{GALAH}. Moreover, to ensure that the field samples do not contain stars belonging to globular clusters, we remove known \textsl{APOGEE} globular cluster stars from the value added catalog \citet[][]{Schiavon2024} and the catalog from \citet{Horta2020}.

For samples A--C (i.e., those including only RGB field stars), we run \textsl{Lux} training and testing on twelve stellar labels (namely, $T_{\mathrm{eff}}$, $\log~g$, [Fe/H], [C/Fe], [N/Fe], [O/Fe], [Mg/Fe], [Al/Fe], [Si/Fe], [Ca/Fe], [Mn/Fe], [Ni/Fe]). For sample D (High-SNR OC RGB-test sample), we only test four labels: $T_{\mathrm{eff}}$, $\log~g$, [Fe/H], [Mg/Fe]. For samples E and F (those that contain RGB, MS, and dwarf stars), we run \textsl{Lux} using the following labels: $T_{\mathrm{eff}}$, $\log~g$, [Fe/H], [Mg/Fe], $v_{\mathrm{micro}}$ (microturbulent velocity), $v_{\mathrm{sin}i}$ (stellar rotation); these last two labels are included to enable the model to differentiate between an RGB, MS, and dwarf star. Lastly, for samples G and H (containing solely overlapping giant stars between the \textsl{GALAH} and \textsl{APOGEE} surveys), we train and test \textsl{Lux} using the following \textsl{GALAH} labels: $T_{\mathrm{eff}}, \log~g$, [Fe/H], [Li/Fe], [Na/Fe], [O/Fe], [Mg/Fe], [Y/Fe], [Ce/Fe], [Ba/Fe], and [Eu/Fe].

\section{The \textsl{Lux} model}
\label{sec_model}
In this Section we lay out the framework of \textsl{Lux} and discuss the choices we make for this implementation and demonstration of the model.
Our approach aims to infer a latent vector representation (embedding) $\boldsymbol{z}$ for each $n^{\mathrm{th}}$ star that is observed through transformation into stellar labels and spectral data (the outputs).
These transformations (from latent vector to outputs) can be arbitrarily complex functions with flexible parametrizations that are also inferred during the application of the model to data.
In the most general case, there may even be multiple label and spectral outputs to represent data from different surveys or data sources.
There could even be other representations such as broad-band photometry or kinematic information.
Here, however, we restrict to a model structure with a single label representation and a single spectral representation with linear transformations from latent vectors to these outputs.
In this form, the model has similar structure to an autoencoder \citep{bank2021}, but with no encoder and two decoders (that ``decode'' the latent representation into either stellar labels or spectral flux).
This model can also be thought of as a multi-task latent variable model \citep{Zhang2008}.

\subsection{Model structure}
\label{sec_model_struc}
In our fiducial implementation of \textsl{Lux}, we use linear transformations to compute the model predicted label values $\bs{\ell}$ and spectral fluxes $\bs{f}$.
Under this formulation, the observed stellar labels are generated as
\begin{align}
\label{eq_labels}
    \boldsymbol{\ell}_n &= \boldsymbol{A} \, \boldsymbol{z}_n + \textrm{noise}
\end{align}
where $\boldsymbol{\ell}_n$ represents the vector of labels (of length $M$) and $\boldsymbol{z}_n$ the latent parameters (of length $P$) for the $n^{\mathrm{th}}$ star.
Similarly, the observed stellar spectra (flux values) are generated as
\begin{align}
\label{eq_spectra}
    \boldsymbol{f}_n = \boldsymbol{B} \, \boldsymbol{z}_n + \textrm{noise}
\end{align}
where $\boldsymbol{f}_n$ represents the set of fluxes (of length $\Lambda$) for the $n^{\mathrm{th}}$ star.
For both outputs (labels and spectral flux), we assume that the noise is Gaussian with known variances.

For the stellar labels, this means that the likelihood of the observed label data for a star is
\begin{equation}
    p(\bs{\ell}_n \given \bs{A}, \bs{z}_n) =
        \mathcal{N}(\bs{\ell}_n \given \bs{A} \, \bs{z}_n, \sigma_{\bs{\ell}, n}^2) \label{eq:lAz}
\end{equation}
where $\mathcal{N}(x \given \mu, \sigma^2)$ represents the normal distribution over a variable $x$ with mean $\mu$ and variance $\sigma^2$, and $\sigma_{\bs{\ell}, n}$ represents the (\textsl{ASPCAP}) catalog reported uncertainties on the labels $\bs{\ell}$ for the $n^{\mathrm{th}}$ star.
For the spectral fluxes, the likelihood is similarly Gaussian such that
\begin{equation}
    p(\bs{f}_n \given \bs{B}, \bs{z}_n, \bs{s}_f) =
        \mathcal{N}(\bs{f}_n \given \bs{B} \, \bs{z}_n, \sigma_{\bs{f}, n}^2 + \bs{s}_f^2) \label{eq:fBz}
\end{equation}
where here $\sigma_{\bs{f}, n}$ represents the (\textsl{APOGEE}) per-pixel flux uncertainties and we include an additional variance per pixel $\bs{s}_f$ as a set of free parameters in the likelihood that is meant to capture the intrinsic scatter and any uncharacterized systematic errors in the spectral data (e.g., from sky lines).
In principle, we could add a similar ``extra variance'' to the stellar labels but from experimentation we have found this to be unneeded. However, the \textsl{Lux} software has been developed to be able to include per label scatter terms if wanted. 

Figure~\ref{fig:gpm} shows a graphical model representation of \textsl{Lux}.
To reiterate, $\boldsymbol{A}$ and $\boldsymbol{B}$ are the matrices that project the latent vectors, $\boldsymbol{z}_n$, onto stellar labels and stellar spectra for every $n^{\mathrm{th}}$ star, respectively. $\boldsymbol{A}$ and $\boldsymbol{B}$ are both rectangular matrices with dimensions $\boldsymbol{A} = [M \times P]$ and $\boldsymbol{B}=[\Lambda \times P]$. In this sense, $\boldsymbol{A}$ and $\boldsymbol{z}$ together contain all the information for inferring the stellar labels for all stars, $\boldsymbol{\ell}$. Similarly, $\boldsymbol{B}$ and $\boldsymbol{z}$ jointly contain all the information for producing the flux (spectra) for all stars, $\boldsymbol{f}$.

As we will show in the following Sections, this linear form for the transformations performs well in our demonstrative applications.
However, more complex transformations (e.g., Gaussian process or a neural network) would be more flexible and could be necessary for predicting other forms of output data.
We have formulated the \textsl{Lux} software so that it is straightforward to use more complex output transformation functions in future work.

\begin{table*}
	\centering
	\caption{Definitions, dimensionalities, and initializations for the parameters in the \textsl{Lux} model shown in Figure~\ref{fig:gpm}.}
	\begin{tabular}{lccccr} 
  		\hline
		Parameter & Definition & Dimensionality& Initialization\\
  		\hline
            \textit{M} & Stellar label dimensionality\\
            $\Lambda$ & Spectral flux dimensionality\\
            \textit{N} & Number of point sources, in this case stars\\
            \textit{P} & Latent variable dimensionality\\
            $\boldsymbol{\ell}$ & Stellar labels for all stars & [$M \times N$]&  \\
            $\boldsymbol{f}$ & Stellar fluxes for all stars & [$\Lambda \times N$]&  \\
            $\boldsymbol{\sigma_{\ell}}$ & Stellar label uncertainties for all stars & [$M \times N$]&  \\
            $\boldsymbol{\sigma_{f}}$ & Stellar flux uncertainties for all stars & [$\Lambda \times N$]&  \\
            $\boldsymbol{A}$ & Matrix that projects the latent parameters into stellar labels & [$M \times P$]& uniformly random $U(0,1)$\\
            $\boldsymbol{B}$ & Matrix that projects the latent parameters into stellar fluxes& [$\Lambda \times P$]& uniformly random $U(0,1)$\\
            $\boldsymbol{z}$ & Latent parameters & [$P \times N$]& using re-scaled label values, see Equation~\ref{eq_init}\\
            $\boldsymbol{s}$ & Vector of scatters in the model fit at every flux wavelength & [$\Lambda$] & $\ln \boldsymbol{s} = -8$\\

\hline
	\end{tabular}
 \label{tab:definitions}
\end{table*}

The full likelihood of the joint data (stellar labels and flux) for a given star $n$ is then the product of Equations~\ref{eq:lAz}--\ref{eq:fBz},
\begin{equation}
    p(\bs{\ell}_n, \bs{f}_n \given \bs{A}, \bs{B}, \bs{z}_n, \bs{s}_f) =
        p(\bs{\ell}_n \given \bs{A}, \bs{z}_n) \, p(\bs{f}_n \given \bs{B}, \bs{z}_n, \bs{s}_f)
\end{equation}
and we assume that the likelihood is conditionally independent per star, so the likelihood for a set of $N$ stars is the product of the per-star likelihoods
\begin{equation}
    \mathcal{L}(\bs{A}, \bs{B}, \{\bs{z}_n\}_N, \bs{s}_f) = p(\{\bs{\ell}_n\}_N, \{\bs{f}_n\}_N \given \bs{A}, \bs{B}, \{\bs{z}_n\}_N, \bs{s}_f) = \prod_n^N p(\bs{\ell}_n, \bs{f}_n \given \bs{A}, \bs{B}, \bs{z}_n, \bs{s}_f) \quad . \label{eq:full_like}
\end{equation}
At this point, we have specified a likelihood function for a sample of data (stellar labels and spectral fluxes), and we have the option to proceed probabilistically (i.e. by specifying prior probability distribution functions, \pdfs, over all parameters and working with the posterior \pdf) or to optimize the likelihood directly.

Whether optimizing the \textsl{Lux} likelihood or using it within a probabilistic setting, an important (and yet unspecified) hyperparameter of the model is the dimensionality of the latent space, $P$.
This parameter will ultimately control the flexibility of the model: With too small of a value, the model will not be able to represent the data even with arbitrarily complex transform matrices $\bs{A}$ and $\bs{B}$, but with too large of a value, the model risks over-fitting.
Anecdotally, we have found that values for $P$ that are larger than the label dimensionality $M$ but smaller than the number of pixels in your stellar spectra $\Lambda$ (i.e. $M < P < \Lambda$) seem to perform well.
We discuss how to set this parameter using cross-validation in our application of the \textsl{Lux} model below.

\subsection{Inferring parameters of a \textsl{Lux} model}

Given the large number of parameters in \textsl{Lux}, a standard approach is to optimize the likelihood (Equation~\ref{eq:full_like}).
In this context, we optimize the likelihood on a set of training data and then apply the model to held-out test data.
That is, we use the training data to infer parameters $\bs{A}$, $\bs{B}$, and $\bs{s}$ (and the latent vectors $\bs{z}_n$ for the training set stars), and then use the model with a test data set in which we use the stellar fluxes to infer latent vectors and project into stellar labels, or vice versa.
This ends up being an efficient way of using the model to determine stellar labels for test set stars and is analogous to how models like the \textsl{Cannon} operate.
However, as mentioned above, \textsl{Lux} is a generative model and we could instead have put prior \pdfs on all parameters and hyper-parameters and proceeded with all available data by approaching the model training and application simultaneously as a hierarchical inference.
This approach is substantially more computationally intensive, and we therefore leave this for future exploration.

In our experiments with this form of the \textsl{Lux} model, we have found it helpful to include a regularization term in our optimization of the log-likelihood function.
We have found that \textsl{Lux} performs better on held-out test data if we optimize the log-likelihood of the training data with an L2 regularization term on the latent vectors $\bs{z}_n$, so that our objective function over all parameters, $g$, is
\begin{equation}
    g(\bs{A}, \bs{B}, \{\bs{z}_n\}_N, \bs{s}_f) = \ln\mathcal{L} + \Omega \sum_n^N \sum_p^P z_{pn}^2
\end{equation}
where the sum in the regularization term is done over the $P$ latent vector values for all $N$ stars with regularization strength $\Omega$.
Expanding the log-likelihood function, our objective function is
\begin{equation}
    g(\bs{A}, \bs{B}, \{\bs{z}_n\}_N, \bs{s}_f) =
    \sum_n^N\left[
        \ln p(\bs{\ell}_n \given \bs{A}, \bs{z}_n)
        + \ln p(\bs{f}_n \given \bs{B}, \bs{z}_n, \bs{s}_f)
        + \Omega \sum_p^P z_{pn}^2
    \right] \label{eq:full-objective}
\end{equation}
We choose L2 over L1 regularization because L1 is known to favor stricter sparsity in the regularized parameters, and we want to instead encourage sparsity in the mapping matrices $\bs{A}$ and $\bs{B}$.
In more detail, if a given latent vector dimension does not interact with either the labels or fluxes, the model optimization can enforce this by either setting relevant elements of the matrices $\bs{A}$ or $\bs{B}$ to zero, or by nulling out values in the latent vector $\bs{z}$.
To weaken this degeneracy, we instead opt for L2 regularization, which can also prefer sparsity but tends instead to make parameters more equal in scale.

\section{Results: An application with \textsl{APOGEE} data}\label{sec_model_res}

In this Section, we apply \textsl{Lux} to \textsl{APOGEE} data to showcase the model's capacity for determination of precise stellar labels and spectra across a wide range of stellar label space.
As mentioned above, we proceed here by optimizing the \textsl{Lux} likelihood given a training set of data and then use the model to predict stellar labels or spectra for a test set, to assess performance.

In more detail, we first train two \textsl{Lux} models. The first is trained on the high-SNR field RGB-train sample (5,000 RGB stars) and tested on the high-SNR field RGB-test sample, the low-SNR field RGB-test sample, and the high-SNR OC RGB-test sample (also RGB stars), see Section~\ref{sec_data}. The second model is trained on the high-SNR field all-train sample (4,000 RGB, MS, and dwarf stars), and is tested on the high-SNR field all-test sample (also RGB, MS, and dwarf stars). The aim of this exercise is to assess: 1) how well our model is able to determine stellar labels for a given stellar type across multiple SNR regimes (tests on the high-SNR field RGB-test and low-SNR field RGB-test samples); 2) how well our model compares to stars benchmark objects like open clusters (test on the high-SNR OC RGB-test sample); 3) how well our model is able to simultaneously infer stellar labels across different stellar types (samples high-SNR field all-train and high-SNR field all-test).
For the first model, we use twelve stellar labels: $T_{\mathrm{eff}}$, $\log~g$, [Fe/H], [C/Fe], [N/Fe], [O/Fe], [Mg/Fe], [Al/Fe], [Si/Fe], [Ca/Fe], [Mn/Fe], and [Ni/Fe].
For the second model, we use $T_{\mathrm{eff}}$, $\log~g$, [Fe/H], [Mg/Fe], $v_{\mathrm{micro}}$, and $v_{\mathrm{sin}i}$.

The choice of the latent dimensionality, $P$, and regularization strength, $\Omega$, are hyper-parameters of \textsl{Lux}. For this application, we set these values with a $K$-fold cross validation. We have tested $P=[1,2,4,8] \times M$, where $M$ is the number of labels, and $\Omega = [1, 10^{1}, 10^{2}, 10^{3}]$ (see Section~\ref{app_kfold} for details). After performing the $K$-fold cross-validation, we choose to adopt $P=4\,M$ and $\Omega=10^{3}$.

\subsection{Initialization of the latent parameters}

In order to optimize the parameters in \textsl{Lux}, $\boldsymbol{A}, \boldsymbol{B}, \boldsymbol{z}_n$, and scatter in the flux pixels, $\boldsymbol{s}$, we must first initialize them.
We initialize $\boldsymbol{A}$ and $\boldsymbol{B}$ randomly from a uniform distribution over $[0,1]$ with shapes $\boldsymbol{A} = [M \times P]$ and $\boldsymbol{B} = [\Lambda \times P]$.
For the latent vectors $\boldsymbol{z}_n$, we initialize following a similar procedure to the pre-computation of feature vectors described in the \textsl{Cannon} model \citep[][]{Ness2015}.
In more detail, we resize all the labels by a given centroid and scale equal to the $50^{\mathrm{th}}$ and $(97.5^{\mathrm{th}} - 2.5^{\mathrm{th}})/4$ percentile values of the sample distribution, respectively\footnote{We divide $(97.5^{\mathrm{th}} - 2.5^{\mathrm{th}})/4$ by four as the $95^{\mathrm{th}}$ percentile range is approximately $\sim4~\sigma$.}. This is appropriate for two main reasons: $i$) it is safe numerically, as some labels may have scales around $10^{3}$ whilst others may have relevant scales around $10^{-1}$. This way, all values are around unity; $ii$) because we are assuming a linear relationship between (functions of) the stellar labels and the spectral fluxes (as done in the \textsl{Cannon} \citep[][]{Ness2015} that is very successful in these kinds of tasks). Using these re-scaled label values, we initialize the latent vectors for each star, $\boldsymbol{z}_n$, as
\begin{equation}
\label{eq_init}
    \boldsymbol{z}_n = [1, (\ell_{m_{1}} - c_{m_{1}})/d_{m_{1}}, \cdots, (\ell_{m_{M}} - c_{m_{M}})/d_{m_{M}}, 0, 0, \cdots, 0]
\end{equation}
where the first element will permit a linear offset, and $c_{m_{1}}$ and $d_{m_{1}}$ are the centroid and scale of each label and training data set, respectively.
This initialization of $\boldsymbol{z}_n$ requires that $P~\geq~M +1$ (i.e. the latent space is always larger than the label space, where the $+1$ corresponds to the unity value in Equation~\ref{eq_init}).
We set all values of $\boldsymbol{z}_n$ beyond the dimension of $M$ to 0.
Lastly, we initialize the scatters at all fluxes/pixels, $\boldsymbol{s}$, as a very small number (namely, $\ln~\boldsymbol{s} = -8$). A summary of the model parameter definitions, dimensionalities, and initializations are provided in Table~\ref{tab:definitions}.

\subsection{Training step} \label{sec:apogee-train}

\begin{figure}
    \centering
    \includegraphics[width=0.5\columnwidth]{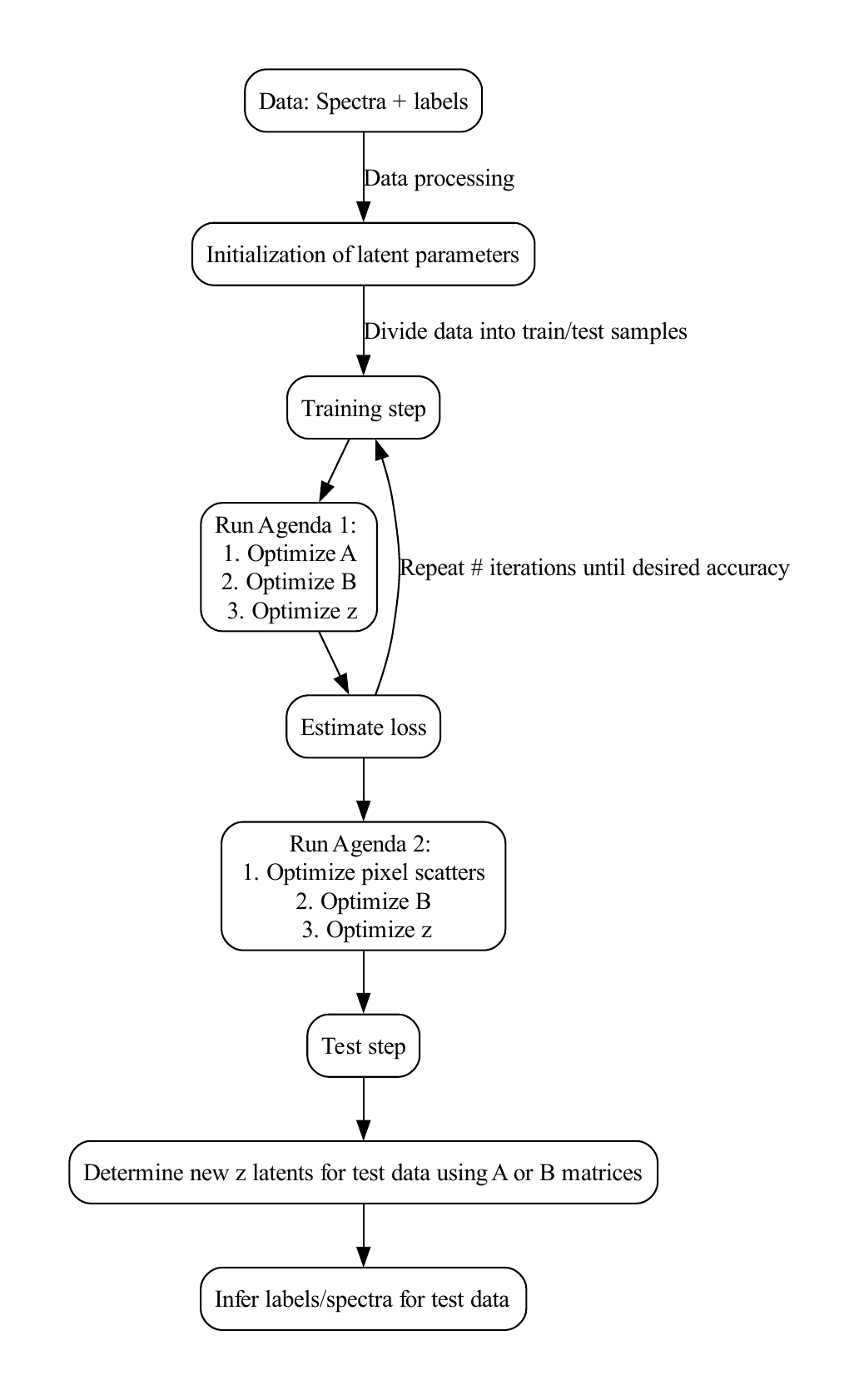}
    \caption{
    A Flowchart summarizing the optimization scheme for our application of the linear \textsl{Lux} model to \textsl{APOGEE} data, described in Section~\ref{sec:apogee-train}.
    \label{fig:flowchart}
    }
\end{figure}

The training step of \textsl{Lux} consists of running a two-part procedure.
In the first part (Agenda 1), we optimize the parameters $\boldsymbol{A}, \boldsymbol{B}, \boldsymbol{z}_n$, without any regularization.
We do this using a custom multi-step optimization scheme.
The first step (the $a$-step) optimizes $\boldsymbol{A}$ using the stellar label data at fixed $\boldsymbol{z}_n$; the second step ($b$-step) optimizes $\boldsymbol{B}$ using stellar flux data at fixed $\boldsymbol{z}_n$; the third step ($z$-step) then optimizes $\boldsymbol{z}_n$ using the stellar label and stellar spectra data at fixed (and newly optimized) $\boldsymbol{A}$ and $\boldsymbol{B}$. Here, the optimization in all three steps is performed assuming there is no scatter in the fluxes/pixels and no regularization.
In the case of a linear \textsl{Lux} model (as we use here), these optimization steps can be done using closed-form least-squares solutions.
However, for future generalizations, we instead use a Gauss--Newton least squares solver for each step (using the \texttt{GaussNewton} solver in \texttt{JAXopt}; \citealt[][]{jaxopt_implicit_diff}).\footnote{Even though it is overkill to use a nonlinear least-squares solver for this particular model form, we have found that the solutions converge very fast here because the Hessian is tractable and can be computed exactly with \texttt{JAX}.}
A run through the $a$, $b$, and $z$ steps completes one iteration. After testing different numbers of iterations and inspecting the accuracy of the model, we have found that the model reaches a plateau after five iterations\footnote{Model accuracy is calculated by computing a $\chi^{2}$ metric summed over all labels, fluxes, and stars.}. Thus, we run this first agenda for five iterations.

In the second part (Agenda 2), we first optimize the pixel (flux) scatters, $\boldsymbol{s}$, at fixed $\boldsymbol{B}$ and $\boldsymbol{z}_n$ using stellar flux data.
We then again optimize $\boldsymbol{B}$ and $\boldsymbol{z}_n$ to account for noise in the stellar spectral fit by re-optimizing $\boldsymbol{B}$ at fixed $\boldsymbol{z}_n$ and $\boldsymbol{s}$ using stellar flux data, and optimizing $\boldsymbol{z}_n$ at fixed $\boldsymbol{A}$, $\boldsymbol{B}$, and $\boldsymbol{s}$ using the stellar flux and stellar label data.
When performing this final optimization, we add an L2 regularization (Equation~\ref{eq:full-objective}).
A run through the optimization of $\boldsymbol{s}$ and the updated $\boldsymbol{B}$ and $\boldsymbol{z}_n$ latent variables completes a run through the second agenda.
For this step, we use the \texttt{LBFGS} solver (also in \texttt{JAXopt}) including the L2 regularization from Equation~\ref{eq:full-objective}; we switch to \texttt{LBFGS} because the problem is no longer a least squares problem with varied flux scatters $s$.
We set the following hyperparameters in the \texttt{LBFGS} optimizer: \texttt{tol}, \texttt{maxiter}, and \texttt{max$_{-}$stepsize} to $1^{-6}, 3\times10^{3}$, and $1\times10^{3}$, respectively.
We choose to only run through this second agenda once, but in principle this step could also be iterated.

\textsl{Lux} has very large capacity and strong degeneracies between the transform parameters and the latent vector values, so we have found that this two-step, highly structured optimization scheme leads to model parameters that predict well on held-out data, as we describe next.
A flowchart of this optimization scheme is depicted in Figure~\ref{fig:flowchart}.

\subsection{Test step}

Once \textsl{Lux} parameters ($\boldsymbol{A}, \boldsymbol{B}, \boldsymbol{z}_n$) and the scatter in the fluxes, $\boldsymbol{s}$, are optimized using the training set data, we can use \textsl{Lux} to predict labels (Equation~\ref{eq_labels}) or predict spectra (Equation~\ref{eq_spectra}) given the optimized latent vectors for the training set $\bs{z}_n$.
To use \textsl{Lux} with a test set (i.e. data not included in the training) with held out labels or spectra, we must first determine the corresponding latent vectors for the test set stars.

For evaluating the performance of \textsl{Lux} on the test data, we have several options.
One option is to use the spectra or labels to determine the latent vectors of the test set, and then use the latent vectors to again predict the spectra or labels.
Interestingly, due to the multi-task nature of the \textsl{Lux} model, we could also instead use the spectra to determine the latent vectors and then evaluate the accuracy of the predicted labels, or vice versa.

To determine the stellar labels using the test set stellar fluxes, we optimize the latent vectors $\boldsymbol{z}_n$ for stars in the test data set at fixed $\bs{B}$ and $\bs{s}$ (from the training step), using the fluxes $\boldsymbol{f}$ and uncertainties $\boldsymbol{\sigma}_{f}$ for the test set to optimize only the $\boldsymbol{z}_n$ term in our objective function.
That is, we find the latent vectors for the test set by optimizing the objective
\begin{equation}
    g(\{\bs{z}_n\}_N) = \sum_n^N \ln p(\bs{f}_n \given \bs{B}, \bs{z}_n, \bs{s}_f)
    \label{eq_teststep_B} \quad .
\end{equation}
We perform this optimization again using the \texttt{LBFGS} optimizer.
With the latent vectors for the test set, we can then predict stellar labels for the test set using Equation~\ref{eq_labels}.
We can alternatively optimize for the latent vectors from the stellar labels and then use the trained $\bs{A}$ to predict stellar spectra, by optimizing
\begin{equation}
    g(\{\bs{z}_n\}_N) = \sum_n^N \ln p(\bs{\ell}_n \given \bs{A}, \bs{z}_n)
    \label{eq_teststep_A}
\end{equation}
which operates more like a spectral emulator. In our test set evaluations below, we perform tests in both directions.

\subsection{Accuracy of predicted stellar spectra}
\label{sec_res_spectra}

\begin{figure*}
    \centering
    \includegraphics[width=1\textwidth]{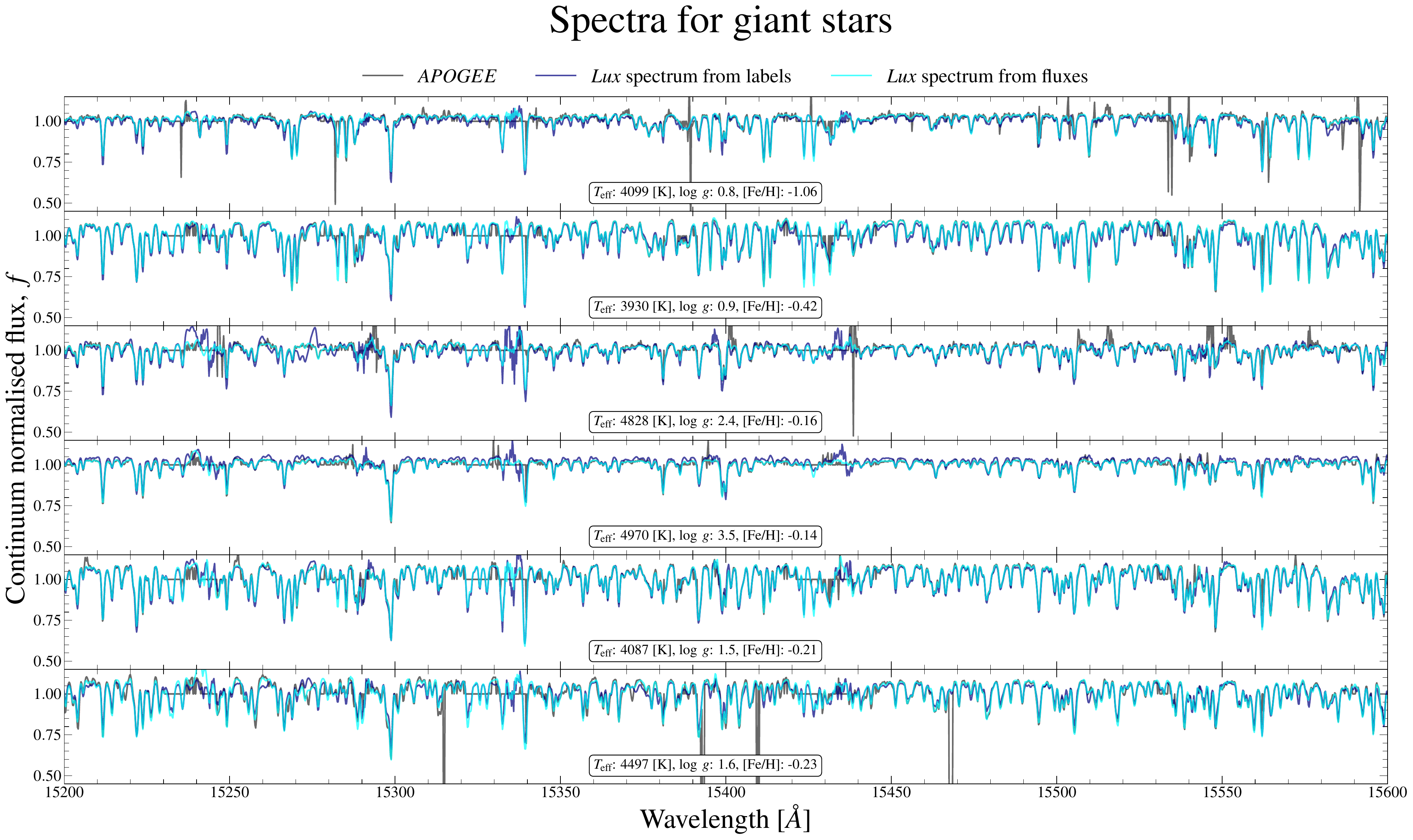}
    \caption{
    A comparison of \textsl{APOGEE} spectra (black lines) and predicted spectra from \textsl{Lux} (navy and cyan lines) for six stars (each panel is for one star).
    We use \textsl{Lux} to predict stellar spectra in two ways.
    First, we use the stellar labels for each star to infer latent representations ($\boldsymbol{z}_n$) for the test objects, and then use the model to transform latent vectors to stellar flux (navy lines).
    Then, as a self-test of the model, we use the spectral fluxes for each star to infer latent representations again and transform back to spectral fluxes (cyan lines).
    The six stars chosen are from the high-SNR fiend RGB-test sample: \texttt{2M00204904-7133587}, \texttt{2M00055986+6850184}, \texttt{2M00002926+5737333}, \texttt{2M00101602+0154424}, \texttt{2M00052512+5642535}, and \texttt{2M01233689+6233205}). Each row includes stars with either similar $T_{\mathrm{eff}}, \log~g$, or [Fe/H]. 
    These results demonstrate that \textsl{Lux} is able to accurately predict (and impute) stellar spectra for test objects across a wide range of stellar parameter space, either when using the stellar labels or stellar spectra in the test set.}

    \label{fig:spectra}
\end{figure*}

\begin{figure}
    \centering
    \includegraphics[width=0.5\columnwidth]{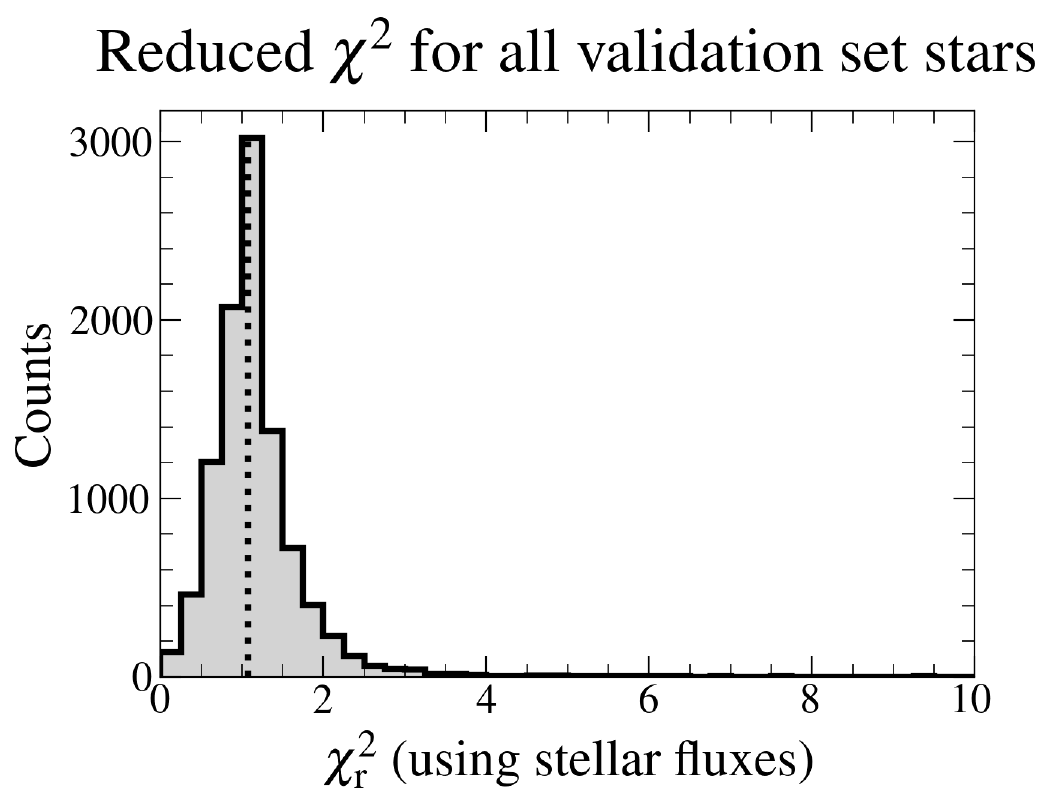}
    \caption{Reduced $\chi^{2}$ value for all stars in the validation/test set using the stellar fluxes. The median value (dotted line) is 1.07. These stellar fluxes are generated from the latent vectors inferred from the stellar spectra themselves. $\chi^{2}_{\mathrm{r}}$ is computed as $\chi^{2}$ divided by the number of pixels in the spectrum. The majority of our test objects yield a $\chi^{2}_{\mathrm{r}}$ around unity, indicating that the \textsl{Lux} model is a good fit and the extent of the match between the observed (\textsl{APOGEE}) stellar spectra and estimates from the \textsl{Lux} model are in accord with the error variance.}
    \label{fig:reduced_chi2}
\end{figure}

\begin{figure*}
    \centering
    \includegraphics[width=1\textwidth]{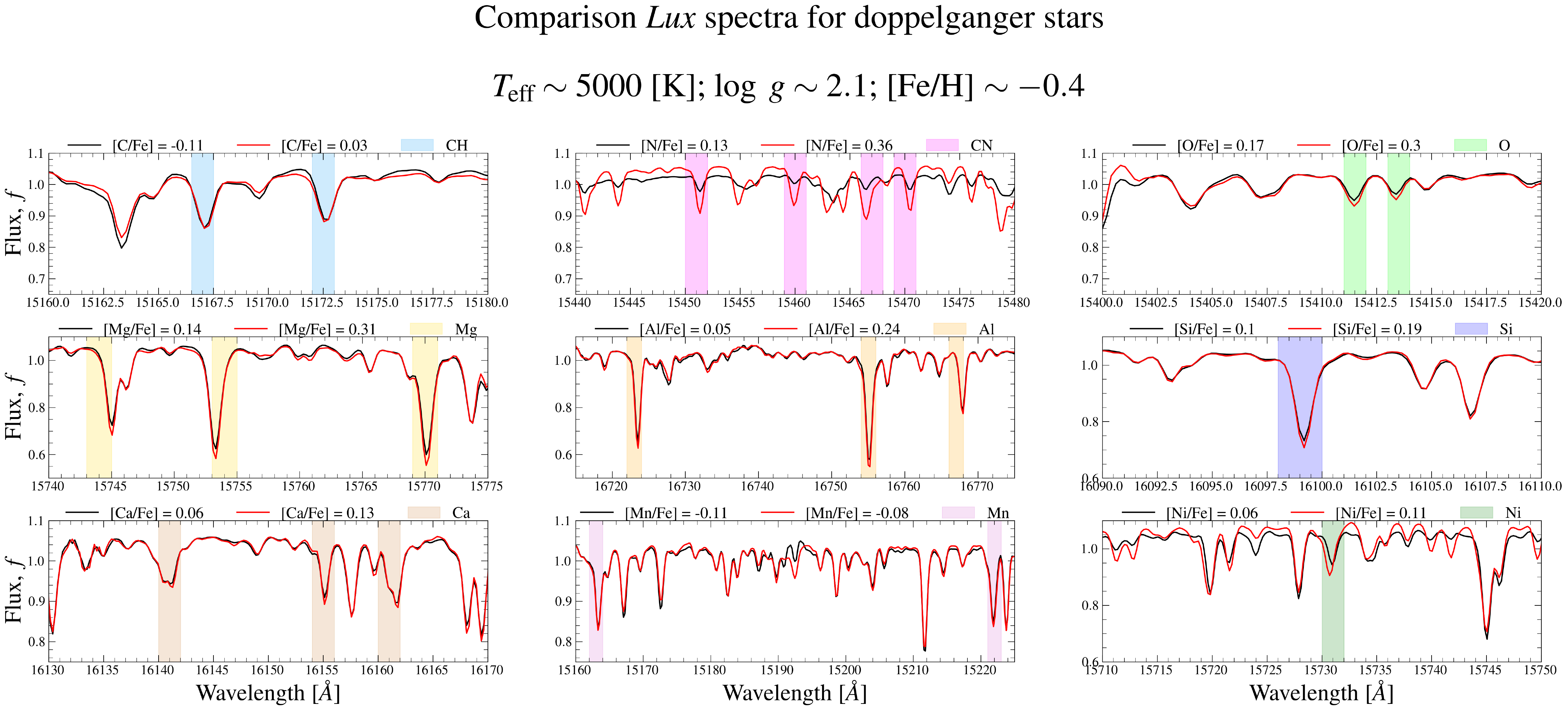}
    \caption{Comparison of portions of the spectrum for two stars (see legend) in the high-SNR field RGB-test sample with similar $T_{\mathrm{eff}}, \log~g$, and [Fe/H] but different individual chemical abundance ratios. The stellar spectra shown in each panel are the \textsl{Lux} model spectra for those stars. To determine these spectra, we use the spectral fluxes for each star to infer the latent representations ($\boldsymbol{z}_n$), and then impute spectral fluxes (using Equation~\ref{eq_spectra}). For the case of [C/Fe] and [N/Fe], as these elements are determined from the CH and CN molecular lines, we also constrain the comparison to two stars with similar [N/Fe] abundance when examining [C/Fe], and two stars with similar [C/Fe] when examining [N/Fe]. For all element abundance ratios examined, stars with a higher [X/Fe] abundance present deeper absorption lines when compared to their doppelganger star with lower [X/Fe] at the location of individual atomic/molecular lines. These results illustrate how \textsl{Lux} is able to accurately determine spectra for doppelganger stars (i.e., approximately the same $T_{\mathrm{eff}}, \log~g$, and [Fe/H]) with different individual abundances, and thus accurately identifies the spectral features associated with a given stellar label.}
    \label{fig:comp-spectra}
\end{figure*}

\begin{figure*}
    \centering
    \includegraphics[width=1\textwidth]{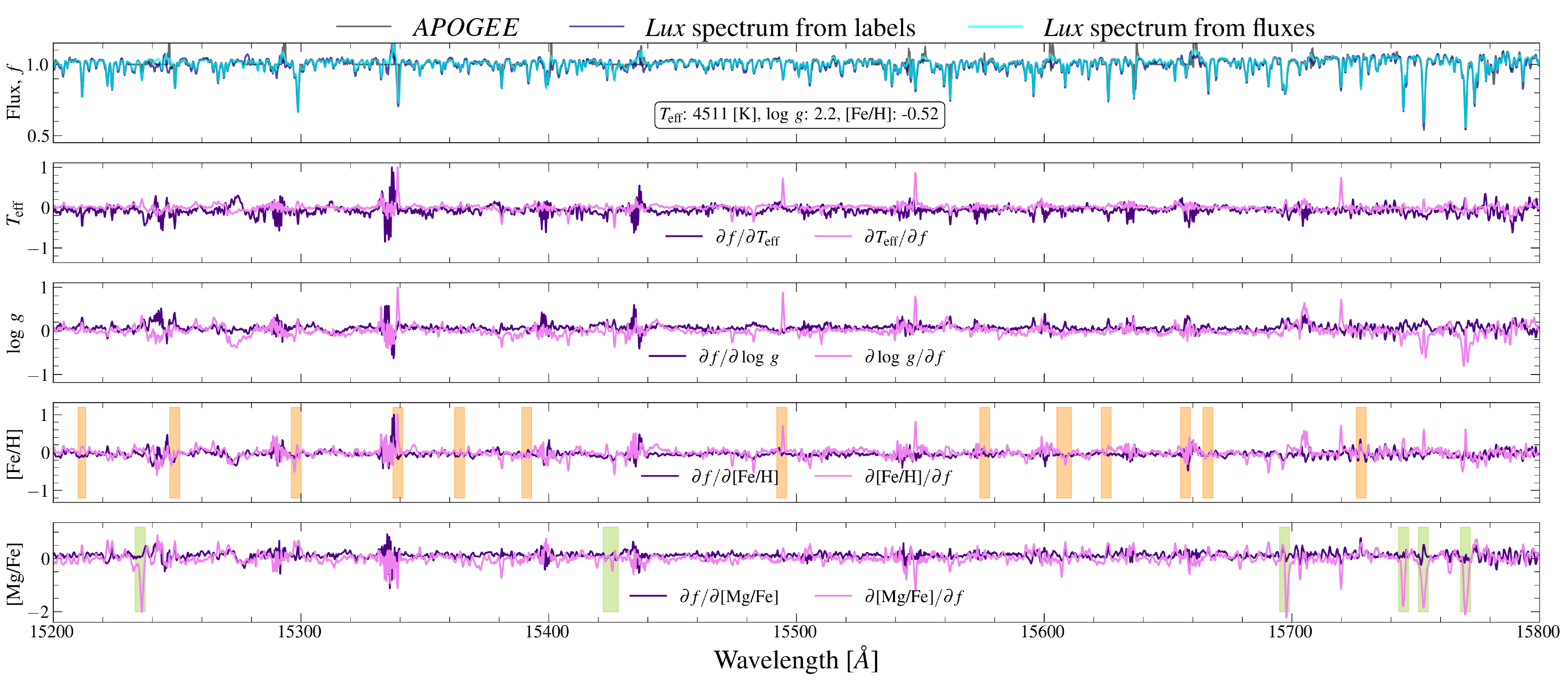}
    \caption{\textbf{Top}: Portion of the \textsl{APOGEE} spectrum (black) and inferred \textsl{Lux} spectra for a random star (\texttt{2M00011399+8408446}) from our High-SNR field RGB-test set. The meaning of the cyan and navy spectra shown in the top panel is the same as in Figure~\ref{fig:spectra}. \textbf{Other rows}: derivatives of the spectrum w.r.t. $T_{\mathrm{eff}}, \log~g$, [Fe/H], and [Mg/Fe], computed using Eq.~\ref{eq:dfdl} and Eq.~\ref{eq:dldf}. Highlighted with colored bands in the bottom two rows are the main atomic Fe I (orange) and Mg I (green) lines in this portion of the spectrum. If one focuses on the bottom two rows, the $\partial\,$[X/Fe]/$\partial\,f$ derivative shows large values around the atomic lines for that particular element. This result illustrates how the \textsl{Lux} model is learning the correct portions of the spectrum for a corresponding label (see text in Section~\ref{sec_res_spectra} for further details). We note that some of the bad sky subtraction features that have been seen before in this region of the \textsl{APOGEE} spectra are visible (e.g., $15343~\AA$, \citealp[][]{mckinnon2024}).}
    \label{fig:spectra-derivatives}
\end{figure*}

Figure~\ref{fig:spectra} shows a comparison between the observed \textsl{APOGEE} spectral data (black) and the spectra predicted by \textsl{Lux} using fluxes(labels) as cyan(navy).
That is, for one case we use the spectral fluxes to determine the latent vectors of the test set, and then use the trained $\bs{B}$ to project back into flux (cyan line).
In the other case, we use the stellar labels to determine the latent vectors of the test set, and then again project into spectral flux (navy line).
Here, we show the data and the model spectra for six stars from the High-SNR field RGB-test sample; the top two rows are two stars with similar $T_{\mathrm{eff}}$ and $\log~g$ but different [Fe/H], the middle two rows are two stars with similar $T_{\mathrm{eff}}$ and [Fe/H] but different $\log~g$, and the bottom two rows are two stars with similar $\log~g$ and [Fe/H] but different $T_{\mathrm{eff}}$. Overall, \textsl{Lux} yields realistic spectra that matches well the observed spectra for a wide range of stars across the Kiel diagram; this is the case when the spectra are determined either using the stellar fluxes or stellar labels of the test set. Interestingly, as with other data-driven spectral models, \textsl{Lux} is able to impute stellar spectra in particular wavelength windows where the observed \textsl{APOGEE} spectra show strong sky lines or missing data.

To quantify how well \textsl{Lux} is able to generate stellar spectra, we compute the reduced $\chi^{2}$ value across all stellar fluxes for each star in the test set (i.e., $\chi^{2}$/number of pixels in the spectrum), shown in Figure~\ref{fig:reduced_chi2}.
For this test, we use spectra generated from latent vectors inferred from the stellar spectra themselves. We find that the majority of the values are around unity, indicating that \textsl{Lux} model is a good fit to the data and the extent of the match between observed (\textsl{APOGEE}) stellar spectra and estimates from the \textsl{Lux} model is in accord with the error variance.

Along those lines, the inferred \textsl{Lux} spectra capture well the information in the stellar labels, at least visually. Figure~\ref{fig:comp-spectra} illustrates a comparison of spectra for two random doppelganger stars (i.e. stars with similar stellar labels) in the high-SNR field RGB-test sample. Each panel shows a portion of the spectrum for the two stars that have similar $T_{\mathrm{eff}}, \log~g$, and [Fe/H], but different particular chemical abundances, from [C/Fe] in the top left to [Ni/Fe] in the bottom right\footnote{For the case of [C/Fe] and [N/Fe], as these elements are determined from the CH and CN molecular lines, we also constrain the comparison to two stars with similar [N/Fe] abundance when examining [C/Fe], and two stars with similar [C/Fe] when examining [N/Fe].}. Our aim with this illustration is to show how \textsl{Lux} is able to accurately determine spectra for doppelganger stars with different individual element abundance ratios. Each panel of Figure~\ref{fig:comp-spectra} shows the portion of the spectrum where some of the main atomic/molecular lines are used in \textsl{ASPCAP} to determine the species on the numerator of the element abundance ratio. We find that at the location of individual atomic/molecular lines, the absorption line in the \textsl{Lux} spectrum corresponding to the star with enhanced [X/Fe] is deeper than for the star with lower [X/Fe]. This result highlights how \textsl{Lux} is able to accurately identify the spectral features associated with a given stellar label.

Furthermore, in Figure~\ref{fig:spectra-derivatives} we show the derivative spectrum for one random star from our high-SNR field RGB-test set with respect to four labels: $T_{\mathrm{eff}}, \log~g$, [Fe/H], and [Mg/Fe]. For completeness, in the top row we also show its \textsl{APOGEE} spectrum (black), its \textsl{Lux} spectrum determined using stellar fluxes to infer the latent representations (cyan), and its \textsl{Lux} spectrum determined using stellar labels to infer the latent representations (navy). Illustrated in this figure in the bottom two rows are also the main Fe I and Mg I atomic lines for this wavelength range of the spectrum.
This figure shows that the inferred \textsl{Lux} spectra, determined using either fluxes or labels, captures well the atomic absorption lines, as there are large derivative values at the location of individual atomic lines.
We note that the \textsl{ASPCAP} line windows are conservative, in that there could be other lines for a given species along the spectral dimension that are blended or overlap with other lines that do not show up as line windows (which may explain some of the other spectral variations and structure in the derivatives).

The derivative spectra provide one means for interpreting how the model is learning dependencies between the spectral fluxes and labels.
Naively, we want to inspect derivatives of the spectral flux with respect to stellar labels to see if \textsl{Lux} learns that certain regions of the spectrum depend strongly on given labels (e.g., the trained model should have larger derivatives around spectral lines of a given species when looking at the derivatives with respect to element abundance ratios).
However, unlike \textsl{The Cannon}, in which the flux values are predicted directly as a function of the labels, \textsl{Lux} generates both fluxes and labels from the inferred latent vectors $\boldsymbol{z}_n$.
To compute the derivatives of interest, we therefore want to inspect rows of the derivative matrix
\begin{equation}
    \frac{\partial \bs{f}}{\partial\bs{\ell}} =
        \frac{\partial\bs{f}}{\partial\bs{z}} \, \frac{\partial\bs{z}}{\partial\bs{\ell}}
    = \boldsymbol{B} \cdot \boldsymbol{A}^{+} \label{eq:dfdl}
\end{equation}
where $\bs{A}^{+}$ is the pseudoinverse of $\bs{A}$.
We have found that this path towards estimating the derivatives is unstable due to the pseudoinverse of $\bs{A}$: this matrix compresses the latent vectors into labels (i.e. $M < P$), so the inverse mapping attempts to expand from the label dimensionality $M$ up to the latent dimensionality $P$.
We therefore instead compute the derivatives in the other direction,
\begin{equation}
    \frac{\partial \bs{\ell}}{\partial\bs{f}} =
        \frac{\partial\bs{\ell}}{\partial\bs{z}} \, \frac{\partial\bs{z}}{\partial\bs{f}}
    = \boldsymbol{A} \cdot \boldsymbol{B}^{+}
    \label{eq:dldf}
\end{equation}
which instead involves the pseudoinverse of $\bs{B}$; We expect this to preserve the information flow better because $\Lambda > P$.
We therefore visualize columns of this Jacobian matrix (Equation~\ref{eq:dldf}).
In the following three rows of Figure~\ref{fig:spectra-derivatives} we show these derivatives corresponding to four labels ($T_{\mathrm{eff}}$, $\log~g$, [Fe/H], and [Mg/Fe]) as pink lines.
Encouragingly, the derivative spectra show features with the correct signs at the Fe I lines (fourth panel from top) and  for the Mg I lines (bottom panel). We have also checked this is the case with other elements (e.g., Al and Mn).
This suggests that the \textsl{Lux} model is correctly learning the locations in the spectral flux data that are relevant to each stellar label.

\subsection{Accuracy of predicted stellar labels}
\label{sec_res_labels}

\begin{figure*}
    \centering
    \includegraphics[width=1\textwidth]{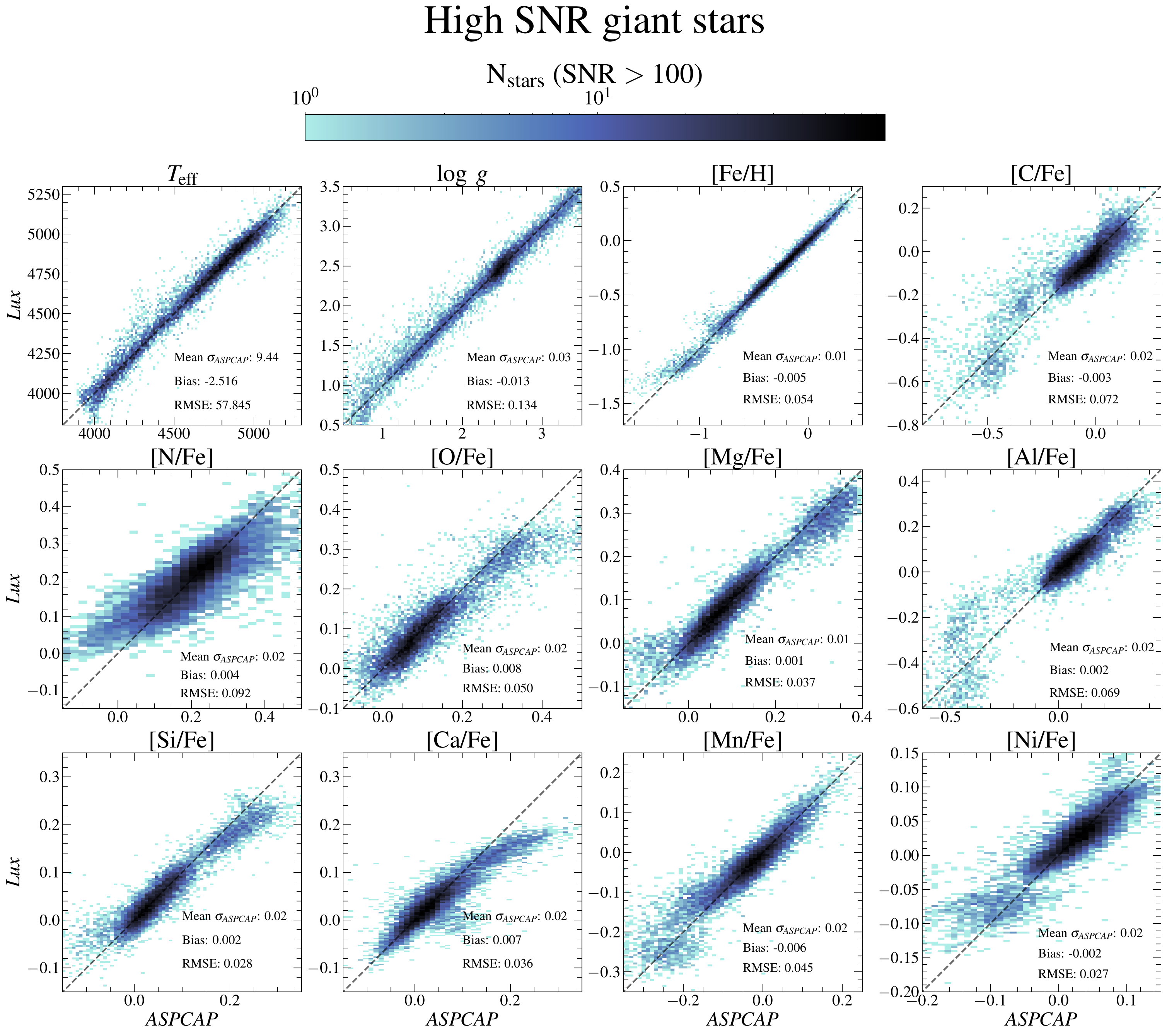}
    \caption{Validation for the high-SNR field RGB-test set stars using the twelve labels trained/tested on. 
    Each panel shows the \textsl{Lux}-predicted label values plotted against the \textsl{ASPCAP} values. The \textsl{Lux} labels are determined by optimizing the test set latent representations using each star's spectral fluxes. Each panel also lists the mean \textsl{ASPCAP} uncertainty for that stellar label ($\sigma_{\rm ASPCAP}$), the mean difference between the \textsl{Lux} model labels and the \textsl{ASPCAP} labels (bias), and the root-mean-squared error between the labels (RMSE). For all labels, the bias is smaller than the average reported \textsl{ASPCAP} uncertainty, indicating that the \textsl{Lux} model is emulating the \textsl{ASPCAP} pipeline well. Similarly, the RMSE value is small across all labels, and for most labels is comparable to the average \textsl{ASPCAP} uncertainty; this indicates that the \textsl{Lux} model is able to robustly estimate stellar labels for high signal-to-noise RGB stars. See text in Section~\ref{sec_res_labels} for further details.}
    \label{fig:cv_highsnr}
\end{figure*}

\begin{figure*}
    \centering
    \includegraphics[width=1\textwidth]{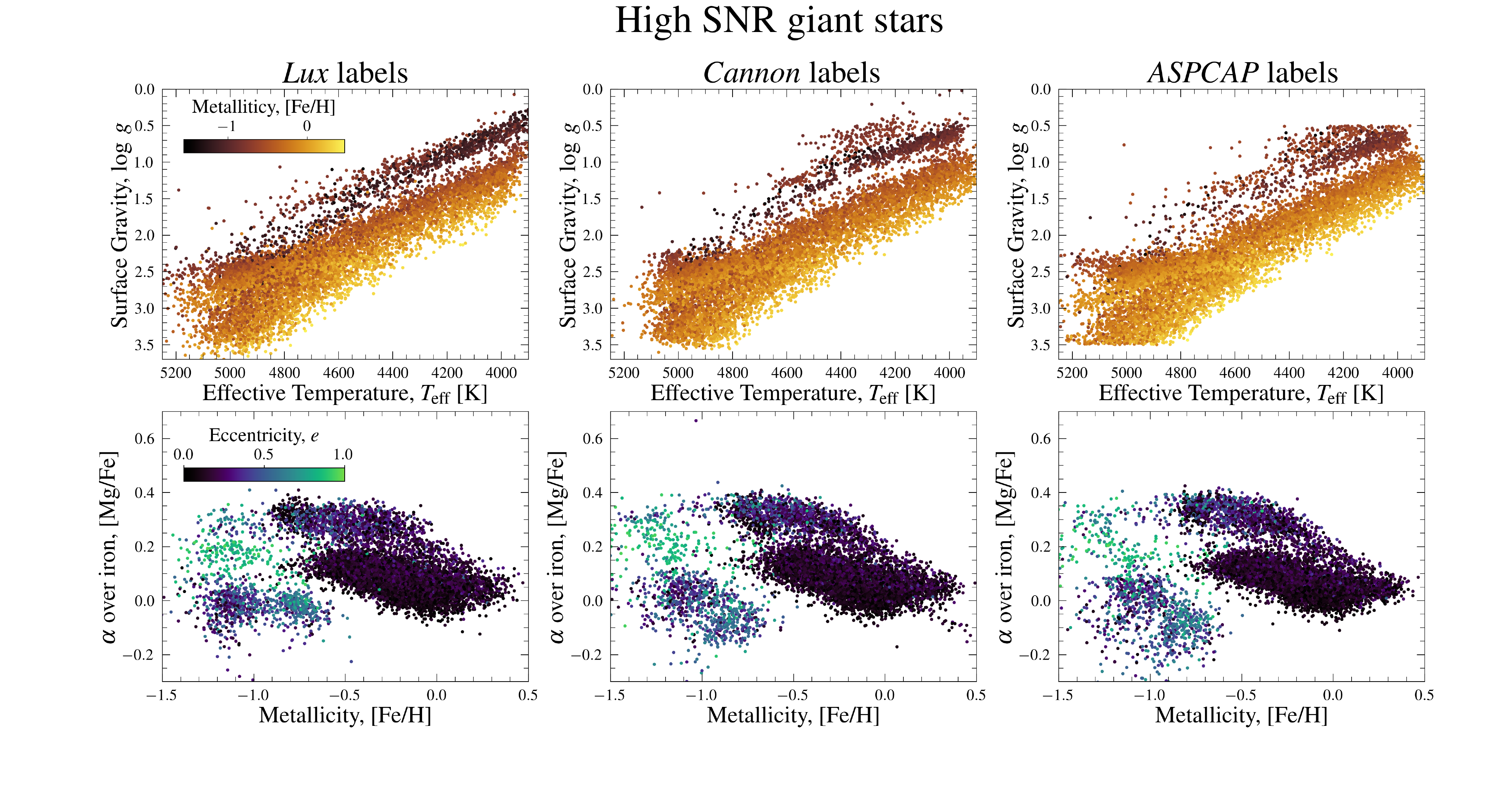}
    \caption{Kiel (top) and Tinsley-Wallerstein (bottom) diagrams for stars in the high-SNR field RGB-test sample. We show \textsl{Lux} labels (left), the labels computed using the \textsl{Cannon} \citep{Ness2015} assuming a quadratic relationship in the labels (middle), and the labels from \textsl{ASPCAP} (right). The \textsl{Lux} labels are determined by optimizing the test set latent representations using each star's spectral fluxes. The samples look qualitatively similar, highlighting that our model works well. However, small differences can be seen in the details; \textsl{Lux} labels appear tighter and show less scatter than the \textsl{Cannon} labels, which in turn appear tighter and show less scatter again than the \textsl{ASPCAP} labels. See text in Section~\ref{sec_res_labels} for further details.}
    \label{fig:kiel-tinsley}
\end{figure*}

Another test we perform is to assess how well \textsl{Lux} is able to predict stellar labels of high signal-to-noise ratio (SNR) red giant branch (RGB) stars from the \textsl{APOGEE} catalog. To do so, we train a \textsl{Lux} model with 12 labels on the high-SNR field RGB-train sample, following the method described above. Given our $K$-fold cross validation test (see Appendix~\ref{app_kfold}), we again set $P=4M$ and $\Omega=10^{3}$. We then use the $\boldsymbol{A}$ and $\boldsymbol{B}$ matrices from this training set, and determine the $\boldsymbol{z}_n$ latent vectors for stars in the high-SNR field RGB-test sample, using the training set's $\boldsymbol{B}$ matrix and the test's set spectral flux and associated flux error. We finally predict stellar labels in the high-SNR field RGB-test sample using the test-set $\boldsymbol{z}_n$ latent parameters and the training set's $\boldsymbol{A}$ matrix via Equation~\ref{eq_labels}.

Figure~\ref{fig:cv_highsnr} shows the one-to-one comparison of the predicted labels for stars in the high-SNR field RGB-test sample from \textsl{Lux} as compared to those determined from \textsl{ASPCAP}. We note here that none of these stars were used in the training of the model, that was trained on the high-SNR field RGB-train sample. In each panel, we also compute the bias and RMSE value, and show the mean \textsl{ASPCAP} stellar label uncertainty, $\sigma_{ASPCAP}$. Overall, we can see that this linear \textsl{Lux} model is able to robustly determine stellar labels for a wide variety of parameters and parameter ranges.
The estimated bias for all labels is low (e.g., $\sim10$ K for $T_{\mathrm{eff}}$ and $\sim10^{-3}$ for element abundance ratios), and is approximately equal to the mean uncertainty from \textsl{ASPCAP} in each stellar label. Of particular importance is the fact that this simple model is able to capture well the label space for metal-poor stars ([Fe/H] $<-1$); for example, these stars are known to have depleted [Al/Fe] and [C/Fe], which the model captures surprisingly well despite the sample of [Fe/H] $<-1$ stars comprising a small fraction of the data set ($\sim7\%$).
We suspect that this is because of the linear nature of our model: linear models can extrapolate well compared to some heavily parametrized alternatives, and this is something we will explore in future work.

Interestingly, we see that some labels show deviations from the one-to-one line. This can be seen at high [Ca/Fe] abundance ratios, for example. This has been noted before in the literature \citep[][]{Ness2016_cannon}, and occurs when the model is not flexible enough to capture the extremes in the data. If we repeat the exercise narrowing the range in the label to the region where the deviations occur, we find that the model is able to capture the data well. While this feature can be solved with this temporary fix (i.e., narrowing the stellar label range), this phenomenon is a limitation in our model that in practice could be resolved by making the \textsl{Lux} model more complex. Nonetheless, it is a feature to be aware of when using \textsl{Lux} for determining a subset of the labels.

In order to visualize how \textsl{Lux} labels compare to those derived from other methods, in Figure~\ref{fig:kiel-tinsley} we show the Kiel ($\log~g$ vs. $T_{\mathrm{eff}}$) and Tinsley--Wallerstein ([Mg/Fe] vs. [Fe/H]) diagrams for sets of labels computed using \textsl{ASPCAP} (right) and the \textsl{Cannon} (middle); here the Kiel diagram is color-coded by metallicity, [Fe/H], and the Tinsley--Wallerstein diagram is color-coded by each star's galactic orbital eccentricity, computed using the \texttt{MilkyWayPotential2022} in the \texttt{gala} package \citep[]{Price2017}.
If one focuses on the Kiel diagram (top row), one can see that while the overall distribution of \textsl{Lux}, \textsl{Cannon}, and \textsl{ASPCAP} labels appear similar, there are subtle differences. For example, if one examines closely the metal-poor star sequence in \textsl{Lux} labels in the Kiel diagram, one can see that the sequence breaks up into two: a metal-poor RGB sequence (black), and an AGB sequence (dark brown), following two separate trends. This difference is less pronounced and has higher scatter in the \textsl{ASPCAP/Cannon} labels. The fact we see this separation in \textsl{Lux} labels and not either in the \textsl{ASPCAP} or \textsl{Cannon} labels may be because \textsl{Lux} yields more precise and less biased stellar labels (although note that the \textsl{Cannon} labels are a closer match to the \textsl{Lux} ones).

In the Tinsley--Wallerstein diagram (bottom row), we see that \textsl{Lux} labels show a tighter correlation between [Mg/Fe] and [Fe/H] in the metal-poor regime (low [Fe/H]) than the \textsl{Cannon} or \textsl{ASPCAP} labels.
The high eccentricity, $e$, (halo) stars show a wider scatter at fixed [Fe/H] in the \textsl{Cannon} labels, and then a higher scatter still in the \textsl{ASPCAP} labels. These distributions appear much tighter in \textsl{Lux} labels, as expected for stars originating from a single system (i.e., the LMC \citealp[][]{Nidever2020}, or the stellar halo debris.
In this region --- where stellar label uncertainties are generally larger --- we expect \textsl{Lux} to perform (in terms of label precision) better than both other comparisons.
We expect \textsl{Lux} labels to have less scatter than the \textsl{Cannon} and \textsl{ASPCAP} because \textsl{Lux} uses the label uncertainties to deconvolve the intrinsic distribution of the labels (in the latent vector space).
We also expect \textsl{Lux} to be more precise than \textsl{ASPCAP} because we use the full spectrum, whereas \textsl{ASPCAP} uses particular windows and spectral ranges to determine these stellar parameters.
On the other hand, the abundance values in the metal rich end seem to have less definition in \textsl{Lux} labels as compared to \textsl{ASPCAP}.
All of these properties are encouraging and warrant further investigation to understand the full capacity of \textsl{Lux} for improving and interpreting stellar label distributions.

Further examples illustrating the labels determined by our \textsl{Lux} model are shown in Appendix~\ref{app_plots} in Figure~\ref{fig:all-abun}, where we show stars from the high-SNR field RGB-test sample in the Kiel diagram as well as every element abundance modeled as a function of metallicity.

\subsection{Tests on lower signal-to-noise spectra}
\label{sec_lowsnr}
\begin{figure*}
    \centering
    \includegraphics[width=1\textwidth]{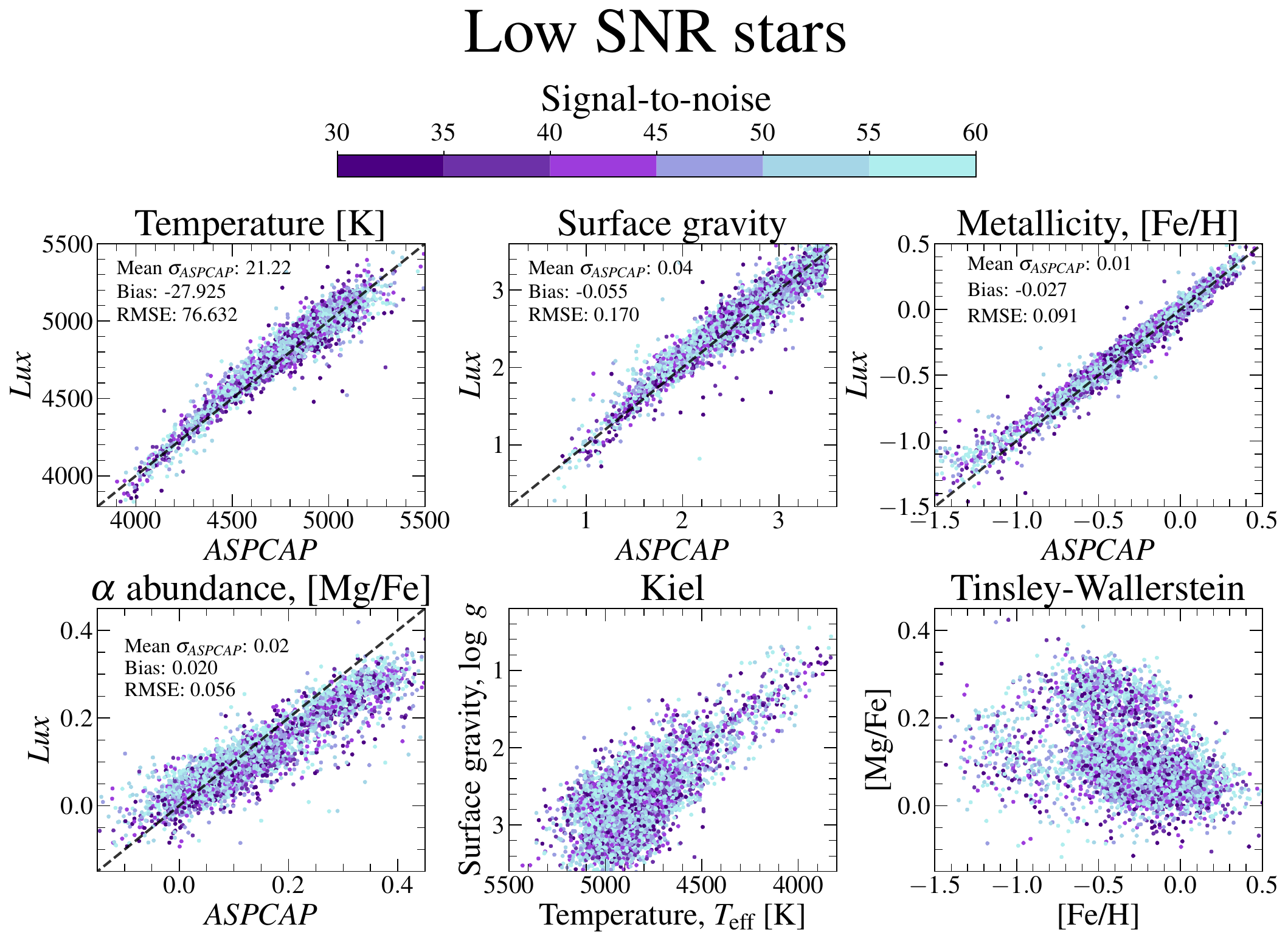}
    \caption{Validation results for RGB stars at lower signal-to-noise (low-SNR field RGB-test). Here, we have chosen a signal-to-noise range that is expected for the \textsl{SDSS-V Galactic Genesis} survey \citep[][]{Kollmeier2017}. As in Figure~\ref{fig:cv_highsnr}, we show the mean \textsl{ASPCAP} uncertainty, bias, and RMSE values for each label in each of the first four panels. In the last two panels we also show the \textsl{Lux} labels for these stars in the Kiel and Tinsley-Wallerstein diagrams. The \textsl{Lux} labels are determined by optimizing the test set latent representations using each star's spectral fluxes. Overall, the RMSE values obtained are reasonably low and the bias values are approximately equal to the average \textsl{ASPCAP} uncertainty, indicating that the \textsl{Lux} model is able to infer stellar labels at reasonable precision for lower SNR stars. The full validation for all labels is shown in Figure~\ref{fig:cv_lowsnr} in Appendix~\ref{app_plots}. See text in Section~\ref{sec_lowsnr} for further details.}
    \label{fig:low-snr}
\end{figure*}

An important aspect of any data-driven model for stellar spectra is its ability to determine precise stellar labels for spectra with lower signal-to-noise than those on which it is trained on. Figure~\ref{fig:low-snr} shows the validation results for a test on the low-SNR field RGB-test sample. This sample contains 5,000 RGB stars with lower SNR, in the range of $30 < $ SNR $< 60$. We choose this range of SNR to match what is expected for the \textsl{SDSS-V Galactic Genesis} survey \citep[][]{Kollmeier2017}, that will deliver over $\approx3$ million (near-infrared) spectra for Milky Way stars. Overall, our \textsl{Lux} model is able to infer a wide range of stellar labels at lower signal-to-noise. We are able to recover labels with a precision that is comparable to the higher signal-to-noise stars (Figure~\ref{fig:low-snr} and Figure~\ref{fig:cv_lowsnr} in Appendix~\ref{app_plots}). We do note however that we observe a larger scatter/RMSE for some elements (N, O, Ca, and Ni, for example). Despite this, our ability to separate the high-/low-$\alpha$ disks, as well as accreted halo populations (bottom right panel of Figure~\ref{fig:low-snr}) illustrates that, for important labels, our model is able to infer well stellar labels at lower signal-to-noise.

\begin{figure*}
    \centering
    \includegraphics[width=1\textwidth]{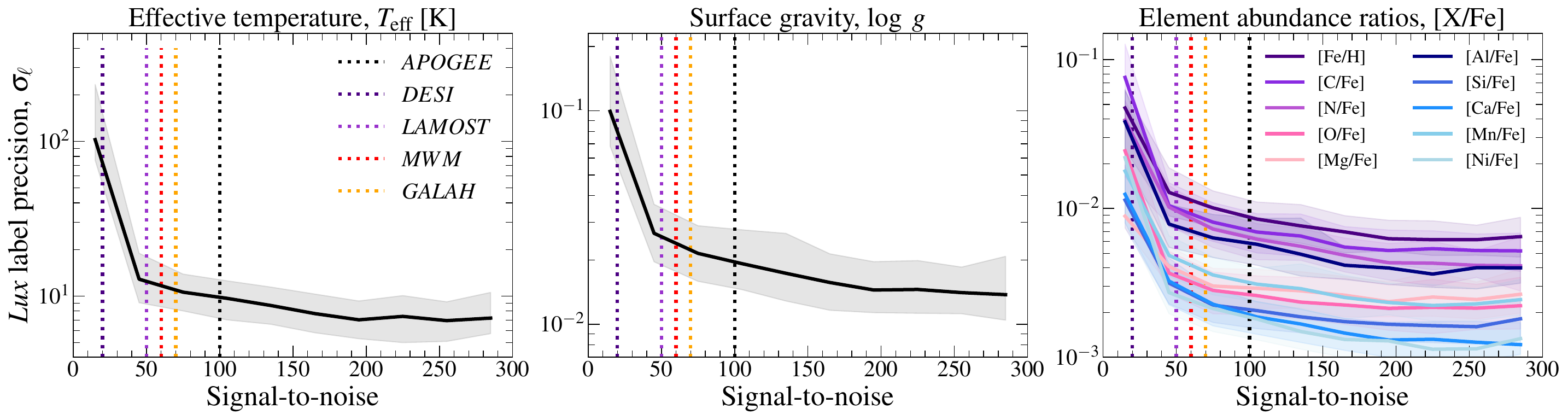}
    \caption{The \textsl{Lux} model label precision as a function of signal-to-noise for effective temperature (left), surface gravity (middle), and element abundance ratios (right). The median (solid) and ($16^{th}$, $84^{th}$) inter-quartile range (shaded) are computed by imputing ten realizations of each stellar label for 2,000 randomly selected stars from the high-SNR field RGB-test and low-SNR field RGB-test sets; this is accomplished by sampling ten realizations of the spectrum of these 2,000 stars and using these stellar fluxes to infer ten realizations of the latent representations for each star (see text in Section~\ref{sec_lowsnr} for further details). Overall, the precision on the \textsl{Lux} model is quite remarkable for all stellar labels examined, even at low signal-to-noise. This test illustrates how precisely \textsl{Lux} is able to infer stellar labels for \textsl{APOGEE} stars across signal-to-noise. In each panel we also show as vertical dotted lines the average signal-to-noise expected for some large-scale stellar surveys of interest.}
    \label{fig:precision}
\end{figure*}

In order to assess how well the \textsl{Lux} model is able to infer stellar labels as a function of SNR, in Figure~\ref{fig:precision} we show the \textsl{Lux} model uncertainty value for each label as a function of signal-to-noise for 2,000 random stars from the high-SNR field RGB-test and low-SNR field RGB-test samples. This value is computed by taking ten random realizations of the spectrum of each star drawing from a normal distribution with mean(standard deviation) equal to the flux(flux error). We then compute ten realizations of the $\boldsymbol{z}$ latent parameters for each star using these sampled spectral fluxes, and respectively compute ten realizations of (\textsl{Lux}) labels; using these ten realizations of the labels for each star, we compute the [$5^{th}$, $50^{th}$, $95^{th}$] percentiles as a function of signal-to-noise, which we show as a solid line ($50^{th}$) and shaded regions ($5^{th}$, $95^{th}$) in Figure~\ref{fig:precision}\footnote{This procedure is equivalent to computing the inverse of the Fisher information matrix for $\boldsymbol{z}$ and computing the uncertainties analytically.}. Overall, the precision on \textsl{Lux} labels is quite remarkable, even at low signal-to-noise. This test illustrates how precisely \textsl{Lux} is able to infer stellar labels for \textsl{APOGEE} stars. For $T_{\mathrm{eff}}$, precision is on the order of $\sigma_{T_{\mathrm{eff}}}< 20$ [K] down to SNR $\gtrsim40$. At the same SNR, $\log~g$ precision is $\sigma_{\log~g}< 0.1$, and individual element abundance ratio precision is on the order of $\sigma_{\mathrm{[X/Fe]}}\lesssim0.05$ dex.

In summary, \textsl{Lux}'s ability to robustly infer stellar labels across different SNRs is telling that this model can precisely determine stellar labels (and stellar spectra, not shown). This is likely thanks to \textsl{Lux}'s ability to use the entire stellar spectrum of a star to infer a particular label, which is much richer in information when compared to using particular spectral lines.

\begin{figure*}
    \centering
    \includegraphics[width=1\textwidth]{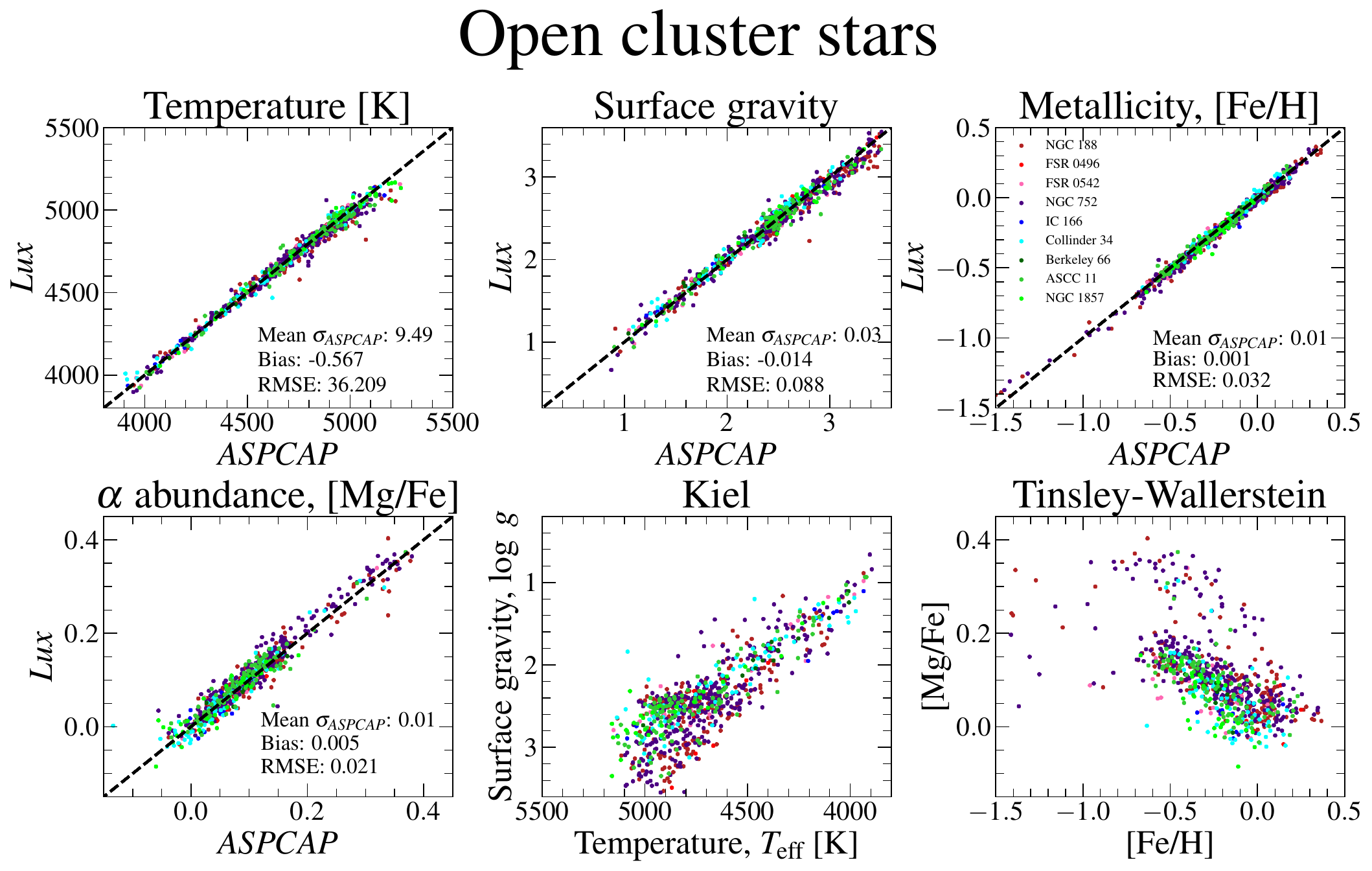}
    \caption{Replica of Figure~\ref{fig:low-snr} but for open cluster stars in the High-SNR OC RGB-test set. The \textsl{Lux} labels are determined by optimizing the test set latent representations using each star's spectral fluxes. Overall, the RMSE values obtained are low and the bias values are smaller than the average \textsl{ASPCAP} uncertainty, indicating that the \textsl{Lux} model is able to infer stellar labels accurately for benchmark stars in open clusters. See text in Section~\ref{sec_oc} for further details.}
    \label{fig:occam_cv}
\end{figure*}

\subsection{Validation on open clusters}
\label{sec_oc}
As a further test of the model's ability to emulate \textsl{ASPCAP} labels, we apply \textsl{Lux} to open cluster stars. Using stars from the \textsl{OCCAM} value-added catalog \citep[][]{Myers2022} in \textsl{APOGEE} DR17, we estimate four stellar labels ($T_{\mathrm{eff}}$, $\log~g$, [Fe/H], and [Mg/Fe]) for 790 stars across nine different open clusters (the high-SNR OC RGB-test sample).

Figure~\ref{fig:occam_cv} shows the comparison between the predicted \textsl{Lux} labels and those derived from \textsl{ASPCAP} for benchmark stars in open clusters. We find excellent agreement and no trends between \textsl{Lux} and \textsl{ASPCAP} labels across all parameters. This is visualized both in one-to-one comparisons and in the Kiel and Tinsley--Wallerstein diagrams (bottom panels). The ability of \textsl{Lux} to accurately recover labels for coeval stellar populations across a range of stellar parameters demonstrates that the model has successfully learned the underlying mapping between spectra and labels, even for these benchmark objects.

\subsection{Tests on different stellar types}
\label{sec_allstars}
\begin{figure*}
    \centering
    \includegraphics[width=1\textwidth]{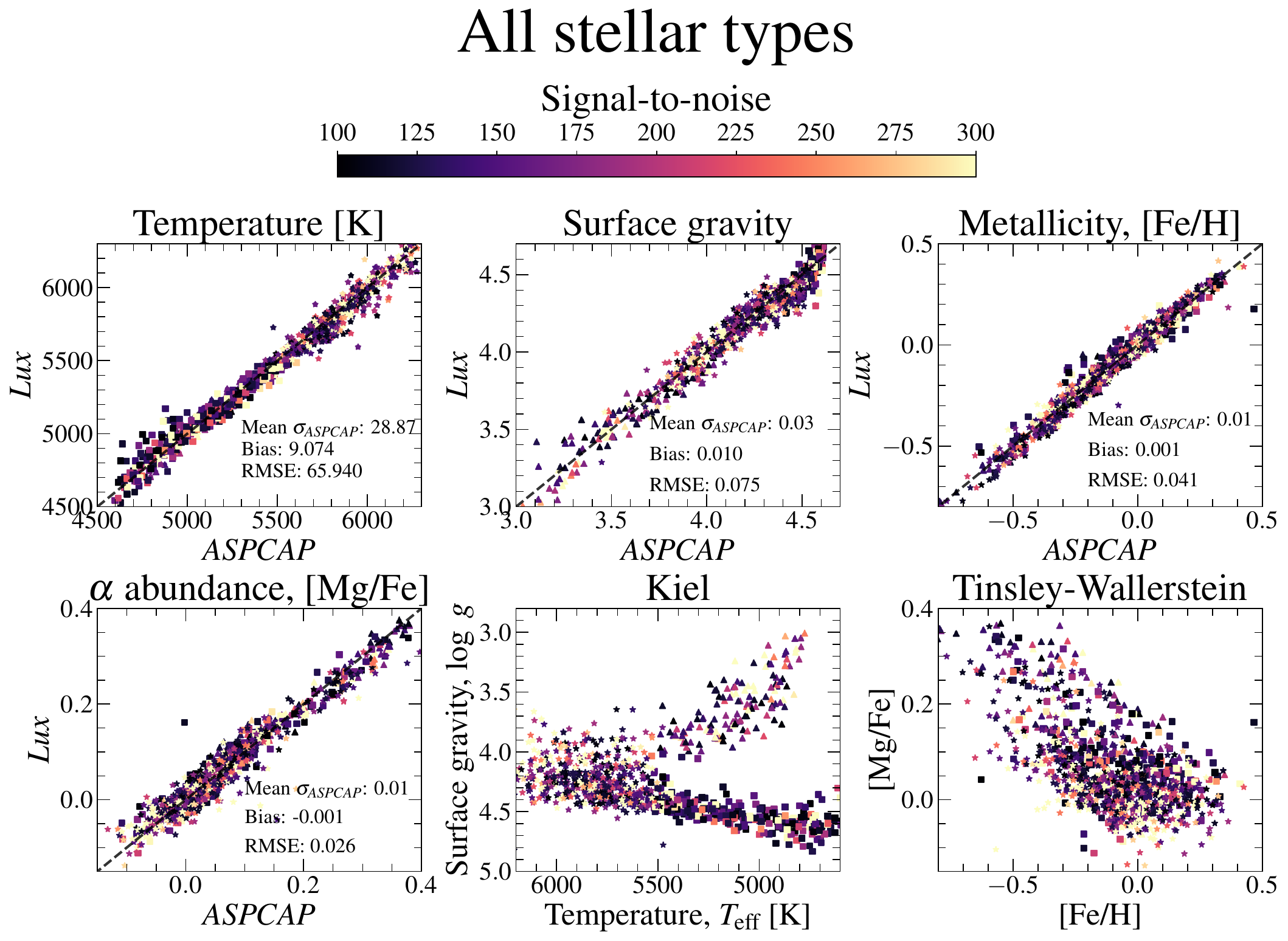}
    \caption{Replica of Figure~\ref{fig:low-snr} but for high SNR RGB (triangles), dwarf (squares) and other (star symbol) stars from the high-SNR field all-test sample. The \textsl{Lux} model used to infer these labels was trained on the high-SNR field all-train sample. The \textsl{Lux} labels shown are determined by optimizing the test set latent representations using each star's spectral fluxes. Overall, the RMSE values obtained are reasonably low and the bias values are smaller than the average \textsl{ASPCAP} uncertainty, indicating that the \textsl{Lux} model is able to infer stellar labels accurately for samples including all stellar types. See text in Section~\ref{sec_allstars} for further details.}
    \label{fig:giants_dwarfs}
\end{figure*}

As a final test of \textsl{Lux} model's ability to infer stellar labels, we evaluate its performance across different stellar types. We train an \textsl{Lux} model using 4,000 RGB, MS, and dwarf stars from the high-SNR field all-train sample. The model predicts six labels: $T_{\mathrm{eff}}$, $\log~g$, [Fe/H], [Mg/Fe], $v_{\mathrm{micro}}$, and $v_{\mathrm{sin}i}$, with the latter two parameters providing additional constraints on stellar type classification\footnote{We note that it is not vital to include these last two parameters for distinguishing stellar types, as $\log~g$ should suffice to distinguish e.g., giant stars from dwarfs. However, during testing we found that including $v_{\mathrm{micro}}$ and $v_{\mathrm{sin}i}$ did help increase label precision.}. We validate the model using 1,000 stars from the high-SNR field all-test sample.

The results for $T_{\mathrm{eff}}$, $\log~g$, [Fe/H], and [Mg/Fe] are shown in Figure~\ref{fig:giants_dwarfs}, along with the corresponding Kiel and Tinsley--Wallerstein diagrams. Here, the \textsl{Lux} labels are determined by optimizing the test set latent representations using each star's spectral fluxes.
The model demonstrates robust performance in simultaneously inferring stellar parameters across RGB, MS, and dwarf populations, successfully recovering all four primary stellar labels. These results confirm the \textsl{Lux} model's capability to deliver precise stellar parameters across the full extent of the Kiel diagram.

\section{Results: Label transfer between \textsl{APOGEE} and \textsl{GALAH}}
\label{sec_galah}

In the previous demonstration, we use \textsl{APOGEE} spectra and \textsl{APOGEE} stellar labels to train and test the \textsl{Lux} model. In this section, we demonstrate \textsl{Lux}'s ability to transfer labels between different surveys. In particular, we train a model using \textsl{APOGEE} DR17 spectra and \textsl{GALAH} DR3 labels for stars that are common between the two surveys to attempt to infer \textsl{GALAH} labels for \textsl{APOGEE} spectra without \textsl{GALAH} observations. This exercise is particularly interesting because the two surveys observe in different wavelength regimes, with \textsl{APOGEE} observing in the near-infrared and \textsl{GALAH} in the optical.

In detail, we identify 5,000 overlapping red giant branch stars between the \textsl{APOGEE} and \textsl{GALAH} surveys. We then divide this parent sample into a training set comprised of 4,000 stars, and a validation set of 1,000 stars. As described in Section~\ref{sec_data}, we will label these as \textsl{GALAH-APOGEE} field giants-train and \textsl{GALAH-APOGEE} field giants-test, respectively.

Using the \textsl{GALAH-APOGEE} field giants-train set of 4,000 stars, we train a \textsl{Lux} model using $P=4M$ and $\Omega=10^{3}$. Specifically, we train this model using the corresponding \textsl{APOGEE} (near-infrared) spectra for these stars and eleven stellar labels determined using the \textsl{GALAH} optical spectra. The stellar labels we use are: $T_{\mathrm{eff}}, \log~g$, [Fe/H], [Li/Fe], [Na/Fe], [O/Fe], [Mg/Fe], [Y/Fe], [Ce/Fe], [Ba/Fe], and [Eu/Fe]. At this point it is worth mentioning that, with the exception of $T_{\mathrm{eff}}, \log~g$, [Fe/H], [O/Fe], and [Mg/Fe], all the other stellar labels trained on are not well determined in \textsl{APOGEE}'s \textsl{ASPCAP}, if determined at all. For example, [Eu/Fe] and [Y/Fe] are (in principle) not possible to be determined at all in \textsl{APOGEE} with \textsl{ASPCAP} because of a lack of spectral lines for Eu and Y in the near-infrared. While this may be a provocative exercise for some readers, we argue that it is interesting to test how well our model is able to infer abundances for \textsl{APOGEE} stars that cannot be determined using \textsl{ASPCAP}, even if these inferred abundances are obtained via correlations with other elements and not causal relations with spectral features. Similarly to the model presented in Section~\ref{sec_model_res}, we use the reported \textsl{GALAH} DR3 errors as the stellar label uncertainties.

\begin{figure*}
    \centering
    \includegraphics[width=1\textwidth]{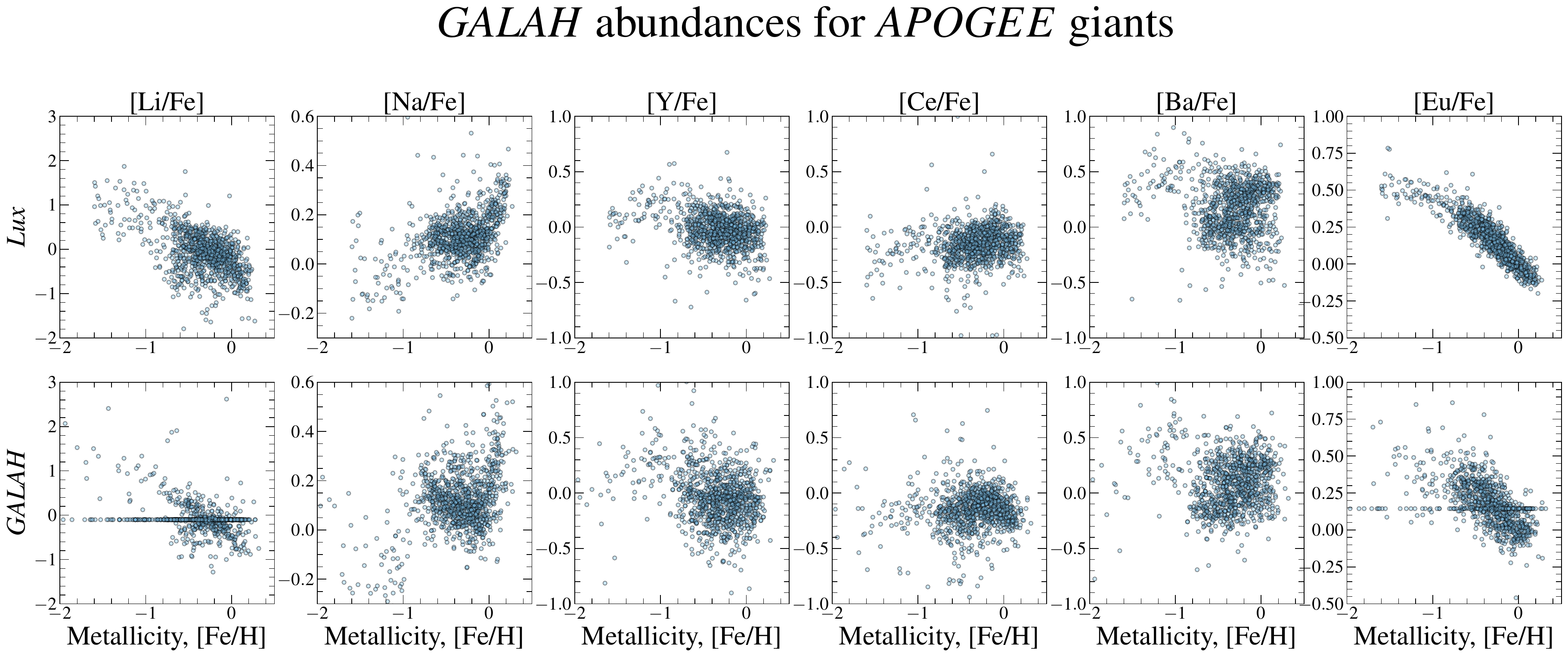}
    \caption{Six \textsl{GALAH} chemical abundances for 1,000 \textsl{APOGEE} giant stars (\textsl{GALAH-APOGEE} field giants-test set), determined using an \textsl{Lux} label tranfer model trained on 4,000 stars \textsl{APOGEE} DR17 spectra and \textsl{GALAH} DR3 labels ((\textsl{GALAH-APOGEE} field giants-train set)). Here \textsl{Lux} labels are inferred by optimizing the latent representations for test set stars using each star's spectral fluxes. All these abundances, with the exception of [Na/Fe] and [Ce/Fe], cannot be determined in \textsl{APOGEE}'s \textsl{ASPCAP} but can be determined using our \textsl{Lux} label transfer model. However, while \textsl{Lux} is able to determine these stellar labels, care should be taken to treat these results as causal rather than a correlation with other label information (see text in Section~\ref{sec_galah} for further details). The horizontal streak of \textsl{GALAH} [Li/Fe] and [Eu/Fe] abundances are for stars that \textsl{GALAH} reported a \texttt{NaN} value and \textsl{Lux} set their abundance value equal to the median of the distribution (and its uncertainty inflated to 9999); this ensures these stars do not influence the training and testing.}
    \label{fig:abun_galah}
\end{figure*}

Figure~\ref{fig:abun_galah} shows the distribution of seven stellar labels in the [X/Fe]-[Fe/H] plane for the 1,000 stars in the \textsl{GALAH-APOGEE} field giants-test set; here we only show the stellar labels that are not possible to be determined in \textsl{APOGEE}\footnote{[Na/Fe] and [Ce/Fe] has been shown to be possible to be determined in \textsl{ASPCAP}, albeit for a relatively small number of the \textsl{APOGEE} data set. Moreover, these are chemical abundance ratios that \textsl{ASPCAP} struggles to determine precisely due to the weak atomic lines in the $1.5-1.7~\mu$m regime.}. We show the estimated \textsl{Lux} labels in the top row, and the corresponding \textsl{GALAH} labels in the bottom row. Here, the \textsl{Lux} labels are determined by optimizing the latent representations for the test set stars using each star's spectral fluxes, and projecting those to stellar labels using $\bs{A}$ through Equation~\ref{eq_labels}. We find that \textsl{Lux} model is able to infer stellar labels determined in \textsl{GALAH} for \textsl{APOGEE} stars, and in some cases, is able to estimate an abundance for stars that do not report a \textsl{GALAH} stellar label\footnote{These are the stars that appear as a horizontal stripe in Figure~\ref{fig:abun_galah}, which have been set to the median value of the distribution by our model due to \textsl{GALAH} not being able to provide a measurement.}; this is one of the main advantages of \textsl{Lux} framework, as it is able to operate with partial missing labels. The full test set validation is shown in Figure~\ref{fig:cv_galah} in Appendix~\ref{app_plots}.

We caution that while the training and validation set performance implies that we successfully label the \textsl{APOGEE} stars with \textsl{GALAH} abundances, for many of these elements (e.g. Li, Ba, Eu) the \textsl{APOGEE} wavelength region is not known to have spectral features corresponding to these elements. As we allow the model to use all wavelength values for each label's inference, the prediction may be based on correlation rather than causation, i.e. inferred not directly from the element variation in the flux, but rather how that element varies with stellar parameters or other labels in the training set as expressed in the flux. The model may therefore fail to correctly infer these abundances for stars with different label-correlation behaviors. This is always the case when one allows the full wavelength region to be leveraged for abundances, and is in many cases well motivated (e.g., for elements blended with molecules and those that impact the entire spectral region; \citealp[e.g.,][]{Ting2018}). To restrict the model's learning, it is straightforward to implement wavelength ``masking'' to limit the model to learn particular element labels from specified regions.
In the case where the full spectral region is used for element inference and the element absorption features being inferred are present in the spectra, the generative nature of the model enables a fidelity test of the labels. Thus, the generated spectral model can be used to calculate a goodness of fit between the generated spectral model and observed spectra for the element lines being inferred. This would be a possible way to verify that specific absorption lines are being reproduced by the corresponding element labels (e.g. see Figure~\ref{fig:comp-spectra} and \citealp[][]{Manea2024}). However, this is not possible fo Li, for example, which does not have any absorption lines in the \textsl{APOGEE} wavelength region. This label prediction is likely inherited from the mapping between stellar parameters and this label in the training set. The exercise of label transfer between different surveys means that the model is useful as a tool of information or (absorption) line discovery \citep[][]{Hasselquist2016}, by examining the origin of the information pertaining to each label inference (see Figure~\ref{fig:spectra-derivatives}).

In summary, the results presented in this Section show that \textsl{Lux} is able to perform label transfer between different stellar surveys, even when the spectral range is different.
It is also able to recover stellar label measurements for stars with no stellar labels. It would be interesting in the future to push this exercise to also be able to perform label transfer between surveys at different resolutions simultaneously (i.e. increase the number of outputs of \textsl{Lux}).

\section{Discussion}\label{sec_discussion}

In the sections below, we summarize some novel aspects of \textsl{Lux}, discuss
possible extensions to the model, and present some potential applications of the
\textsl{Lux} model.

\subsection{Lux's model structure}

\textsl{Lux} is a model framework built around a generative latent-variable model structure that is designed to support a range of tasks in the analysis and use of astronomical data.
The framework is quite general, but here we present it in the context of stellar spectra and stellar labels (stellar parameters) from large spectroscopic surveys.
We have shown that \textsl{Lux} is able to emulate pipelines that determine stellar labels from stellar spectra (here, \textsl{APOGEE}'s \textsl{ASPCAP} pipeline) and to perform label transfer between different surveys (here, \textsl{APOGEE} and \textsl{GALAH}).
The model implementation we use in this work is (in some sense) bi-linear, but the framework is general and can be extended to use more complex transformations from latent representations to output data.

As a multi-output latent-variable model, \textsl{Lux} is related to other machine-learning models that perform data embedding or compression, such as encoder-decoder networks.
With the L2 regularization on the latent parameters, \textsl{Lux} even resembles a variational autoencoder \citep{autoencoders} with linear decoders and linear pseudo-inverse encoders.
However, \textsl{Lux} is different in that it is a generative model and can be used in probabilistic contexts, and it is able to generate multiple outputs simultaneously.
All the output quantities, at training time, constrain the embedded representation of the objects.

Our current implementation of \textsl{Lux} makes a specific choice about where to situate the model flexibility.
In this work, we have chosen to make the mapping linear and the latent dimensionality larger than that of the stellar labels (but smaller than the spectral dimensionality).
We could have instead chosen to make the mapping non-linear, for example using a multi-layer perceptron or Gaussian process layers, and then fixed the latent dimensionality to a much smaller size.
This structure would allow the model to learn more complex relationships between the latent parameters and the output data but potentially keep the embedded representations simple.
We found that the benefits of using a linear mapping (for computation and simplicity) made our current implementation a good choice for the tasks we have demonstrated here, but we envision constructing future versions of \textsl{Lux} that are non-linear for tasks that require more flexibility or capacity.
For example, in the case of label transfer between surveys, a non-linear model might be able to better capture the differences in the spectral features between the surveys and provide more accurate label transfer.
Or, one may want to include an output data block that predicts kinematic or non-intrinsic properties of the sources, such as distance or extinction, which involves more physics, and probably requires a more complex mapping from the latent parameters to the output data.

Thus, even though in this work we have only tested how \textsl{Lux} can handle stellar spectroscopic data, the model is equipped to be able to handle other type of astronomical data. For example, we could have instead chosen to feed \textsl{Lux} photometric or astrometric data from \textsl{Gaia}. $G$-band magnitudes, $G_{\rm BP}-G_{\rm RP}$ colors, and parallaxes could have instead been fed into the model to train the latent variables to then infer extinction coefficients, for example. Along those lines, one could envision training an \textsl{Lux} model that deals with both spectroscopic data \textit{and} photometric \textit{and} astrometric data, simultaneously. This example could be achieved by adding plates to Figure~\ref{fig:gpm} to include \textsl{Gaia} $G$-band magnitudes and parallaxes, for example, plus perhaps the associated Galactic phase-space variables. Such a model would be useful, for example, for inferring data-driven spectro-photometric distances of stars, or for inferring stellar luminosities.

\subsection{Applications of \textsl{Lux}}

The \textsl{Lux} framework is designed to enable a range of tasks in astronomy, with a particular focus for stellar spectroscopy and survey science. Here we have demonstrated how it can be used to emulate the stellar parameter pipeline used for \textsl{APOGEE} spectra, and to transfer labels between the \textsl{APOGEE} and \textsl{GALAH} surveys. Below we outline three broad categories of applications enabled by the \textsl{Lux} framework: stellar label inference, multi-survey translation, and classification.

\subsubsection{Stellar label inference}
\textsl{Lux} can be used to infer stellar parameters from spectroscopic survey data by learning from a training set with known parameters. This is valuable for efficiently determining parameters for large stellar surveys (e.g., \textsl{SDSS}: \citealp{Blanton2017, Kollmeier2017}, \textsl{Gaia}: \citealp{Gaia2020}, \textsl{LAMOST}: \citealp{Zhao2012}, \textsl{GALAH}: \citealp{Freeman2012}, \textsl{DESI}: \citealp{Cooper_2023}, \textsl{4MOST}: \citealp{DeJong2012}, \textsl{WEAVE}: \citealp{Jin_2023}) by emulating more costly pipeline runs. One immediate application is determining stellar parameters and abundances for stars in the \textsl{SDSS-V Galactic Genesis} Survey \citep[][]{Kollmeier2017}, which is collecting millions of \textsl{APOGEE} spectra. \textsl{Lux} could also be used to determine spectro-photometric distances from a reliable training set \citep[][]{Hogg2019}, or compile catalogs of stellar ages for giants stars based on [C/N] abundances and asteroseismology \citep[][]{Ness2016}.

\subsubsection{Multi-survey translation}
\textsl{Lux} enables translation between the notoriously different stellar parameter outputs of different surveys and instruments by training on overlapping sources. This allows determination of parameters that may be difficult or impossible to measure directly in one survey but are well-measured in another. For example, stellar parameters and abundances could potentially be determined for the vast set of \textsl{Gaia} XP spectra by training on stars that overlap with \textsl{APOGEE} \citep[e.g.,][]{Andrae2023,Li2024}. Similarly, \textsl{APOGEE}-quality stellar labels could potentially be determined for \textsl{BOSS} spectra using overlapping stars from \textsl{SDSS-V Milky Way Mapper}. However, care must be taken to validate that the translated parameters reflect genuine spectral features rather than just correlations in the training set.

\subsubsection{Classification}
The \textsl{Lux} framework could also enable classification tasks in a way that properly handles uncertainties on input data. This is similar to parameter inference but with discrete parameters.
One application would be identifying chemically peculiar stars in large spectroscopic surveys. After training on a set of stars with known peculiar abundance patterns, one could use \textsl{Lux} to compute latent representations for all target sources as a means to efficiently search for similar objects in surveys like \textsl{Gaia} XP, based solely on their spectra.

\section{Summary and Conclusions}
\label{sec_conclusions}

We present in this work the first and simplest version of \textsl{Lux}, a multi-task generative latent-variable model for data-driven stellar label and spectral inference. We have demonstrated that this model is successful at inferring precise stellar labels and stellar spectra for a wide range of \textsl{APOGEE} stars. We have also shown that the \textsl{Lux} model can be used for label transfer tasks. The main strengths and novel aspects of \textsl{Lux} are:
\begin{description}
    \item[1. A multi-output generative model permitting noisy data] \textsl{Lux} is a generative model of both stellar labels and spectral fluxes (and potentially any additional data added as outputs to the model). This enables the model to properly handle uncertainties in the stellar labels and fluxes during training so that \textsl{Lux} is able to account for imperfect stellar labels. This is important, as current data-driven models (e.g., the \textsl{Cannon} and the \textsl{Payne}) require assuming that the stellar labels for the training set are perfectly known, which places severe quality limits on the training data and is not the case in detail for even the highest signal-to-noise spectra.
    This aspect of \textsl{Lux} also enables the model to handle missing data (e.g., missing pixels in some spectra or missing labels for some stars) in a principled way.
    This facilitates label-transfer and emulation between different data sets, as typically one data set may have robust measurements of one stellar label that is not in the test set and vice versa.
    Finally, the generative nature of the model allows for the model to be used in fully probabilistic contexts, where the distinction between training and test data is no longer necessary.
    \item[2. Computationally fast] \textsl{Lux} is written with \texttt{JAX} \citep[][]{jax2018github}, and has very simple model structure. For these reasons it is computationally fast. For reference, the training step of the model used in this paper took approximately $\approx30$ minutes to train on 5,000 stars using one CPU of a high-end laptop, while the test step on 10,000 stars took $\approx20$ minutes.
    \item[3. Flexible model form] In our current demonstration, we use a version of \textsl{Lux} with two outputs (stellar labels and spectral flux) with linear transformations from the latent vectors and these output data.
    However, our implementation is written such that more complex transformations from latent vectors to outputs can be used (e.g., a multi-layer perceptron or layers of Gaussian process), and more outputs can be added (e.g., to simultaneously operate on multiple surveys or data types).
\end{description}

\textsl{Lux} is a powerful new frameworkfor data-driven stellar label and spectra inference, multi-survey translation, and classification. We have demonstrated how \textsl{Lux} can be used to infer precise stellar labels and stellar spectra for \textsl{APOGEE} stars using only linear model transforms, and how it can be used to transfer labels between different surveys. We have also discussed how \textsl{Lux} model can be used for classification tasks. We hope that the \textsl{Lux} model will be a useful tool for data driven modeling of stellar and galactic data, especially in the realm of spectroscopic data.

\section*{Acknowledgements}
The authors would like to thank Adam Wheeler for providing the \textsl{Korg} spectra, Julianne Dalcanton for enlightening conversations about future prospects of \textsl{Lux}, Carrie Filion for all the help and support, Catherine Manea, David Nidever, Andrew Saydjari, Greg Green, Hans-Walter Rix, and the CCA stellar spectroscopy, CCA Astronomical Data, and CCA Nearby Universe groups for helpful discussions. The authors would also like to thank the anonymous reviewer for a helpful report on earlier versions of this manuscript. DH would also like to thank Sue, Alex, and Debra for everything they do. The Flatiron Institute is a division of the Simons Foundation.

\facilities{\textsl{SDSS-IV} \citep{Blanton2017}, Apache Point Observatory \citep{Gunn2006}, Las Campanas Observatory \citep{BowenVaughan1973}}

\software{
    \texttt{matplotlib} \citep{Hunter:2007},
    \texttt{numpy} \citep{NumPy},
    \texttt{Gala} \citep{Price2017},
    \texttt{JAX} \citep[][]{jax2018github},
    \texttt{JAXOpt} \citep[][]{jaxopt_implicit_diff}
}

\bibliography{refs}

\begin{thebibliography}{}
\expandafter\ifx\csname natexlab\endcsname\relax\def\natexlab#1{#1}\fi
\providecommand{\url}[1]{\href{#1}{#1}}
\providecommand{\dodoi}[1]{doi:~\href{http://doi.org/#1}{\nolinkurl{#1}}}
\providecommand{\doeprint}[1]{\href{http://ascl.net/#1}{\nolinkurl{http://ascl.net/#1}}}
\providecommand{\doarXiv}[1]{\href{https://arxiv.org/abs/#1}{\nolinkurl{https://arxiv.org/abs/#1}}}

\bibitem[{{Abdurro'uf} {et~al.}(2022){Abdurro'uf}, {Accetta}, {Aerts}, {Silva Aguirre}, {Ahumada}, {Ajgaonkar}, {Filiz Ak}, {Alam}, {Allende Prieto}, {Almeida}, {Anders}, {Anderson}, {Andrews}, {Anguiano}, {Aquino-Ort{\'\i}z}, {Arag{\'o}n-Salamanca}, {Argudo-Fern{\'a}ndez}, {Ata}, {Aubert}, {Avila-Reese}, {Badenes}, {Barb{\'a}}, {Barger}, {Barrera-Ballesteros}, {Beaton}, {Beers}, {Belfiore}, {Bender}, {Bernardi}, {Bershady}, {Beutler}, {Bidin}, {Bird}, {Bizyaev}, {Blanc}, {Blanton}, {Boardman}, {Bolton}, {Boquien}, {Borissova}, {Bovy}, {Brandt}, {Brown}, {Brownstein}, {Brusa}, {Buchner}, {Bundy}, {Burchett}, {Bureau}, {Burgasser}, {Cabang}, {Campbell}, {Cappellari}, {Carlberg}, {Wanderley}, {Carrera}, {Cash}, {Chen}, {Chen}, {Cherinka}, {Chiappini}, {Choi}, {Chojnowski}, {Chung}, {Clerc}, {Cohen}, {Comerford}, {Comparat}, {da Costa}, {Covey}, {Crane}, {Cruz-Gonzalez}, {Culhane}, {Cunha}, {Dai}, {Damke}, {Darling}, {Davidson}, {Davies}, {Dawson}, {De Lee}, {Diamond-Stanic}, {Cano-D{\'\i}az}, {S{\'a}nchez},
  {Donor}, {Duckworth}, {Dwelly}, {Eisenstein}, {Elsworth}, {Emsellem}, {Eracleous}, {Escoffier}, {Fan}, {Farr}, {Feng}, {Fern{\'a}ndez-Trincado}, {Feuillet}, {Filipp}, {Fillingham}, {Frinchaboy}, {Fromenteau}, {Galbany}, {Garc{\'\i}a}, {Garc{\'\i}a-Hern{\'a}ndez}, {Ge}, {Geisler}, {Gelfand}, {G{\'e}ron}, {Gibson}, {Goddy}, {Godoy-Rivera}, {Grabowski}, {Green}, {Greener}, {Grier}, {Griffith}, {Guo}, {Guy}, {Hadjara}, {Harding}, {Hasselquist}, {Hayes}, {Hearty}, {Hern{\'a}ndez}, {Hill}, {Hogg}, {Holtzman}, {Horta}, {Hsieh}, {Hsu}, {Hsu}, {Huber}, {Huertas-Company}, {Hutchinson}, {Hwang}, {Ibarra-Medel}, {Chitham}, {Ilha}, {Imig}, {Jaekle}, {Jayasinghe}, {Ji}, {Johnson}, {Jones}, {J{\"o}nsson}, {Katkov}, {Khalatyan}, {Kinemuchi}, {Kisku}, {Knapen}, {Kneib}, {Kollmeier}, {Kong}, {Kounkel}, {Kreckel}, {Krishnarao}, {Lacerna}, {Lane}, {Langgin}, {Lavender}, {Law}, {Lazarz}, {Leung}, {Leung}, {Lewis}, {Li}, {Li}, {Lian}, {Liang}, {Lin}, {Lin}, {Lin}, {Lintott}, {Long}, {Longa-Pe{\~n}a}, {L{\'o}pez-Cob{\'a}}, {Lu},
  {Lundgren}, {Luo}, {Mackereth}, {de la Macorra}, {Mahadevan}, {Majewski}, {Manchado}, {Mandeville}, {Maraston}, {Margalef-Bentabol}, {Masseron}, {Masters}, {Mathur}, {McDermid}, {Mckay}, {Merloni}, {Merrifield}, {Meszaros}, {Miglio}, {Di Mille}, {Minniti}, {Minsley}, {Monachesi}, {Moon}, {Mosser}, {Mulchaey}, {Muna}, {Mu{\~n}oz}, {Myers}, {Myers}, {Nadathur}, {Nair}, {Nandra}, {Neumann}, {Newman}, {Nidever}, {Nikakhtar}, {Nitschelm}, {O'Connell}, {Garma-Oehmichen}, {Luan Souza de Oliveira}, {Olney}, {Oravetz}, {Ortigoza-Urdaneta}, {Osorio}, {Otter}, {Pace}, {Padilla}, {Pan}, {Pan}, {Parikh}, {Parker}, {Peirani}, {Pe{\~n}a Ram{\'\i}rez}, {Penny}, {Percival}, {Perez-Fournon}, {Pinsonneault}, {Poidevin}, {Poovelil}, {Price-Whelan}, {B{\'a}rbara de Andrade Queiroz}, {Raddick}, {Ray}, {Rembold}, {Riddle}, {Riffel}, {Riffel}, {Rix}, {Robin}, {Rodr{\'\i}guez-Puebla}, {Roman-Lopes}, {Rom{\'a}n-Z{\'u}{\~n}iga}, {Rose}, {Ross}, {Rossi}, {Rubin}, {Salvato}, {S{\'a}nchez}, {S{\'a}nchez-Gallego}, {Sanderson}, {Santana
  Rojas}, {Sarceno}, {Sarmiento}, {Sayres}, {Sazonova}, {Schaefer}, {Schiavon}, {Schlegel}, {Schneider}, {Schultheis}, {Schwope}, {Serenelli}, {Serna}, {Shao}, {Shapiro}, {Sharma}, {Shen}, {Shetrone}, {Shu}, {Simon}, {Skrutskie}, {Smethurst}, {Smith}, {Sobeck}, {Spoo}, {Sprague}, {Stark}, {Stassun}, {Steinmetz}, {Stello}, {Stone-Martinez}, {Storchi-Bergmann}, {Stringfellow}, {Stutz}, {Su}, {Taghizadeh-Popp}, {Talbot}, {Tayar}, {Telles}, {Teske}, {Thakar}, {Theissen}, {Tkachenko}, {Thomas}, {Tojeiro}, {Hernandez Toledo}, {Troup}, {Trump}, {Trussler}, {Turner}, {Tuttle}, {Unda-Sanzana}, {V{\'a}zquez-Mata}, {Valentini}, {Valenzuela}, {Vargas-Gonz{\'a}lez}, {Vargas-Maga{\~n}a}, {Alfaro}, {Villanova}, {Vincenzo}, {Wake}, {Warfield}, {Washington}, {Weaver}, {Weijmans}, {Weinberg}, {Weiss}, {Westfall}, {Wild}, {Wilde}, {Wilson}, {Wilson}, {Wilson}, {Wolf}, {Wood-Vasey}, {Yan}, {Zamora}, {Zasowski}, {Zhang}, {Zhao}, {Zheng}, {Zheng}, \& {Zhu}}]{SDSSDR17}
{Abdurro'uf}, {Accetta}, K., {Aerts}, C., {et~al.} 2022, \apjs, 259, 35, \dodoi{10.3847/1538-4365/ac4414}

\bibitem[{{Allende Prieto} {et~al.}(2006){Allende Prieto}, {Beers}, {Wilhelm}, {Newberg}, {Rockosi}, {Yanny}, \& {Lee}}]{Prieto2006}
{Allende Prieto}, C., {Beers}, T.~C., {Wilhelm}, R., {et~al.} 2006, \apj, 636, 804, \dodoi{10.1086/498131}

\bibitem[{{Andrae} {et~al.}(2023{\natexlab{a}}){Andrae}, {Rix}, \& {Chandra}}]{Andrae2023}
{Andrae}, R., {Rix}, H.-W., \& {Chandra}, V. 2023{\natexlab{a}}, \apjs, 267, 8, \dodoi{10.3847/1538-4365/acd53e}

\bibitem[{{Andrae} {et~al.}(2023{\natexlab{b}}){Andrae}, {Rix}, \& {Chandra}}]{Rene2023}
---. 2023{\natexlab{b}}, \apjs, 267, 8, \dodoi{10.3847/1538-4365/acd53e}

\bibitem[{Bank {et~al.}(2021{\natexlab{a}})Bank, Koenigstein, \& Giryes}]{bank2021}
Bank, D., Koenigstein, N., \& Giryes, R. 2021{\natexlab{a}}, Autoencoders.
\newblock \doarXiv{2003.05991}

\bibitem[{Bank {et~al.}(2021{\natexlab{b}})Bank, Koenigstein, \& Giryes}]{autoencoders}
---. 2021{\natexlab{b}}, Autoencoders.
\newblock \doarXiv{2003.05991}

\bibitem[{{Beaton} {et~al.}(2021){Beaton}, {Oelkers}, {Hayes}, {Covey}, {Chojnowski}, {De Lee}, {Sobeck}, {Majewski}, {Cohen}, {Fernandez-Trincado}, {Longa-Pena}, {O'Connell}, {Santana}, {Stringfellow}, {Zasowski}, {Aerts}, {Anguiano}, {Bender}, {Canas}, {Cunha}, {Fleming}, {Frinchaboy}, {Feuillet}, {Harding}, {Hasselquist}, {Holtzman}, {Johnson}, {Kollmeier}, {Kounkel}, {Mahadevan}, {Price-Whelan}, {Rojas-Arriagada}, {Roman-Zuniga}, {Schlafly}, {Schultheis}, {Shetrone}, {Simon}, {Stassun}, {Stutz}, {Tayar}, {Teske}, {Tkachenko}, {Troup}, {Albareti}, {Bizyaev}, {Bovy}, {Burgasser}, {Comparat}, {Downes}, {Geisler}, {Inno}, {Manchado}, {Ness}, {Pinsonneault}, {Prada}, {Roman-Lopes}, {Simonian}, {Smith}, {Yan}, \& {Zamora}}]{Beaton2021}
{Beaton}, R.~L., {Oelkers}, R.~J., {Hayes}, C.~R., {et~al.} 2021, arXiv e-prints, arXiv:2108.11907.
\newblock \doarXiv{2108.11907}

\bibitem[{{Blanton} {et~al.}(2017){Blanton}, {Bershady}, {Abolfathi}, {Albareti}, {Allende Prieto}, {Almeida}, {Alonso-Garc{\'\i}a}, {Anders}, {Anderson}, {Andrews}, {Aquino-Ort{\'\i}z}, {Arag{\'o}n-Salamanca}, {Argudo-Fern{\'a}ndez}, {Armengaud}, {Aubourg}, {Avila-Reese}, {Badenes}, {Bailey}, {Barger}, {Barrera-Ballesteros}, {Bartosz}, {Bates}, {Baumgarten}, {Bautista}, {Beaton}, {Beers}, {Belfiore}, {Bender}, {Berlind}, {Bernardi}, {Beutler}, {Bird}, {Bizyaev}, {Blanc}, {Blomqvist}, {Bolton}, {Boquien}, {Borissova}, {van den Bosch}, {Bovy}, {Brandt}, {Brinkmann}, {Brownstein}, {Bundy}, {Burgasser}, {Burtin}, {Busca}, {Cappellari}, {Delgado Carigi}, {Carlberg}, {Carnero Rosell}, {Carrera}, {Chanover}, {Cherinka}, {Cheung}, {G{\'o}mez Maqueo Chew}, {Chiappini}, {Choi}, {Chojnowski}, {Chuang}, {Chung}, {Cirolini}, {Clerc}, {Cohen}, {Comparat}, {da Costa}, {Cousinou}, {Covey}, {Crane}, {Croft}, {Cruz-Gonzalez}, {Garrido Cuadra}, {Cunha}, {Damke}, {Darling}, {Davies}, {Dawson}, {de la Macorra}, {Dell'Agli}, {De
  Lee}, {Delubac}, {Di Mille}, {Diamond-Stanic}, {Cano-D{\'\i}az}, {Donor}, {Downes}, {Drory}, {du Mas des Bourboux}, {Duckworth}, {Dwelly}, {Dyer}, {Ebelke}, {Eigenbrot}, {Eisenstein}, {Emsellem}, {Eracleous}, {Escoffier}, {Evans}, {Fan}, {Fern{\'a}ndez-Alvar}, {Fernandez-Trincado}, {Feuillet}, {Finoguenov}, {Fleming}, {Font-Ribera}, {Fredrickson}, {Freischlad}, {Frinchaboy}, {Fuentes}, {Galbany}, {Garcia-Dias}, {Garc{\'\i}a-Hern{\'a}ndez}, {Gaulme}, {Geisler}, {Gelfand}, {Gil-Mar{\'\i}n}, {Gillespie}, {Goddard}, {Gonzalez-Perez}, {Grabowski}, {Green}, {Grier}, {Gunn}, {Guo}, {Guy}, {Hagen}, {Hahn}, {Hall}, {Harding}, {Hasselquist}, {Hawley}, {Hearty}, {Gonzalez Hern{\'a}ndez}, {Ho}, {Hogg}, {Holley-Bockelmann}, {Holtzman}, {Holzer}, {Huehnerhoff}, {Hutchinson}, {Hwang}, {Ibarra-Medel}, {da Silva Ilha}, {Ivans}, {Ivory}, {Jackson}, {Jensen}, {Johnson}, {Jones}, {J{\"o}nsson}, {Jullo}, {Kamble}, {Kinemuchi}, {Kirkby}, {Kitaura}, {Klaene}, {Knapp}, {Kneib}, {Kollmeier}, {Lacerna}, {Lane}, {Lang}, {Law},
  {Lazarz}, {Lee}, {Le Goff}, {Liang}, {Li}, {Li}, {Lian}, {Lima}, {Lin}, {Lin}, {Bertran de Lis}, {Liu}, {de Icaza Lizaola}, {Long}, {Lucatello}, {Lundgren}, {MacDonald}, {Deconto Machado}, {MacLeod}, {Mahadevan}, {Geimba Maia}, {Maiolino}, {Majewski}, {Malanushenko}, {Malanushenko}, {Manchado}, {Mao}, {Maraston}, {Marques-Chaves}, {Masseron}, {Masters}, {McBride}, {McDermid}, {McGrath}, {McGreer}, {Medina Pe{\~n}a}, {Melendez}, {Merloni}, {Merrifield}, {Meszaros}, {Meza}, {Minchev}, {Minniti}, {Miyaji}, {More}, {Mulchaey}, {M{\"u}ller-S{\'a}nchez}, {Muna}, {Munoz}, {Myers}, {Nair}, {Nandra}, {Correa do Nascimento}, {Negrete}, {Ness}, {Newman}, {Nichol}, {Nidever}, {Nitschelm}, {Ntelis}, {O'Connell}, {Oelkers}, {Oravetz}, {Oravetz}, {Pace}, {Padilla}, {Palanque-Delabrouille}, {Alonso Palicio}, {Pan}, {Parejko}, {Parikh}, {P{\^a}ris}, {Park}, {Patten}, {Peirani}, {Pellejero-Ibanez}, {Penny}, {Percival}, {Perez-Fournon}, {Petitjean}, {Pieri}, {Pinsonneault}, {Pisani}, {Poleski}, {Prada}, {Prakash}, {Queiroz},
  {Raddick}, {Raichoor}, {Barboza Rembold}, {Richstein}, {Riffel}, {Riffel}, {Rix}, {Robin}, {Rockosi}, {Rodr{\'\i}guez-Torres}, {Roman-Lopes}, {Rom{\'a}n-Z{\'u}{\~n}iga}, {Rosado}, {Ross}, {Rossi}, {Ruan}, {Ruggeri}, {Rykoff}, {Salazar-Albornoz}, {Salvato}, {S{\'a}nchez}, {Aguado}, {S{\'a}nchez-Gallego}, {Santana}, {Santiago}, {Sayres}, {Schiavon}, {da Silva Schimoia}, {Schlafly}, {Schlegel}, {Schneider}, {Schultheis}, {Schuster}, {Schwope}, {Seo}, {Shao}, {Shen}, {Shetrone}, {Shull}, {Simon}, {Skinner}, {Skrutskie}, {Slosar}, {Smith}, {Sobeck}, {Sobreira}, {Somers}, {Souto}, {Stark}, {Stassun}, {Stauffer}, {Steinmetz}, {Storchi-Bergmann}, {Streblyanska}, {Stringfellow}, {Su{\'a}rez}, {Sun}, {Suzuki}, {Szigeti}, {Taghizadeh-Popp}, {Tang}, {Tao}, {Tayar}, {Tembe}, {Teske}, {Thakar}, {Thomas}, {Thompson}, {Tinker}, {Tissera}, {Tojeiro}, {Hernandez Toledo}, {de la Torre}, {Tremonti}, {Troup}, {Valenzuela}, {Martinez Valpuesta}, {Vargas-Gonz{\'a}lez}, {Vargas-Maga{\~n}a}, {Vazquez}, {Villanova}, {Vivek}, {Vogt},
  {Wake}, {Walterbos}, {Wang}, {Weaver}, {Weijmans}, {Weinberg}, {Westfall}, {Whelan}, {Wild}, {Wilson}, {Wood-Vasey}, {Wylezalek}, {Xiao}, {Yan}, {Yang}, {Ybarra}, {Y{\`e}che}, {Zakamska}, {Zamora}, {Zarrouk}, {Zasowski}, {Zhang}, {Zhao}, {Zheng}, {Zheng}, {Zhou}, {Zhou}, {Zhu}, {Zoccali}, \& {Zou}}]{Blanton2017}
{Blanton}, M.~R., {Bershady}, M.~A., {Abolfathi}, B., {et~al.} 2017, \aj, 154, 28, \dodoi{10.3847/1538-3881/aa7567}

\bibitem[{Blondel {et~al.}(2021)Blondel, Berthet, Cuturi, Frostig, Hoyer, Llinares-L{\'o}pez, Pedregosa, \& Vert}]{jaxopt_implicit_diff}
Blondel, M., Berthet, Q., Cuturi, M., {et~al.} 2021, arXiv preprint arXiv:2105.15183

\bibitem[{{Bowen} \& {Vaughan}(1973)}]{BowenVaughan1973}
{Bowen}, I.~S., \& {Vaughan}, A.~H., J. 1973, \ao, 12, 1430, \dodoi{10.1364/AO.12.001430}

\bibitem[{Bradbury {et~al.}(2018)Bradbury, Frostig, Hawkins, Johnson, Leary, Maclaurin, Necula, Paszke, Vander{P}las, Wanderman-{M}ilne, \& Zhang}]{jax2018github}
Bradbury, J., Frostig, R., Hawkins, P., {et~al.} 2018, {JAX}: composable transformations of {P}ython+{N}um{P}y programs, 0.3.13.
\newblock \url{http://github.com/google/jax}

\bibitem[{{Buck} \& {Schwarz}(2024)}]{Buck2024}
{Buck}, T., \& {Schwarz}, C. 2024, arXiv e-prints, arXiv:2410.16081, \dodoi{10.48550/arXiv.2410.16081}

\bibitem[{{Buder} {et~al.}(2020){Buder}, {Sharma}, {Kos}, {Amarsi}, {Nordlander}, {Lind}, {Martell}, {Asplund}, {Bland-Hawthorn}, {Casey}, {De Silva}, {D'Orazi}, {Freeman}, {Hayden}, {Lewis}, {Lin}, {Schlesinger}, {Simpson}, {Stello}, {Zucker}, {Zwitter}, {Beeson}, {Buck}, {Casagrande}, {Clark}, {Cotar}, {Da Costa}, {de Grijs}, {Feuillet}, {Horner}, {Khanna}, {Kafle}, {Liu}, {Montet}, {Nandakumar}, {Nataf}, {Ness}, {Spina}, {Traven}, {Trepper-Garcia}, {Ting}, {Vogrincic}, {Wittenmyer}, {Zerjal}, \& {the GALAH collaboration}}]{Buder2020}
{Buder}, S., {Sharma}, S., {Kos}, J., {et~al.} 2020, arXiv e-prints, arXiv:2011.02505.
\newblock \doarXiv{2011.02505}

\bibitem[{{Casey} {et~al.}(2016){Casey}, {Hogg}, {Ness}, {Rix}, {Ho}, \& {Gilmore}}]{Casey2016}
{Casey}, A.~R., {Hogg}, D.~W., {Ness}, M., {et~al.} 2016, arXiv e-prints, arXiv:1603.03040, \dodoi{10.48550/arXiv.1603.03040}

\bibitem[{{Ciuca} \& {Ting}(2022)}]{Ciuca2022}
{Ciuca}, I., \& {Ting}, Y.-S. 2022, in Machine Learning for Astrophysics, 17, \dodoi{10.48550/arXiv.2207.02785}

\bibitem[{Cooper {et~al.}(2023)Cooper, Koposov, Allende~Prieto, Manser, Kizhuprakkat, Myers, Dey, Gänsicke, Li, Rockosi, Valluri, Najita, Deason, Raichoor, Wang, Ting, Kim, Carrillo, Wang, Beraldo~e Silva, Han, Ding, Sánchez-Conde, Aguilar, Ahlen, Bailey, Belokurov, Brooks, Cunha, Dawson, de~la Macorra, Doel, Eisenstein, Fagrelius, Fanning, Font-Ribera, Forero-Romero, Gaztañaga, A~Gontcho, Guy, Honscheid, Kehoe, Kisner, Kremin, Landriau, Levi, Martini, Meisner, Miquel, Moustakas, Nie, Palanque-Delabrouille, Percival, Poppett, Prada, Rehemtulla, Schlafly, Schlegel, Schubnell, Sharples, Tarlé, Wechsler, Weinberg, Zhou, \& Zou}]{Cooper_2023}
Cooper, A.~P., Koposov, S.~E., Allende~Prieto, C., {et~al.} 2023, The Astrophysical Journal, 947, 37, \dodoi{10.3847/1538-4357/acb3c0}

\bibitem[{{Cunha} {et~al.}(2017){Cunha}, {Smith}, {Hasselquist}, {Souto}, {Shetrone}, {Allende Prieto}, {Bizyaev}, {Frinchaboy}, {Garc{\'\i}a-Hern{\'a}ndez}, {Holtzman}, {Johnson}, {J{\H{o}}nsson}, {Majewski}, {M{\'e}sz{\'a}ros}, {Nidever}, {Pinsonneault}, {Schiavon}, {Sobeck}, {Skrutskie}, {Zamora}, {Zasowski}, \& {Fern{\'a}ndez-Trincado}}]{Cunha2017}
{Cunha}, K., {Smith}, V.~V., {Hasselquist}, S., {et~al.} 2017, \apj, 844, 145, \dodoi{10.3847/1538-4357/aa7beb}

\bibitem[{{de Jong} {et~al.}(2012){de Jong}, {Bellido-Tirado}, {Chiappini}, {Depagne}, {Haynes}, {Johl}, {Schnurr}, {Schwope}, {Walcher}, {Dionies}, {Haynes}, {Kelz}, {Kitaura}, {Lamer}, {Minchev}, {M{\"u}ller}, {Nuza}, {Olaya}, {Piffl}, {Popow}, {Steinmetz}, {Ural}, {Williams}, {Winkler}, {Wisotzki}, {Ansorge}, {Banerji}, {Gonzalez Solares}, {Irwin}, {Kennicutt}, {King}, {McMahon}, {Koposov}, {Parry}, {Sun}, {Walton}, {Finger}, {Iwert}, {Krumpe}, {Lizon}, {Vincenzo}, {Amans}, {Bonifacio}, {Cohen}, {Francois}, {Jagourel}, {Mignot}, {Royer}, {Sartoretti}, {Bender}, {Grupp}, {Hess}, {Lang-Bardl}, {Muschielok}, {B{\"o}hringer}, {Boller}, {Bongiorno}, {Brusa}, {Dwelly}, {Merloni}, {Nandra}, {Salvato}, {Pragt}, {Navarro}, {Gerlofsma}, {Roelfsema}, {Dalton}, {Middleton}, {Tosh}, {Boeche}, {Caffau}, {Christlieb}, {Grebel}, {Hansen}, {Koch}, {Ludwig}, {Quirrenbach}, {Sbordone}, {Seifert}, {Thimm}, {Trifonov}, {Helmi}, {Trager}, {Feltzing}, {Korn}, \& {Boland}}]{DeJong2012}
{de Jong}, R.~S., {Bellido-Tirado}, O., {Chiappini}, C., {et~al.} 2012, in Society of Photo-Optical Instrumentation Engineers (SPIE) Conference Series, Vol. 8446, Ground-based and Airborne Instrumentation for Astronomy IV, ed. I.~S. {McLean}, S.~K. {Ramsay}, \& H.~{Takami}, 84460T, \dodoi{10.1117/12.926239}

\bibitem[{{Eilers} {et~al.}(2019){Eilers}, {Hogg}, {Rix}, \& {Ness}}]{Eilers2019}
{Eilers}, A.-C., {Hogg}, D.~W., {Rix}, H.-W., \& {Ness}, M.~K. 2019, \apj, 871, 120, \dodoi{10.3847/1538-4357/aaf648}

\bibitem[{{Freeman}(2012)}]{Freeman2012}
{Freeman}, K.~C. 2012, in Astronomical Society of the Pacific Conference Series, Vol. 458, Galactic Archaeology: Near-Field Cosmology and the Formation of the Milky Way, ed. W.~{Aoki}, M.~{Ishigaki}, T.~{Suda}, T.~{Tsujimoto}, \& N.~{Arimoto}, 393

\bibitem[{{Gaia Collaboration} {et~al.}(2020){Gaia Collaboration}, {Brown}, {Vallenari}, {Prusti}, {de Bruijne}, {Babusiaux}, \& {Biermann}}]{Gaia2020}
{Gaia Collaboration}, {Brown}, A.~G.~A., {Vallenari}, A., {et~al.} 2020, arXiv e-prints, arXiv:2012.01533.
\newblock \doarXiv{2012.01533}

\bibitem[{{Gaia Collaboration} {et~al.}(2023){Gaia Collaboration}, {Vallenari}, {Brown}, {Prusti}, {de Bruijne}, {Arenou}, {Babusiaux}, {Biermann}, {Creevey}, {Ducourant}, {Evans}, {Eyer}, {Guerra}, {Hutton}, {Jordi}, {Klioner}, {Lammers}, {Lindegren}, {Luri}, {Mignard}, {Panem}, {Pourbaix}, {Randich}, {Sartoretti}, {Soubiran}, {Tanga}, {Walton}, {Bailer-Jones}, {Bastian}, {Drimmel}, {Jansen}, {Katz}, {Lattanzi}, {van Leeuwen}, {Bakker}, {Cacciari}, {Casta{\~n}eda}, {De Angeli}, {Fabricius}, {Fouesneau}, {Fr{\'e}mat}, {Galluccio}, {Guerrier}, {Heiter}, {Masana}, {Messineo}, {Mowlavi}, {Nicolas}, {Nienartowicz}, {Pailler}, {Panuzzo}, {Riclet}, {Roux}, {Seabroke}, {Sordo}, {Th{\'e}venin}, {Gracia-Abril}, {Portell}, {Teyssier}, {Altmann}, {Andrae}, {Audard}, {Bellas-Velidis}, {Benson}, {Berthier}, {Blomme}, {Burgess}, {Busonero}, {Busso}, {C{\'a}novas}, {Carry}, {Cellino}, {Cheek}, {Clementini}, {Damerdji}, {Davidson}, {de Teodoro}, {Nu{\~n}ez Campos}, {Delchambre}, {Dell'Oro}, {Esquej},
  {Fern{\'a}ndez-Hern{\'a}ndez}, {Fraile}, {Garabato}, {Garc{\'\i}a-Lario}, {Gosset}, {Haigron}, {Halbwachs}, {Hambly}, {Harrison}, {Hern{\'a}ndez}, {Hestroffer}, {Hodgkin}, {Holl}, {Jan{\ss}en}, {Jevardat de Fombelle}, {Jordan}, {Krone-Martins}, {Lanzafame}, {L{\"o}ffler}, {Marchal}, {Marrese}, {Moitinho}, {Muinonen}, {Osborne}, {Pancino}, {Pauwels}, {Recio-Blanco}, {Reyl{\'e}}, {Riello}, {Rimoldini}, {Roegiers}, {Rybizki}, {Sarro}, {Siopis}, {Smith}, {Sozzetti}, {Utrilla}, {van Leeuwen}, {Abbas}, {{\'A}brah{\'a}m}, {Abreu Aramburu}, {Aerts}, {Aguado}, {Ajaj}, {Aldea-Montero}, {Altavilla}, {{\'A}lvarez}, {Alves}, {Anders}, {Anderson}, {Anglada Varela}, {Antoja}, {Baines}, {Baker}, {Balaguer-N{\'u}{\~n}ez}, {Balbinot}, {Balog}, {Barache}, {Barbato}, {Barros}, {Barstow}, {Bartolom{\'e}}, {Bassilana}, {Bauchet}, {Becciani}, {Bellazzini}, {Berihuete}, {Bernet}, {Bertone}, {Bianchi}, {Binnenfeld}, {Blanco-Cuaresma}, {Blazere}, {Boch}, {Bombrun}, {Bossini}, {Bouquillon}, {Bragaglia}, {Bramante}, {Breedt},
  {Bressan}, {Brouillet}, {Brugaletta}, {Bucciarelli}, {Burlacu}, {Butkevich}, {Buzzi}, {Caffau}, {Cancelliere}, {Cantat-Gaudin}, {Carballo}, {Carlucci}, {Carnerero}, {Carrasco}, {Casamiquela}, {Castellani}, {Castro-Ginard}, {Chaoul}, {Charlot}, {Chemin}, {Chiaramida}, {Chiavassa}, {Chornay}, {Comoretto}, {Contursi}, {Cooper}, {Cornez}, {Cowell}, {Crifo}, {Cropper}, {Crosta}, {Crowley}, {Dafonte}, {Dapergolas}, {David}, {David}, {de Laverny}, {De Luise}, {De March}, {De Ridder}, {de Souza}, {de Torres}, {del Peloso}, {del Pozo}, {Delbo}, {Delgado}, {Delisle}, {Demouchy}, {Dharmawardena}, {Di Matteo}, {Diakite}, {Diener}, {Distefano}, {Dolding}, {Edvardsson}, {Enke}, {Fabre}, {Fabrizio}, {Faigler}, {Fedorets}, {Fernique}, {Fienga}, {Figueras}, {Fournier}, {Fouron}, {Fragkoudi}, {Gai}, {Garcia-Gutierrez}, {Garcia-Reinaldos}, {Garc{\'\i}a-Torres}, {Garofalo}, {Gavel}, {Gavras}, {Gerlach}, {Geyer}, {Giacobbe}, {Gilmore}, {Girona}, {Giuffrida}, {Gomel}, {Gomez}, {Gonz{\'a}lez-N{\'u}{\~n}ez},
  {Gonz{\'a}lez-Santamar{\'\i}a}, {Gonz{\'a}lez-Vidal}, {Granvik}, {Guillout}, {Guiraud}, {Guti{\'e}rrez-S{\'a}nchez}, {Guy}, {Hatzidimitriou}, {Hauser}, {Haywood}, {Helmer}, {Helmi}, {Sarmiento}, {Hidalgo}, {Hilger}, {H{\l}adczuk}, {Hobbs}, {Holland}, {Huckle}, {Jardine}, {Jasniewicz}, {Jean-Antoine Piccolo}, {Jim{\'e}nez-Arranz}, {Jorissen}, {Juaristi Campillo}, {Julbe}, {Karbevska}, {Kervella}, {Khanna}, {Kontizas}, {Kordopatis}, {Korn}, {K{\'o}sp{\'a}l}, {Kostrzewa-Rutkowska}, {Kruszy{\'n}ska}, {Kun}, {Laizeau}, {Lambert}, {Lanza}, {Lasne}, {Le Campion}, {Lebreton}, {Lebzelter}, {Leccia}, {Leclerc}, {Lecoeur-Taibi}, {Liao}, {Licata}, {Lindstr{\o}m}, {Lister}, {Livanou}, {Lobel}, {Lorca}, {Loup}, {Madrero Pardo}, {Magdaleno Romeo}, {Managau}, {Mann}, {Manteiga}, {Marchant}, {Marconi}, {Marcos}, {Marcos Santos}, {Mar{\'\i}n Pina}, {Marinoni}, {Marocco}, {Marshall}, {Martin Polo}, {Mart{\'\i}n-Fleitas}, {Marton}, {Mary}, {Masip}, {Massari}, {Mastrobuono-Battisti}, {Mazeh}, {McMillan}, {Messina}, {Michalik},
  {Millar}, {Mints}, {Molina}, {Molinaro}, {Moln{\'a}r}, {Monari}, {Mongui{\'o}}, {Montegriffo}, {Montero}, {Mor}, {Mora}, {Morbidelli}, {Morel}, {Morris}, {Muraveva}, {Murphy}, {Musella}, {Nagy}, {Noval}, {Oca{\~n}a}, {Ogden}, {Ordenovic}, {Osinde}, {Pagani}, {Pagano}, {Palaversa}, {Palicio}, {Pallas-Quintela}, {Panahi}, {Payne-Wardenaar}, {Pe{\~n}alosa Esteller}, {Penttil{\"a}}, {Pichon}, {Piersimoni}, {Pineau}, {Plachy}, {Plum}, {Poggio}, {Pr{\v{s}}a}, {Pulone}, {Racero}, {Ragaini}, {Rainer}, {Raiteri}, {Rambaux}, {Ramos}, {Ramos-Lerate}, {Re Fiorentin}, {Regibo}, {Richards}, {Rios Diaz}, {Ripepi}, {Riva}, {Rix}, {Rixon}, {Robichon}, {Robin}, {Robin}, {Roelens}, {Rogues}, {Rohrbasser}, {Romero-G{\'o}mez}, {Rowell}, {Royer}, {Ruz Mieres}, {Rybicki}, {Sadowski}, {S{\'a}ez N{\'u}{\~n}ez}, {Sagrist{\`a} Sell{\'e}s}, {Sahlmann}, {Salguero}, {Samaras}, {Sanchez Gimenez}, {Sanna}, {Santove{\~n}a}, {Sarasso}, {Schultheis}, {Sciacca}, {Segol}, {Segovia}, {S{\'e}gransan}, {Semeux}, {Shahaf}, {Siddiqui}, {Siebert},
  {Siltala}, {Silvelo}, {Slezak}, {Slezak}, {Smart}, {Snaith}, {Solano}, {Solitro}, {Souami}, {Souchay}, {Spagna}, {Spina}, {Spoto}, {Steele}, {Steidelm{\"u}ller}, {Stephenson}, {S{\"u}veges}, {Surdej}, {Szabados}, {Szegedi-Elek}, {Taris}, {Taylor}, {Teixeira}, {Tolomei}, {Tonello}, {Torra}, {Torra}, {Torralba Elipe}, {Trabucchi}, {Tsounis}, {Turon}, {Ulla}, {Unger}, {Vaillant}, {van Dillen}, {van Reeven}, {Vanel}, {Vecchiato}, {Viala}, {Vicente}, {Voutsinas}, {Weiler}, {Wevers}, {Wyrzykowski}, {Yoldas}, {Yvard}, {Zhao}, {Zorec}, {Zucker}, \& {Zwitter}}]{Gaia2023}
{Gaia Collaboration}, {Vallenari}, A., {Brown}, A.~G.~A., {et~al.} 2023, \aap, 674, A1, \dodoi{10.1051/0004-6361/202243940}

\bibitem[{{Garc{\'\i}a P{\'e}rez} {et~al.}(2016){Garc{\'\i}a P{\'e}rez}, {Allende Prieto}, {Holtzman}, {Shetrone}, {M{\'e}sz{\'a}ros}, {Bizyaev}, {Carrera}, {Cunha}, {Garc{\'\i}a-Hern{\'a}ndez}, {Johnson}, {Majewski}, {Nidever}, {Schiavon}, {Shane}, {Smith}, {Sobeck}, {Troup}, {Zamora}, {Weinberg}, {Bovy}, {Eisenstein}, {Feuillet}, {Frinchaboy}, {Hayden}, {Hearty}, {Nguyen}, {O'Connell}, {Pinsonneault}, {Wilson}, \& {Zasowski}}]{Perez2015}
{Garc{\'\i}a P{\'e}rez}, A.~E., {Allende Prieto}, C., {Holtzman}, J.~A., {et~al.} 2016, \aj, 151, 144, \dodoi{10.3847/0004-6256/151/6/144}

\bibitem[{{Gilmore} {et~al.}(2012){Gilmore}, {Randich}, {Asplund}, {Binney}, {Bonifacio}, {Drew}, {Feltzing}, {Ferguson}, {Jeffries}, {Micela}, {Negueruela}, {Prusti}, {Rix}, {Vallenari}, {Alfaro}, {Allende-Prieto}, {Babusiaux}, {Bensby}, {Blomme}, {Bragaglia}, {Flaccomio}, {Fran{\c{c}}ois}, {Irwin}, {Koposov}, {Korn}, {Lanzafame}, {Pancino}, {Paunzen}, {Recio-Blanco}, {Sacco}, {Smiljanic}, {Van Eck}, {Walton}, {Aden}, {Aerts}, {Affer}, {Alcala}, {Altavilla}, {Alves}, {Antoja}, {Arenou}, {Argiroffi}, {Asensio Ramos}, {Bailer-Jones}, {Balaguer-Nunez}, {Bayo}, {Barbuy}, {Barisevicius}, {Barrado y Navascues}, {Battistini}, {Bellas Velidis}, {Bellazzini}, {Belokurov}, {Bergemann}, {Bertelli}, {Biazzo}, {Bienayme}, {Bland-Hawthorn}, {Boeche}, {Bonito}, {Boudreault}, {Bouvier}, {Brandao}, {Brown}, {de Bruijne}, {Burleigh}, {Caballero}, {Caffau}, {Calura}, {Capuzzo-Dolcetta}, {Caramazza}, {Carraro}, {Casagrande}, {Casewell}, {Chapman}, {Chiappini}, {Chorniy}, {Christlieb}, {Cignoni}, {Cocozza}, {Colless}, {Collet},
  {Collins}, {Correnti}, {Covino}, {Crnojevic}, {Cropper}, {Cunha}, {Damiani}, {David}, {Delgado}, {Duffau}, {Edvardsson}, {Eldridge}, {Enke}, {Eriksson}, {Evans}, {Eyer}, {Famaey}, {Fellhauer}, {Ferreras}, {Figueras}, {Fiorentino}, {Flynn}, {Folha}, {Franciosini}, {Frasca}, {Freeman}, {Fremat}, {Friel}, {Gaensicke}, {Gameiro}, {Garzon}, {Geier}, {Geisler}, {Gerhard}, {Gibson}, {Gomboc}, {Gomez}, {Gonzalez-Fernandez}, {Gonzalez Hernandez}, {Gosset}, {Grebel}, {Greimel}, {Groenewegen}, {Grundahl}, {Guarcello}, {Gustafsson}, {Hadrava}, {Hatzidimitriou}, {Hambly}, {Hammersley}, {Hansen}, {Haywood}, {Heber}, {Heiter}, {Held}, {Helmi}, {Hensler}, {Herrero}, {Hill}, {Hodgkin}, {Huelamo}, {Huxor}, {Ibata}, {Jackson}, {de Jong}, {Jonker}, {Jordan}, {Jordi}, {Jorissen}, {Katz}, {Kawata}, {Keller}, {Kharchenko}, {Klement}, {Klutsch}, {Knude}, {Koch}, {Kochukhov}, {Kontizas}, {Koubsky}, {Lallement}, {de Laverny}, {van Leeuwen}, {Lemasle}, {Lewis}, {Lind}, {Lindstrom}, {Lobel}, {Lopez Santiago}, {Lucas}, {Ludwig},
  {Lueftinger}, {Magrini}, {Maiz Apellaniz}, {Maldonado}, {Marconi}, {Marino}, {Martayan}, {Martinez-Valpuesta}, {Matijevic}, {McMahon}, {Messina}, {Meyer}, {Miglio}, {Mikolaitis}, {Minchev}, {Minniti}, {Moitinho}, {Momany}, {Monaco}, {Montalto}, {Monteiro}, {Monier}, {Montes}, {Mora}, {Moraux}, {Morel}, {Mowlavi}, {Mucciarelli}, {Munari}, {Napiwotzki}, {Nardetto}, {Naylor}, {Naze}, {Nelemans}, {Okamoto}, {Ortolani}, {Pace}, {Palla}, {Palous}, {Parker}, {Penarrubia}, {Pillitteri}, {Piotto}, {Posbic}, {Prisinzano}, {Puzeras}, {Quirrenbach}, {Ragaini}, {Read}, {Read}, {Reyle}, {De Ridder}, {Robichon}, {Robin}, {Roeser}, {Romano}, {Royer}, {Ruchti}, {Ruzicka}, {Ryan}, {Ryde}, {Santos}, {Sanz Forcada}, {Sarro Baro}, {Sbordone}, {Schilbach}, {Schmeja}, {Schnurr}, {Schoenrich}, {Scholz}, {Seabroke}, {Sharma}, {De Silva}, {Smith}, {Solano}, {Sordo}, {Soubiran}, {Sousa}, {Spagna}, {Steffen}, {Steinmetz}, {Stelzer}, {Stempels}, {Tabernero}, {Tautvaisiene}, {Thevenin}, {Torra}, {Tosi}, {Tolstoy}, {Turon}, {Walker},
  {Wambsganss}, {Worley}, {Venn}, {Vink}, {Wyse}, {Zaggia}, {Zeilinger}, {Zoccali}, {Zorec}, {Zucker}, {Zwitter}, \& {Gaia-ESO Survey Team}}]{Gilmore2012}
{Gilmore}, G., {Randich}, S., {Asplund}, M., {et~al.} 2012, The Messenger, 147, 25

\bibitem[{{Guiglion} {et~al.}(2024){Guiglion}, {Nepal}, {Chiappini}, {Khoperskov}, {Traven}, {Queiroz}, {Steinmetz}, {Valentini}, {Fournier}, {Vallenari}, {Youakim}, {Bergemann}, {M{\'e}sz{\'a}ros}, {Lucatello}, {Sordo}, {Fabbro}, {Minchev}, {Tautvai{\v{s}}ien{\.{e}}}, {Mikolaitis}, \& {Montalb{\'a}n}}]{Guiglion2024}
{Guiglion}, G., {Nepal}, S., {Chiappini}, C., {et~al.} 2024, \aap, 682, A9, \dodoi{10.1051/0004-6361/202347122}

\bibitem[{{Gunn} {et~al.}(2006){Gunn}, {Siegmund}, {Mannery}, {Owen}, {Hull}, {Leger}, {Carey}, {Knapp}, {York}, {Boroski}, {Kent}, {Lupton}, {Rockosi}, {Evans}, {Waddell}, {Anderson}, {Annis}, {Barentine}, {Bartoszek}, {Bastian}, {Bracker}, {Brewington}, {Briegel}, {Brinkmann}, {Brown}, {Carr}, {Czarapata}, {Drennan}, {Dombeck}, {Federwitz}, {Gillespie}, {Gonzales}, {Hansen}, {Harvanek}, {Hayes}, {Jordan}, {Kinney}, {Klaene}, {Kleinman}, {Kron}, {Kresinski}, {Lee}, {Limmongkol}, {Lindenmeyer}, {Long}, {Loomis}, {McGehee}, {Mantsch}, {Neilsen}, {Neswold}, {Newman}, {Nitta}, {Peoples}, {Pier}, {Prieto}, {Prosapio}, {Rivetta}, {Schneider}, {Snedden}, \& {Wang}}]{Gunn2006}
{Gunn}, J.~E., {Siegmund}, W.~A., {Mannery}, E.~J., {et~al.} 2006, \aj, 131, 2332, \dodoi{10.1086/500975}

\bibitem[{{Gustafsson} {et~al.}(2008){Gustafsson}, {Edvardsson}, {Eriksson}, {J{\o}rgensen}, {Nordlund}, \& {Plez}}]{Gustafsson2008}
{Gustafsson}, B., {Edvardsson}, B., {Eriksson}, K., {et~al.} 2008, \aap, 486, 951, \dodoi{10.1051/0004-6361:200809724}

\bibitem[{{Hall} {et~al.}(2019){Hall}, {Davies}, {Elsworth}, {Miglio}, {Bedding}, {Brown}, {Khan}, {Hawkins}, {Garc{\'\i}a}, {Chaplin}, \& {North}}]{Hall2019}
{Hall}, O.~J., {Davies}, G.~R., {Elsworth}, Y.~P., {et~al.} 2019, \mnras, 486, 3569, \dodoi{10.1093/mnras/stz1092}

\bibitem[{{Hasselquist} {et~al.}(2016){Hasselquist}, {Shetrone}, {Cunha}, {Smith}, {Holtzman}, {Lawler}, {Allende Prieto}, {Beers}, {Chojnowski}, {Fern{\'a}ndez-Trincado}, {Garc{\'\i}a-Hern{\'a}ndez}, {Hearty}, {Majewski}, {Pereira}, {Placco}, {Villanova}, \& {Zamora}}]{Hasselquist2016}
{Hasselquist}, S., {Shetrone}, M., {Cunha}, K., {et~al.} 2016, \apj, 833, 81, \dodoi{10.3847/1538-4357/833/1/81}

\bibitem[{{Ho} {et~al.}(2017{\natexlab{a}}){Ho}, {Rix}, {Ness}, {Hogg}, {Liu}, \& {Ting}}]{Ho2017b}
{Ho}, A. Y.~Q., {Rix}, H.-W., {Ness}, M.~K., {et~al.} 2017{\natexlab{a}}, \apj, 841, 40, \dodoi{10.3847/1538-4357/aa6db3}

\bibitem[{{Ho} {et~al.}(2017{\natexlab{b}}){Ho}, {Ness}, {Hogg}, {Rix}, {Liu}, {Yang}, {Zhang}, {Hou}, \& {Wang}}]{Ho2017}
{Ho}, A. Y.~Q., {Ness}, M.~K., {Hogg}, D.~W., {et~al.} 2017{\natexlab{b}}, \apj, 836, 5, \dodoi{10.3847/1538-4357/836/1/5}

\bibitem[{Hogg {et~al.}(2019)Hogg, Eilers, \& Rix}]{Hogg2019}
Hogg, D.~W., Eilers, A.-C., \& Rix, H.-W. 2019, The Astronomical Journal, 158, 147, \dodoi{10.3847/1538-3881/ab398c}

\bibitem[{{Horta} {et~al.}(2020){Horta}, {Schiavon}, {Mackereth}, {Beers}, {Fern{\'a}ndez-Trincado}, {Frinchaboy}, {Garc{\'\i}a-Hern{\'a}ndez}, {Geisler}, {Hasselquist}, {J{\"o}nsson}, {Lane}, {Majewski}, {M{\'e}sz{\'a}ros}, {Bidin}, {Nataf}, {Roman-Lopes}, {Nitschelm}, {Vargas-Gonz{\'a}lez}, \& {Zasowski}}]{Horta2020}
{Horta}, D., {Schiavon}, R.~P., {Mackereth}, J.~T., {et~al.} 2020, \mnras, 493, 3363, \dodoi{10.1093/mnras/staa478}

\bibitem[{Hunter(2007)}]{Hunter:2007}
Hunter, J.~D. 2007, Computing In Science \& Engineering, 9, 90, \dodoi{10.1109/MCSE.2007.55}

\bibitem[{Jin {et~al.}(2023)Jin, Trager, Dalton, Aguerri, Drew, Falcón-Barroso, Gänsicke, Hill, Iovino, Pieri, Poggianti, Smith, Vallenari, Abrams, Aguado, Antoja, Aragón-Salamanca, Ascasibar, Babusiaux, Balcells, Barrena, Battaglia, Belokurov, Bensby, Bonifacio, Bragaglia, Carrasco, Carrera, Cornwell, Domínguez-Palmero, Duncan, Famaey, Fariña, Gonzalez, Guest, Hatch, Hess, Hoskin, Irwin, Knapen, Koposov, Kuchner, Laigle, Lewis, Longhetti, Lucatello, Méndez-Abreu, Mercurio, Molaeinezhad, Monguió, Morrison, Murphy, Peralta de Arriba, Pérez, Pérez-Ràfols, Picó, Raddi, Romero-Gómez, Royer, Siebert, Seabroke, Som, Terrett, Thomas, Wesson, Worley, Alfaro, Allende Prieto, Alonso-Santiago, Amos, Ashley, Balaguer-Núñez, Balbinot, Bellazzini, Benn, Berlanas, Bernard, Best, Bettoni, Bianco, Bishop, Blomqvist, Boeche, Bolzonella, Bonoli, Bosma, Britavskiy, Busarello, Caffau, Cantat-Gaudin, Castro-Ginard, Couto, Carbajo-Hijarrubia, Carter, Casamiquela, Conrado, Corcho-Caballero, Costantin, Deason,
  de Burgos, De Grandi, Di Matteo, Domínguez-Gómez, Dorda, Drake, Dutta, Erkal, Feltzing, Ferré-Mateu, Feuillet, Figueras, Fossati, Franciosini, Frasca, Fumagalli, Gallazzi, García-Benito, Gentile Fusillo, Gebran, Gilbert, Gledhill, González Delgado, Greimel, Guarcello, Guerra, Gullieuszik, Haines, Hardcastle, Harris, Haywood, Helmi, Hernandez, Herrero, Hughes, Iršič, Jablonka, Jarvis, Jordi, Kondapally, Kordopatis, Krogager, La Barbera, Lam, Larsen, Lemasle, Lewis, Lhomé, Lind, Lodi, Longobardi, Lonoce, Magrini, Maíz Apellániz, Marchal, Marco, Martin, Matsuno, Maurogordato, Merluzzi, Miralda-Escudé, Molinari, Monari, Morelli, Mottram, Naylor, Negueruela, Oñorbe, Pancino, Peirani, Peletier, Pozzetti, Rainer, Ramos, Read, Rossi, Röttgering, Rubiño-Martín, Sabater, San Juan, Sanna, Schallig, Schiavon, Schultheis, Serra, Shimwell, Simón-Díaz, Smith, Sordo, Sorini, Soubiran, Starkenburg, Steele, Stott, Stuik, Tolstoy, Tortora, Tsantaki, Van der Swaelmen, van Weeren, Vergani, Verheijen,
  Verro, Vink, Vioque, Walcher, Walton, Wegg, Weijmans, Williams, Wilson, Wright, Xylakis-Dornbusch, Youakim, Zibetti, \& Zurita}]{Jin_2023}
Jin, S., Trager, S.~C., Dalton, G.~B., {et~al.} 2023, Monthly Notices of the Royal Astronomical Society, 530, 2688–2730, \dodoi{10.1093/mnras/stad557}

\bibitem[{{Jofr{\'e}} {et~al.}(2014){Jofr{\'e}}, {Heiter}, {Soubiran}, {Blanco-Cuaresma}, {Worley}, {Pancino}, {Cantat-Gaudin}, {Magrini}, {Bergemann}, {Gonz{\'a}lez Hern{\'a}ndez}, {Hill}, {Lardo}, {de Laverny}, {Lind}, {Masseron}, {Montes}, {Mucciarelli}, {Nordlander}, {Recio Blanco}, {Sobeck}, {Sordo}, {Sousa}, {Tabernero}, {Vallenari}, \& {Van Eck}}]{Jofre2014}
{Jofr{\'e}}, P., {Heiter}, U., {Soubiran}, C., {et~al.} 2014, \aap, 564, A133, \dodoi{10.1051/0004-6361/201322440}

\bibitem[{Koblischke \& Bovy(2024)}]{Nolan2024}
Koblischke, N., \& Bovy, J. 2024, SpectraFM: Tuning into Stellar Foundation Models.
\newblock \doarXiv{2411.04750}

\bibitem[{{Kollmeier} {et~al.}(2017){Kollmeier}, {Zasowski}, {Rix}, {Johns}, {Anderson}, {Drory}, {Johnson}, {Pogge}, {Bird}, {Blanc}, {Brownstein}, {Crane}, {De Lee}, {Klaene}, {Kreckel}, {MacDonald}, {Merloni}, {Ness}, {O'Brien}, {Sanchez-Gallego}, {Sayres}, {Shen}, {Thakar}, {Tkachenko}, {Aerts}, {Blanton}, {Eisenstein}, {Holtzman}, {Maoz}, {Nandra}, {Rockosi}, {Weinberg}, {Bovy}, {Casey}, {Chaname}, {Clerc}, {Conroy}, {Eracleous}, {G{\"a}nsicke}, {Hekker}, {Horne}, {Kauffmann}, {McQuinn}, {Pellegrini}, {Schinnerer}, {Schlafly}, {Schwope}, {Seibert}, {Teske}, \& {van Saders}}]{Kollmeier2017}
{Kollmeier}, J.~A., {Zasowski}, G., {Rix}, H.-W., {et~al.} 2017, arXiv e-prints, arXiv:1711.03234, \dodoi{10.48550/arXiv.1711.03234}

\bibitem[{{Kordopatis} {et~al.}(2013){Kordopatis}, {Gilmore}, {Steinmetz}, {Boeche}, {Seabroke}, {Siebert}, {Zwitter}, {Binney}, {de Laverny}, {Recio-Blanco}, {Williams}, {Piffl}, {Enke}, {Roeser}, {Bijaoui}, {Wyse}, {Freeman}, {Munari}, {Carrillo}, {Anguiano}, {Burton}, {Campbell}, {Cass}, {Fiegert}, {Hartley}, {Parker}, {Reid}, {Ritter}, {Russell}, {Stupar}, {Watson}, {Bienaym{\'e}}, {Bland-Hawthorn}, {Gerhard}, {Gibson}, {Grebel}, {Helmi}, {Navarro}, {Conrad}, {Famaey}, {Faure}, {Just}, {Kos}, {Matijevi{\v{c}}}, {McMillan}, {Minchev}, {Scholz}, {Sharma}, {Siviero}, {de Boer}, \& {{\v{Z}}erjal}}]{Kordopatis2013}
{Kordopatis}, G., {Gilmore}, G., {Steinmetz}, M., {et~al.} 2013, \aj, 146, 134, \dodoi{10.1088/0004-6256/146/5/134}

\bibitem[{Leung \& Bovy(2018)}]{Leung_2018_astroNN}
Leung, H.~W., \& Bovy, J. 2018, Monthly Notices of the Royal Astronomical Society, \dodoi{10.1093/mnras/sty3217}

\bibitem[{{Lewis} {et~al.}(2002){Lewis}, {Cannon}, {Taylor}, {Glazebrook}, {Bailey}, {Baldry}, {Barton}, {Bridges}, {Dalton}, {Farrell}, {Gray}, {Lankshear}, {McCowage}, {Parry}, {Sharples}, {Shortridge}, {Smith}, {Stevenson}, {Straede}, {Waller}, {Whittard}, {Wilcox}, \& {Willis}}]{Lewis2002}
{Lewis}, I.~J., {Cannon}, R.~D., {Taylor}, K., {et~al.} 2002, \mnras, 333, 279, \dodoi{10.1046/j.1365-8711.2002.05333.x}

\bibitem[{{Li} {et~al.}(2024){Li}, {Wong}, {Hogg}, {Rix}, \& {Chandra}}]{Li2024}
{Li}, J., {Wong}, K. W.~K., {Hogg}, D.~W., {Rix}, H.-W., \& {Chandra}, V. 2024, \apjs, 272, 2, \dodoi{10.3847/1538-4365/ad2b4d}

\bibitem[{{Majewski} {et~al.}(2017){Majewski}, {Schiavon}, {Frinchaboy}, {Allende Prieto}, {Barkhouser}, {Bizyaev}, {Blank}, {Brunner}, {Burton}, {Carrera}, {Chojnowski}, {Cunha}, {Epstein}, {Fitzgerald}, {Garc{\'\i}a P{\'e}rez}, {Hearty}, {Henderson}, {Holtzman}, {Johnson}, {Lam}, {Lawler}, {Maseman}, {M{\'e}sz{\'a}ros}, {Nelson}, {Nguyen}, {Nidever}, {Pinsonneault}, {Shetrone}, {Smee}, {Smith}, {Stolberg}, {Skrutskie}, {Walker}, {Wilson}, {Zasowski}, {Anders}, {Basu}, {Beland}, {Blanton}, {Bovy}, {Brownstein}, {Carlberg}, {Chaplin}, {Chiappini}, {Eisenstein}, {Elsworth}, {Feuillet}, {Fleming}, {Galbraith-Frew}, {Garc{\'\i}a}, {Garc{\'\i}a-Hern{\'a}ndez}, {Gillespie}, {Girardi}, {Gunn}, {Hasselquist}, {Hayden}, {Hekker}, {Ivans}, {Kinemuchi}, {Klaene}, {Mahadevan}, {Mathur}, {Mosser}, {Muna}, {Munn}, {Nichol}, {O'Connell}, {Parejko}, {Robin}, {Rocha-Pinto}, {Schultheis}, {Serenelli}, {Shane}, {Silva Aguirre}, {Sobeck}, {Thompson}, {Troup}, {Weinberg}, \& {Zamora}}]{Majewski2017}
{Majewski}, S.~R., {Schiavon}, R.~P., {Frinchaboy}, P.~M., {et~al.} 2017, \aj, 154, 94, \dodoi{10.3847/1538-3881/aa784d}

\bibitem[{{Manea} {et~al.}(2024){Manea}, {Hawkins}, {Ness}, {Buder}, {Martell}, \& {Zucker}}]{Manea2024}
{Manea}, C., {Hawkins}, K., {Ness}, M.~K., {et~al.} 2024, \apj, 972, 69, \dodoi{10.3847/1538-4357/ad58d9}

\bibitem[{{Martell} {et~al.}(2017){Martell}, {Sharma}, {Buder}, {Duong}, {Schlesinger}, {Simpson}, {Lind}, {Ness}, {Marshall}, {Asplund}, {Bland-Hawthorn}, {Casey}, {De Silva}, {Freeman}, {Kos}, {Lin}, {Zucker}, {Zwitter}, {Anguiano}, {Bacigalupo}, {Carollo}, {Casagrande}, {Da Costa}, {Horner}, {Huber}, {Hyde}, {Kafle}, {Lewis}, {Nataf}, {Navin}, {Stello}, {Tinney}, {Watson}, \& {Wittenmyer}}]{Martell2017}
{Martell}, S.~L., {Sharma}, S., {Buder}, S., {et~al.} 2017, \mnras, 465, 3203, \dodoi{10.1093/mnras/stw2835}

\bibitem[{McKinnon {et~al.}(2024)McKinnon, Ness, Rockosi, \& Guhathakurta}]{mckinnon2024}
McKinnon, K.~A., Ness, M.~K., Rockosi, C.~M., \& Guhathakurta, P. 2024, Data-driven Discovery of Diffuse Interstellar Bands with APOGEE Spectra.
\newblock \doarXiv{2307.05706}

\bibitem[{{M{\'e}sz{\'a}ros} {et~al.}(2013){M{\'e}sz{\'a}ros}, {Holtzman}, {Garc{\'\i}a P{\'e}rez}, {Allende Prieto}, {Schiavon}, {Basu}, {Bizyaev}, {Chaplin}, {Chojnowski}, {Cunha}, {Elsworth}, {Epstein}, {Frinchaboy}, {Garc{\'\i}a}, {Hearty}, {Hekker}, {Johnson}, {Kallinger}, {Koesterke}, {Majewski}, {Martell}, {Nidever}, {Pinsonneault}, {O'Connell}, {Shetrone}, {Smith}, {Wilson}, \& {Zasowski}}]{Meszaros2013}
{M{\'e}sz{\'a}ros}, S., {Holtzman}, J., {Garc{\'\i}a P{\'e}rez}, A.~E., {et~al.} 2013, \aj, 146, 133, \dodoi{10.1088/0004-6256/146/5/133}

\bibitem[{{Myers} {et~al.}(2022){Myers}, {Donor}, {Spoo}, {Frinchaboy}, {Cunha}, {Price-Whelan}, {Majewski}, {Beaton}, {Zasowski}, {O'Connell}, {Ray}, {Bizyaev}, {Chiappini}, {Garc{\'\i}a-Hern{\'a}ndez}, {Geisler}, {J{\"o}nsson}, {Lane}, {Longa-Pe{\~n}a}, {Minchev}, {Minniti}, {Nitschelm}, \& {Roman-Lopes}}]{Myers2022}
{Myers}, N., {Donor}, J., {Spoo}, T., {et~al.} 2022, \aj, 164, 85, \dodoi{10.3847/1538-3881/ac7ce5}

\bibitem[{{Ness} {et~al.}(2015){Ness}, {Hogg}, {Rix}, {Ho}, \& {Zasowski}}]{Ness2015}
{Ness}, M., {Hogg}, D.~W., {Rix}, H.~W., {Ho}, A. Y.~Q., \& {Zasowski}, G. 2015, \apj, 808, 16, \dodoi{10.1088/0004-637X/808/1/16}

\bibitem[{{Ness} {et~al.}(2016){Ness}, {Hogg}, {Rix}, {Martig}, {Pinsonneault}, \& {Ho}}]{Ness2016_cannon}
{Ness}, M., {Hogg}, D.~W., {Rix}, H.~W., {et~al.} 2016, \apj, 823, 114, \dodoi{10.3847/0004-637X/823/2/114}

\bibitem[{{Ness} \& {Lang}(2016)}]{Ness2016}
{Ness}, M., \& {Lang}, D. 2016, \aj, 152, 14, \dodoi{10.3847/0004-6256/152/1/14}

\bibitem[{{Ness} {et~al.}(2024){Ness}, {Mendel}, {Buder}, {Wheeler}, {Ji}, {Mijnarends}, {Venn}, {Starkenburg}, {Leaman}, {Grasha}, \& {Aquilina}}]{Ness2024}
{Ness}, M.~K., {Mendel}, J.~T., {Buder}, S., {et~al.} 2024, arXiv e-prints, arXiv:2407.17661, \dodoi{10.48550/arXiv.2407.17661}

\bibitem[{{Nidever} {et~al.}(2015){Nidever}, {Holtzman}, {Allende Prieto}, {Beland}, {Bender}, {Bizyaev}, {Burton}, {Desphande}, {Fleming}, {Garc{\'\i}a P{\'e}rez}, {Hearty}, {Majewski}, {M{\'e}sz{\'a}ros}, {Muna}, {Nguyen}, {Schiavon}, {Shetrone}, {Skrutskie}, {Sobeck}, \& {Wilson}}]{Nidever2015}
{Nidever}, D.~L., {Holtzman}, J.~A., {Allende Prieto}, C., {et~al.} 2015, \aj, 150, 173, \dodoi{10.1088/0004-6256/150/6/173}

\bibitem[{Nidever {et~al.}(2020)Nidever, Hasselquist, Hayes, Hawkins, Povick, Majewski, Smith, Anguiano, Stringfellow, Sobeck, \& et~al.}]{Nidever2020}
Nidever, D.~L., Hasselquist, S., Hayes, C.~R., {et~al.} 2020, The Astrophysical Journal, 895, 88, \dodoi{10.3847/1538-4357/ab7305}

\bibitem[{{Nolan} {et~al.}(2006){Nolan}, {Harva}, {Kab{\'a}n}, \& {Raychaudhury}}]{Nolan2006}
{Nolan}, L.~A., {Harva}, M.~O., {Kab{\'a}n}, A., \& {Raychaudhury}, S. 2006, \mnras, 366, 321, \dodoi{10.1111/j.1365-2966.2005.09868.x}

\bibitem[{Oliphant(2006--)}]{NumPy}
Oliphant, T. 2006--, {NumPy}: A guide to {NumPy}, USA: Trelgol Publishing.
\newblock \url{http://www.numpy.org/}

\bibitem[{Piskunov \& Valenti(2016)}]{Piskunov2016}
Piskunov, N., \& Valenti, J.~A. 2016, Astronomy $\&$ Astrophysics, 597, A16, \dodoi{10.1051/0004-6361/201629124}

\bibitem[{{Price-Whelan}(2017)}]{Price2017}
{Price-Whelan}, A.~M. 2017, The Journal of Open Source Software, 2, 388, \dodoi{10.21105/joss.00388}

\bibitem[{{R{\'o}{\.z}a{\'n}ski} {et~al.}(2024){R{\'o}{\.z}a{\'n}ski}, {Ting}, \& {Jab{\l}o{\'n}ska}}]{rozansnki2024}
{R{\'o}{\.z}a{\'n}ski}, T., {Ting}, Y.-S., \& {Jab{\l}o{\'n}ska}, M. 2024, arXiv e-prints, arXiv:2407.05751, \dodoi{10.48550/arXiv.2407.05751}

\bibitem[{{Santana} {et~al.}(2021){Santana}, {Beaton}, {Covey}, {O'Connell}, {Longa-Pe{\~n}a}, {Cohen}, {Fern{\'a}ndez-Trincado}, {Hayes}, {Zasowski}, {Sobeck}, {Majewski}, {Chojnowski}, {De Lee}, {Oelkers}, {Stringfellow}, {Almeida}, {Anguiano}, {Donor}, {Frinchaboy}, {Hasselquist}, {Johnson}, {Kollmeier}, {Nidever}, {Price-Whelan}, {Rojas-Arriagada}, {Schultheis}, {Shetrone}, {Simon}, {Aerts}, {Borissova}, {Drout}, {Geisler}, {Law}, {Medina}, {Minniti}, {Monachesi}, {Mu{\~n}oz}, {Poleski}, {Roman-Lopes}, {Schlaufman}, {Stutz}, {Teske}, {Tkachenko}, {Van Saders}, {Weinberger}, \& {Zoccali}}]{Santana2021}
{Santana}, F.~A., {Beaton}, R.~L., {Covey}, K.~R., {et~al.} 2021, arXiv e-prints, arXiv:2108.11908.
\newblock \doarXiv{2108.11908}

\bibitem[{{Schiavon} {et~al.}(2024){Schiavon}, {Phillips}, {Myers}, {Horta}, {Minniti}, {Allende Prieto}, {Anguiano}, {Beaton}, {Beers}, {Brownstein}, {Cohen}, {Fern{\'a}ndez-Trincado}, {Frinchaboy}, {J{\"o}nsson}, {Kisku}, {Lane}, {Majewski}, {Mason}, {M{\'e}sz{\'a}ros}, \& {Stringfellow}}]{Schiavon2024}
{Schiavon}, R.~P., {Phillips}, S.~G., {Myers}, N., {et~al.} 2024, \mnras, 528, 1393, \dodoi{10.1093/mnras/stad3020}

\bibitem[{{Sheinis} {et~al.}(2015){Sheinis}, {Anguiano}, {Asplund}, {Bacigalupo}, {Barden}, {Birchall}, {Bland-Hawthorn}, {Brzeski}, {Cannon}, {Carollo}, {Case}, {Casey}, {Churilov}, {Warrick}, {Dean}, {De Silva}, {D'Orazi}, {Duong}, {Farrell}, {Fiegert}, {Freeman}, {Gabriella}, {Gers}, {Goodwin}, {Gray}, {Green}, {Heald}, {Heijmans}, {Ireland}, {Jones}, {Kafle}, {Keller}, {Klauser}, {Kondrat}, {Kos}, {Lawrence}, {Lee}, {Mali}, {Martell}, {Mathews}, {Mayfield}, {Miziarski}, {Muller}, {Pai}, {Patterson}, {Penny}, {Orr}, {Schlesinger}, {Sharma}, {Shortridge}, {Simpson}, {Smedley}, {Smith}, {Stafford}, {Staszak}, {Vuong}, {Waller}, {de Boer}, {Xavier}, {Zheng}, {Zhelem}, {Zucker}, \& {Zwitter}}]{Sheinis2015}
{Sheinis}, A., {Anguiano}, B., {Asplund}, M., {et~al.} 2015, Journal of Astronomical Telescopes, Instruments, and Systems, 1, 035002, \dodoi{10.1117/1.JATIS.1.3.035002}

\bibitem[{{Smith} {et~al.}(2021){Smith}, {Bizyaev}, {Cunha}, {Shetrone}, {Souto}, {Allende Prieto}, {Masseron}, {M{\'e}sz{\'a}ros}, {J{\"o}nsson}, {Hasselquist}, {Osorio}, {Garc{\'\i}a-Hern{\'a}ndez}, {Plez}, {Beaton}, {Holtzman}, {Majewski}, {Stringfellow}, \& {Sobeck}}]{Smith2021}
{Smith}, V.~V., {Bizyaev}, D., {Cunha}, K., {et~al.} 2021, \aj, 161, 254, \dodoi{10.3847/1538-3881/abefdc}

\bibitem[{{Steinmetz} {et~al.}(2006){Steinmetz}, {Zwitter}, {Siebert}, {Watson}, {Freeman}, {Munari}, {Campbell}, {Williams}, {Seabroke}, {Wyse}, {Parker}, {Bienaym{\'e}}, {Roeser}, {Gibson}, {Gilmore}, {Grebel}, {Helmi}, {Navarro}, {Burton}, {Cass}, {Dawe}, {Fiegert}, {Hartley}, {Russell}, {Saunders}, {Enke}, {Bailin}, {Binney}, {Bland-Hawthorn}, {Boeche}, {Dehnen}, {Eisenstein}, {Evans}, {Fiorucci}, {Fulbright}, {Gerhard}, {Jauregi}, {Kelz}, {Mijovi{\'c}}, {Minchev}, {Parmentier}, {Pe{\~n}arrubia}, {Quillen}, {Read}, {Ruchti}, {Scholz}, {Siviero}, {Smith}, {Sordo}, {Veltz}, {Vidrih}, {von Berlepsch}, {Boyle}, \& {Schilbach}}]{Steinmetz2006}
{Steinmetz}, M., {Zwitter}, T., {Siebert}, A., {et~al.} 2006, \aj, 132, 1645, \dodoi{10.1086/506564}

\bibitem[{{Ting} {et~al.}(2018){Ting}, {Conroy}, {Rix}, \& {Asplund}}]{Ting2018}
{Ting}, Y.-S., {Conroy}, C., {Rix}, H.-W., \& {Asplund}, M. 2018, \apj, 860, 159, \dodoi{10.3847/1538-4357/aac6c9}

\bibitem[{{Ting} {et~al.}(2019){Ting}, {Conroy}, {Rix}, \& {Cargile}}]{Ting2019}
{Ting}, Y.-S., {Conroy}, C., {Rix}, H.-W., \& {Cargile}, P. 2019, \apj, 879, 69, \dodoi{10.3847/1538-4357/ab2331}

\bibitem[{{Wheeler} {et~al.}(2022){Wheeler}, {Abril-Cabezas}, {Trick}, {Fragkoudi}, \& {Ness}}]{Wheeler.2022}
{Wheeler}, A., {Abril-Cabezas}, I., {Trick}, W.~H., {Fragkoudi}, F., \& {Ness}, M. 2022, \apj, 935, 28, \dodoi{10.3847/1538-4357/ac7da0}

\bibitem[{{Wheeler} {et~al.}(2023){Wheeler}, {Abruzzo}, {Casey}, \& {Ness}}]{Wheeler2023}
{Wheeler}, A.~J., {Abruzzo}, M.~W., {Casey}, A.~R., \& {Ness}, M.~K. 2023, \aj, 165, 68, \dodoi{10.3847/1538-3881/acaaad}

\bibitem[{{Wilson} {et~al.}(2019){Wilson}, {Hearty}, {Skrutskie}, {Majewski}, {Holtzman}, {Eisenstein}, {Gunn}, {Blank}, {Henderson}, {Smee}, {Nelson}, {Nidever}, {Arns}, {Barkhouser}, {Barr}, {Beland}, {Bershady}, {Blanton}, {Brunner}, {Burton}, {Carey}, {Carr}, {Colque}, {Crane}, {Damke}, {Davidson}, {Dean}, {Di Mille}, {Don}, {Ebelke}, {Evans}, {Fitzgerald}, {Gillespie}, {Hall}, {Harding}, {Harding}, {Hammond}, {Hancock}, {Harrison}, {Hope}, {Horne}, {Karakla}, {Lam}, {Leger}, {MacDonald}, {Maseman}, {Matsunari}, {Melton}, {Mitcheltree}, {O'Brien}, {O'Connell}, {Patten}, {Richardson}, {Rieke}, {Rieke}, {Roman-Lopes}, {Schiavon}, {Sobeck}, {Stolberg}, {Stoll}, {Tembe}, {Trujillo}, {Uomoto}, {Vernieri}, {Walker}, {Weinberg}, {Young}, {Anthony-Brumfield}, {Bizyaev}, {Breslauer}, {De Lee}, {Downey}, {Halverson}, {Huehnerhoff}, {Klaene}, {Leon}, {Long}, {Mahadevan}, {Malanushenko}, {Nguyen}, {Owen}, {S{\'a}nchez-Gallego}, {Sayres}, {Shane}, {Shectman}, {Shetrone}, {Skinner}, {Stauffer}, \& {Zhao}}]{Wilson2019}
{Wilson}, J.~C., {Hearty}, F.~R., {Skrutskie}, M.~F., {et~al.} 2019, \pasp, 131, 055001, \dodoi{10.1088/1538-3873/ab0075}

\bibitem[{{Xiang} {et~al.}(2017){Xiang}, {Liu}, {Shi}, {Yuan}, {Huang}, {Chen}, {Wang}, {Tian}, {Wu}, {Yang}, {Zhang}, {Huo}, \& {Ren}}]{Xiang2017}
{Xiang}, M., {Liu}, X., {Shi}, J., {et~al.} 2017, \apjs, 232, 2, \dodoi{10.3847/1538-4365/aa80e4}

\bibitem[{{Xiang} {et~al.}(2019){Xiang}, {Ting}, {Rix}, {Sandford}, {Buder}, {Lind}, {Liu}, {Shi}, \& {Zhang}}]{Xiang2019}
{Xiang}, M., {Ting}, Y.-S., {Rix}, H.-W., {et~al.} 2019, \apjs, 245, 34, \dodoi{10.3847/1538-4365/ab5364}

\bibitem[{{Yanny} {et~al.}(2009){Yanny}, {Rockosi}, {Newberg}, {Knapp}, {Adelman-McCarthy}, {Alcorn}, {Allam}, {Allende Prieto}, {An}, {Anderson}, {Anderson}, {Bailer-Jones}, {Bastian}, {Beers}, {Bell}, {Belokurov}, {Bizyaev}, {Blythe}, {Bochanski}, {Boroski}, {Brinchmann}, {Brinkmann}, {Brewington}, {Carey}, {Cudworth}, {Evans}, {Evans}, {Gates}, {G{\"a}nsicke}, {Gillespie}, {Gilmore}, {Nebot Gomez-Moran}, {Grebel}, {Greenwell}, {Gunn}, {Jordan}, {Jordan}, {Harding}, {Harris}, {Hendry}, {Holder}, {Ivans}, {Ivezi{\v{c}}}, {Jester}, {Johnson}, {Kent}, {Kleinman}, {Kniazev}, {Krzesinski}, {Kron}, {Kuropatkin}, {Lebedeva}, {Lee}, {French Leger}, {L{\'e}pine}, {Levine}, {Lin}, {Long}, {Loomis}, {Lupton}, {Malanushenko}, {Malanushenko}, {Margon}, {Martinez-Delgado}, {McGehee}, {Monet}, {Morrison}, {Munn}, {Neilsen}, {Nitta}, {Norris}, {Oravetz}, {Owen}, {Padmanabhan}, {Pan}, {Peterson}, {Pier}, {Platson}, {Re Fiorentin}, {Richards}, {Rix}, {Schlegel}, {Schneider}, {Schreiber}, {Schwope}, {Sibley}, {Simmons},
  {Snedden}, {Allyn Smith}, {Stark}, {Stauffer}, {Steinmetz}, {Stoughton}, {SubbaRao}, {Szalay}, {Szkody}, {Thakar}, {Sivarani}, {Tucker}, {Uomoto}, {Vanden Berk}, {Vidrih}, {Wadadekar}, {Watters}, {Wilhelm}, {Wyse}, {Yarger}, \& {Zucker}}]{Yanny2009}
{Yanny}, B., {Rockosi}, C., {Newberg}, H.~J., {et~al.} 2009, \aj, 137, 4377, \dodoi{10.1088/0004-6256/137/5/4377}

\bibitem[{{Zasowski} {et~al.}(2013){Zasowski}, {Johnson}, {Frinchaboy}, {Majewski}, {Nidever}, {Rocha Pinto}, {Girardi}, {Andrews}, {Chojnowski}, {Cudworth}, {Jackson}, {Munn}, {Skrutskie}, {Beaton}, {Blake}, {Covey}, {Deshpande}, {Epstein}, {Fabbian}, {Fleming}, {Garcia Hernandez}, {Herrero}, {Mahadevan}, {M{\'e}sz{\'a}ros}, {Schultheis}, {Sellgren}, {Terrien}, {van Saders}, {Allende Prieto}, {Bizyaev}, {Burton}, {Cunha}, {da Costa}, {Hasselquist}, {Hearty}, {Holtzman}, {Garc{\'\i}a P{\'e}rez}, {Maia}, {O'Connell}, {O'Donnell}, {Pinsonneault}, {Santiago}, {Schiavon}, {Shetrone}, {Smith}, \& {Wilson}}]{Zasowski2013}
{Zasowski}, G., {Johnson}, J.~A., {Frinchaboy}, P.~M., {et~al.} 2013, \aj, 146, 81, \dodoi{10.1088/0004-6256/146/4/81}

\bibitem[{{Zasowski} {et~al.}(2017){Zasowski}, {Cohen}, {Chojnowski}, {Santana}, {Oelkers}, {Andrews}, {Beaton}, {Bender}, {Bird}, {Bovy}, {Carlberg}, {Covey}, {Cunha}, {Dell'Agli}, {Fleming}, {Frinchaboy}, {Garc{\'\i}a-Hern{\'a}ndez}, {Harding}, {Holtzman}, {Johnson}, {Kollmeier}, {Majewski}, {M{\'e}sz{\'a}ros}, {Munn}, {Mu{\~n}oz}, {Ness}, {Nidever}, {Poleski}, {Rom{\'a}n-Z{\'u}{\~n}iga}, {Shetrone}, {Simon}, {Smith}, {Sobeck}, {Stringfellow}, {Szigeti{\'a}ros}, {Tayar}, \& {Troup}}]{Zasowski2017}
{Zasowski}, G., {Cohen}, R.~E., {Chojnowski}, S.~D., {et~al.} 2017, \aj, 154, 198, \dodoi{10.3847/1538-3881/aa8df9}

\bibitem[{Zhang {et~al.}(2008)Zhang, Ghahramani, \& Yang}]{Zhang2008}
Zhang, J., Ghahramani, Z., \& Yang, Y. 2008, Machine Learning, 73, 221, \dodoi{10.1007/s10994-008-5050-1}

\bibitem[{{Zhao} {et~al.}(2012){Zhao}, {Zhao}, {Chu}, {Jing}, \& {Deng}}]{Zhao2012}
{Zhao}, G., {Zhao}, Y.-H., {Chu}, Y.-Q., {Jing}, Y.-P., \& {Deng}, L.-C. 2012, Research in Astronomy and Astrophysics, 12, 723, \dodoi{10.1088/1674-4527/12/7/002}

\end{thebibliography}

\appendix

\section{K-fold cross validation for \textsl{Lux} hyperparameters}
\label{app_kfold}
To determine the size of the latent space, $P$, and the strength of the L2 regularization, $\Omega$, we conduct a five-fold cross-validation test using the RGB stars from the high-SNR field RGB-train sample. We split the data into an initial train (4,000) and test (1,000) sample. We then run the \textsl{Lux} model (Figure~\ref{fig:flowchart}) for each of the five $K$-folds varying the size of $P$ each time running the first agenda for five iterations\footnote{We have found that after approximately five iterations, the global $\chi^{2}$ of the model begins to plateau.}, and then running the second agenda (see Figure~\ref{fig:flowchart}) once through. Here, for each $K$-fold and choice of latent size $P$, we also train the model varying $\Omega$. Following, with all the optimized parameters at hand (i.e. $\boldsymbol{A}$, $\boldsymbol{B}$, $\boldsymbol{z}$, and $\boldsymbol{s}$), we set out to estimate the inferred stellar labels and fluxes for the test stars in each $K$-fold. To do so, we must first determine the $\boldsymbol{z}$ latents for the test sample. We do this by optimizing the $test$ star $\boldsymbol{z}$ latent parameters at fixed $\boldsymbol{B}$ and $\boldsymbol{s}$ for a given choice of $P$ and $\Omega$ using the test set spectral fluxes of each star. To compare, we also compute the test $\boldsymbol{z}$ latent parameters at fixed $\boldsymbol{A}$ using the stellar labels of each star. With the $test$ set $\boldsymbol{z}$ latent parameters optimized, we then compute the predicted stellar labels using Equation~\ref{eq_labels} and stellar flux using Equation~\ref{eq_spectra}. 

We assess model performance by computing a $\chi^{2}$ metric on each of the test sets in the $K$-fold cross-validation using the following relation

\begin{equation}
\label{eq_chi2_tot}
\chi^{2} = \sum_{n=1}^{N_{\mathrm{stars}}}\Bigg(\frac{(\boldsymbol{\ell}_n-\boldsymbol{A}\,\boldsymbol{z}_n)^{2}}{\boldsymbol{\sigma}_{\ell_{n}}^{2}} + \frac{(\boldsymbol{f}_n-\boldsymbol{B}\,\boldsymbol{z}_n)^{2}}{\boldsymbol{\sigma}_{f_n}^{2} + \boldsymbol{s}^{2}}\Bigg).
\end{equation}

\begin{figure*}
    \centering
    \includegraphics[width=1\textwidth]{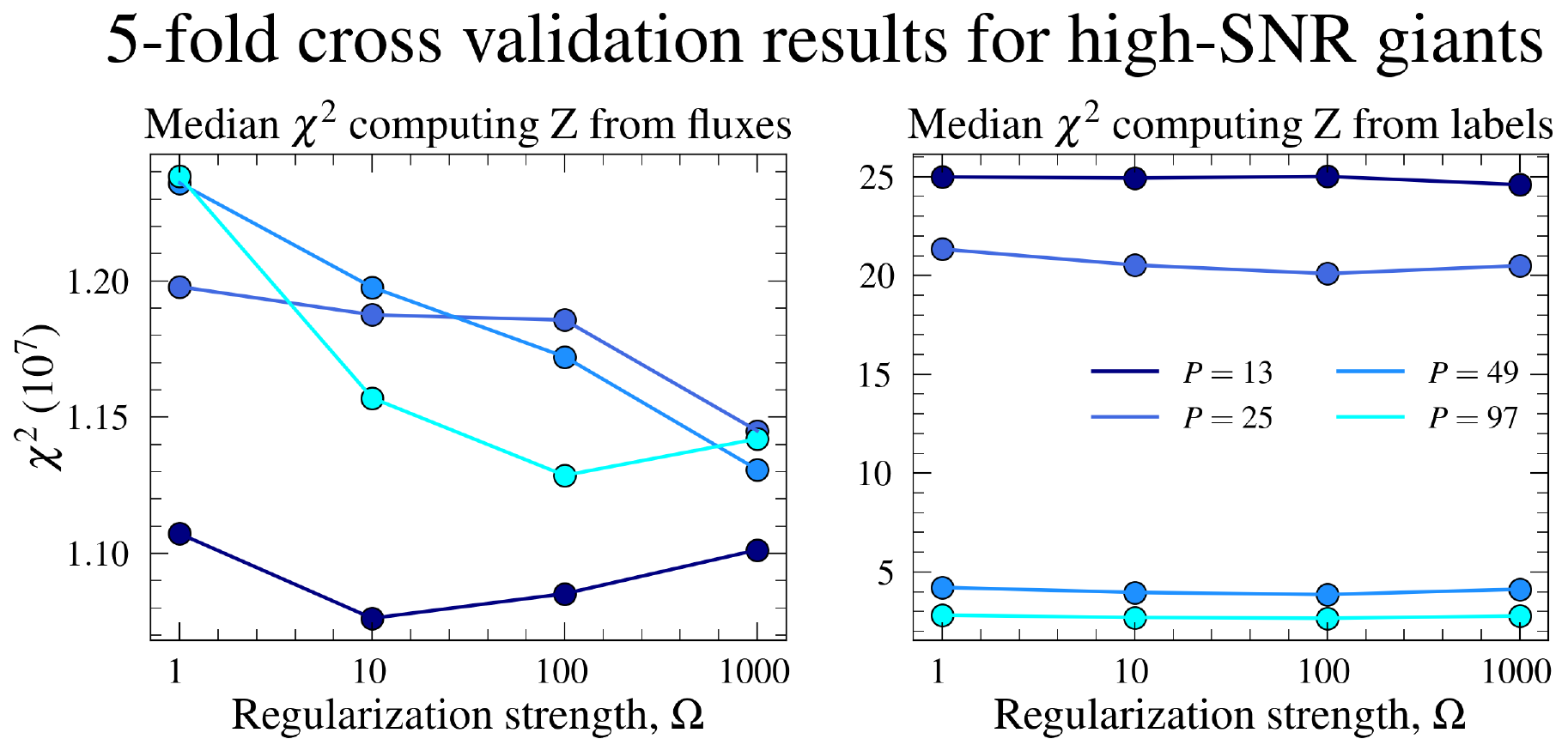}
    \caption{Median resulting total $\chi^{2}$ values from the $K$-fold cross validation test, summed over all wavelengths, all labels, and all stars in the high-SNR field RGB-test set. We show the $\chi^{2}$ metric estimated by computing the test set $\boldsymbol{z}$ latent parameters using each star's stellar fluxes (left) and labels (right). Overall, the model with $P=4\,\times\,M$ and $\Omega=10^{3}$ yields a good trade-off between latent dimensionality and regularization strength.}
    \label{fig:kfold}
\end{figure*}

We test our model varying the size of the latent space in multiples of the stellar label dimension $M$, $P=[M, 2M, 4M, 8M]$, and by varying the strength of the L2 regularization parameter, $\Omega = [1, 10^{1}, 10^{2}, 10^{3}]$. The median $\chi^{2}$ results obtained across all five $K$-folds from this cross-validation exercise are illustrated in Figure~\ref{fig:kfold}. The $K$-fold cross-validation results suggests a model with $P=4M$ and $\Omega=10^{3}$.

\section{Additional tests and validations of our application to \textsl{APOGEE} data}
\label{app_plots}
\begin{figure*}
    \centering
    \includegraphics[width=1\textwidth]{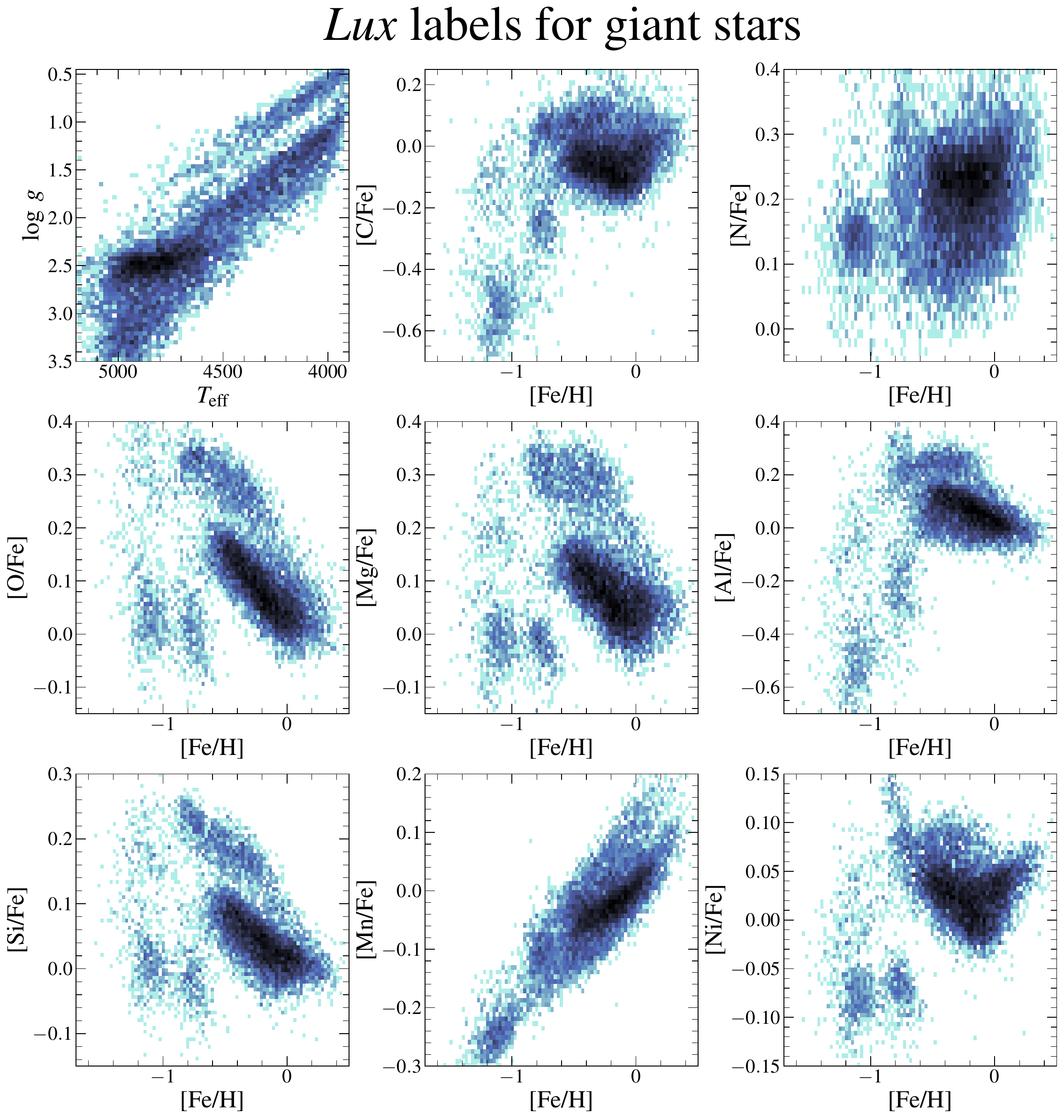}
    \caption{Stellar labels determined using \textsl{Lux} for all 10,000 high signal-to-noise RGB stars in the high-SNR field RGB-test set. Overall, the \textsl{Lux} labels look realistic and do not show any unusual trends. Moreover, the scatter around the labels is small, yielding a tight relation that reveals distinct structures both in stellar parameter and chemical abundance space, especially in the [Fe/H]-poor regime.
    }
    \label{fig:all-abun}
\end{figure*}

Figure~\ref{fig:all-abun} shows stellar labels inferred using \textsl{Lux} applied to test set stars from the high-SNR field RGB-test sample in the Kiel diagram as well as every element abundance modeled as a function of metallicity. To reiterate, these \textsl{Lux} stellar labels are determined by optimizing the latent representations ($\bs{z}_n$) using the spectral fluxes of each star.
We find that the distribution of \textsl{Lux} labels appears realistic, and the trends of [X/Fe] with [Fe/H] appear similar to those derived from \textsl{ASPCAP}. However, as seen in Figure~\ref{fig:kiel-tinsley}, \textsl{Lux} labels show a tighter trend or sequence when compared to the \textsl{ASPCAP} ones. This result illustrates how \textsl{Lux} is not only able to determine precise labels for those element abundances with the strongest lines (e.g., Mg or Fe), but also for other elements (e.g., C, N, Mn, Ni); this is possible across a decently wide range of metallicities ($-1.5 < \mathrm
{[Fe/H]}<0.5$). Of particular importance is the fact that we are able to resolve different metal-poor (halo) populations in different element abundance diagrams. For example, the different sequences at low [O/Fe], [Mg/Fe], [Al/Fe], and [Si/Fe] for metal-poor stars correspond to stars in the LMC and halo debris.

\begin{figure*}
    \centering
    \includegraphics[width=1\textwidth]{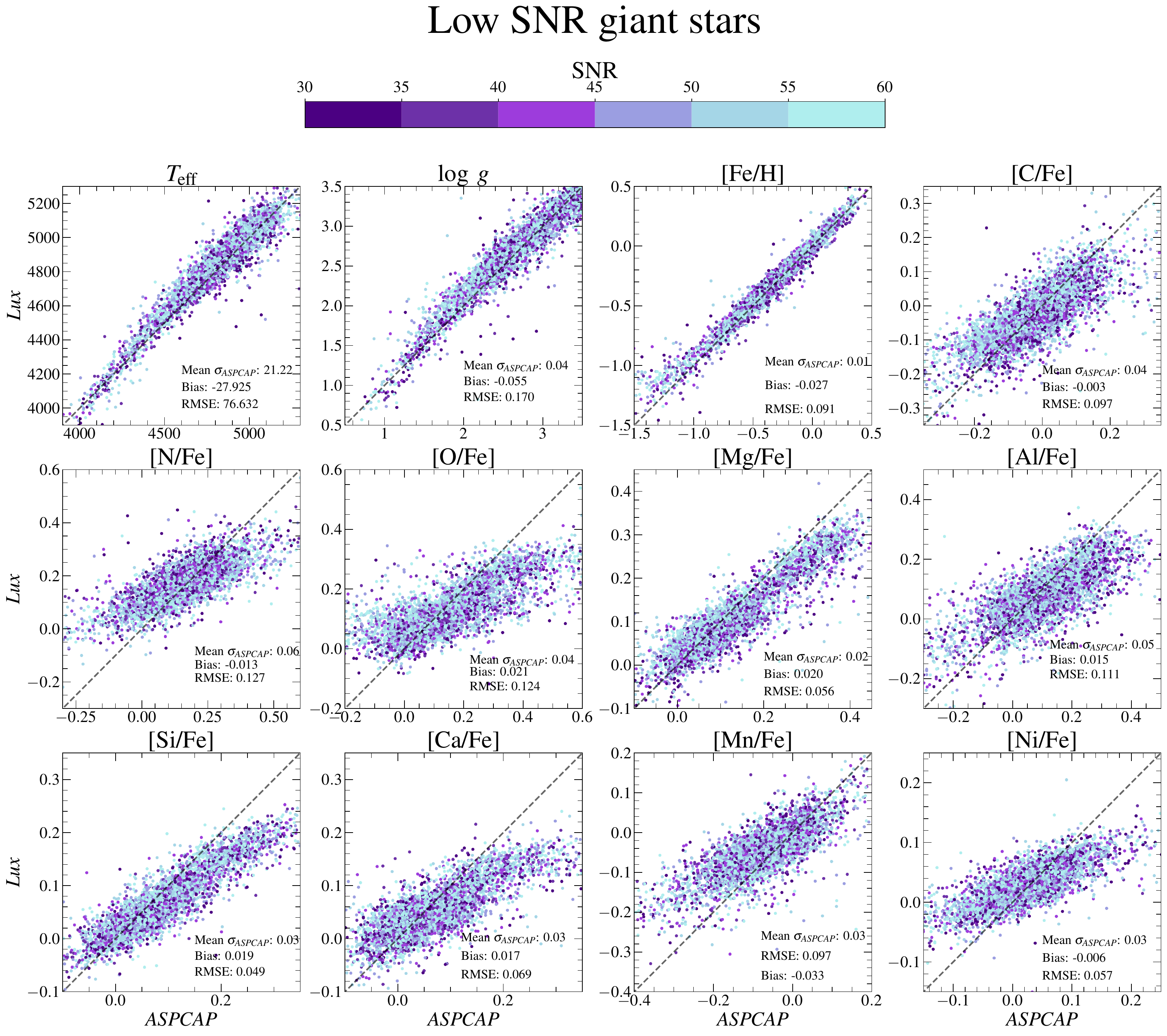}
    \caption{Validation results for RGB stars at lower SNR (low-SNR field RGB-test). The \textsl{Lux} labels shown are determined by optimizing the latent representations for test stars using each star's spectral fluxes. Here, we have chosen a SNR range that is expected for the \textsl{SDSS-V Galactic Genesis} survey. As in Figure~\ref{fig:low-snr}, we show the mean \textsl{ASPCAP} uncertainty, bias, and RMSE values for each stellar label in each panel. Overall, the RMSE values obtained are reasonably low and the bias values are approximately equal to the average \textsl{ASPCAP} uncertainty, indicating that the \textsl{Lux} model is able to infer stellar labels at reasonable precision for lower SNR stars. However, we do note that for some element abundance ratios (e.g., [O/Fe], [Ca/Fe], and [Ni/Fe]), there is a trend from the one-to-one line. See text in Section~\ref{sec_lowsnr} for further details. }
    \label{fig:cv_lowsnr}
\end{figure*}

Similarly, Figure~\ref{fig:cv_lowsnr} shows the full validation results for RGB stars at lower SNR (a continuation of Figure~\ref{fig:low-snr}). \textsl{Lux} is able to infer stellar labels at lower signal-to-noise robustly. However, we do note that there is some higher bias for particular elements (O, Ca, and Ni, for example).

\begin{figure*}
    \centering
    \includegraphics[width=1\textwidth]{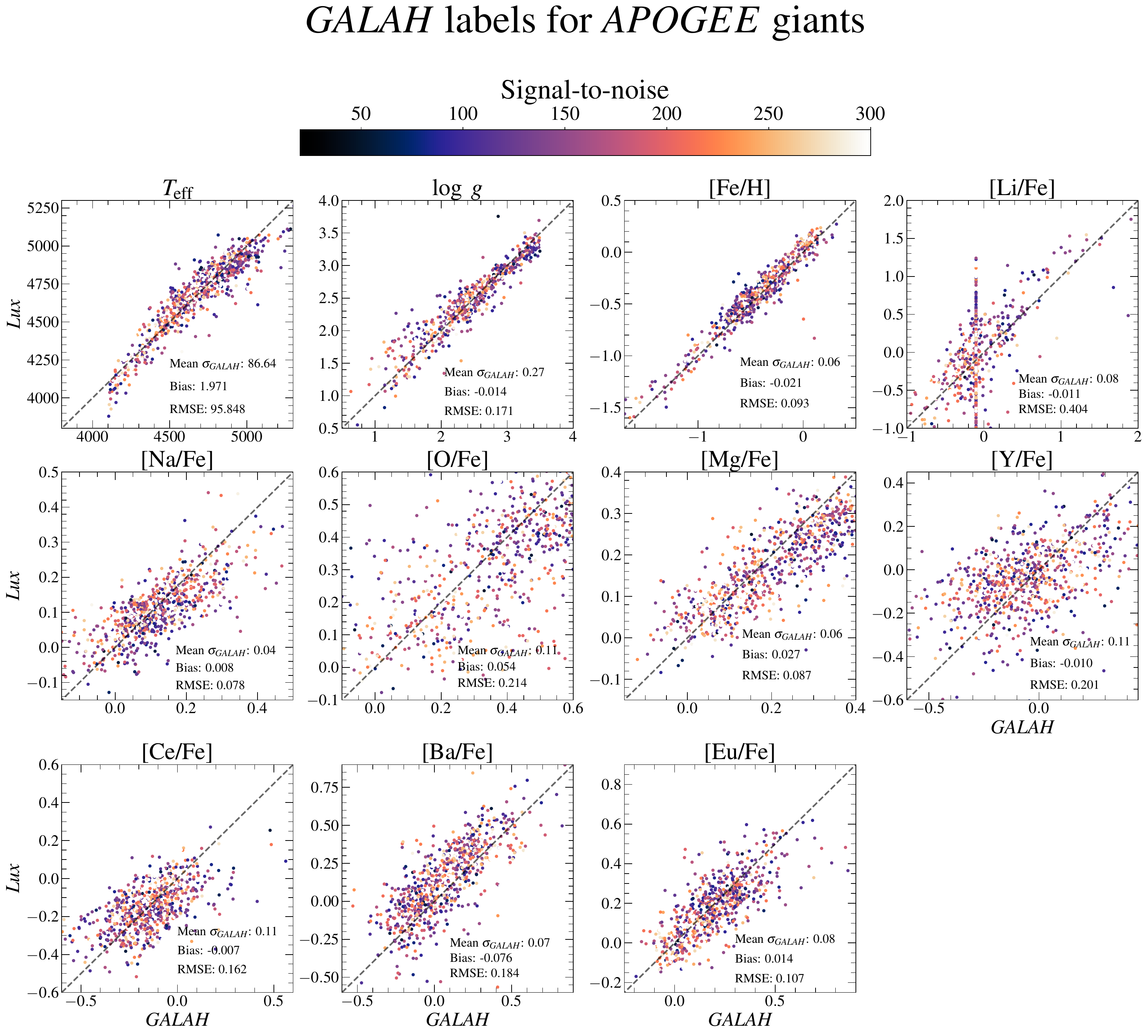}
    \caption{Validation results for RGB stars in the \textsl{GALAH-APOGEE} field giants-test set. The \textsl{Lux} labels shown are determined by optimizing the latent representations for test stars using each star's spectral fluxes. We show the mean reported \textsl{GALAH} label uncertainty, bias, and RMSE values for each stellar label in each panel. Overall, the RMSE values obtained are reasonably low and the bias values are approximately equal to the average \textsl{GALAH} uncertainty, indicating that the \textsl{Lux} model is able to perform successfully label transfer between surveys observing at different wavelengths, such as \textsl{GALAH} and \textsl{APOGEE}. We do note however that some stellar labels show a wide scatter around the one-to-one relation (e.g., [O/Fe]), which may indicate that our simple \textsl{Lux} model is not powerful enough to learn well this stellar label. See text in Section~\ref{sec_galah} for further details.}
    \label{fig:cv_galah}
\end{figure*}

Finally, Figure~\ref{fig:cv_galah} shows the full validation results from our test performing multi-survey translation between the \textsl{APOGEE} and \textsl{GALAH} stars from Section~\ref{sec_galah}. Overall, \textsl{Lux} is able to robustly infer the majority of the stellar labels used in the test. This includes elements which the \textsl{APOGEE} spectral range does not include particular spectral windows ([Li/Fe], [Y/Fe], and [Eu/Fe], for example). We postulate the reason \textsl{Lux} is able to perform well is because the model is likely finding some correlation between the labels that the \textsl{APOGEE} spectral range does include (e.g., Fe, Mg) and the labels it does not. This is likely because we have performed this test using element abundance ratios w.r.t. Fe (i.e., [Li/Fe] instead of [Li/H]). However, we cannot rule out the possibility that \textsl{Lux} is actually inferring these abundances in a causal manner, using weak or hidden spectral lines in the \textsl{APOGEE} spectral data. It would be interesting to follow up this exercise to ascertain if the model is inferring these abundances from variations in the spectral fluxes or via correlation with other elements.

\section{Testing $Lux$ by training on synthetic model spectra} \label{app_korg}

\begin{figure*}
    \centering
    \includegraphics[width=1\textwidth]{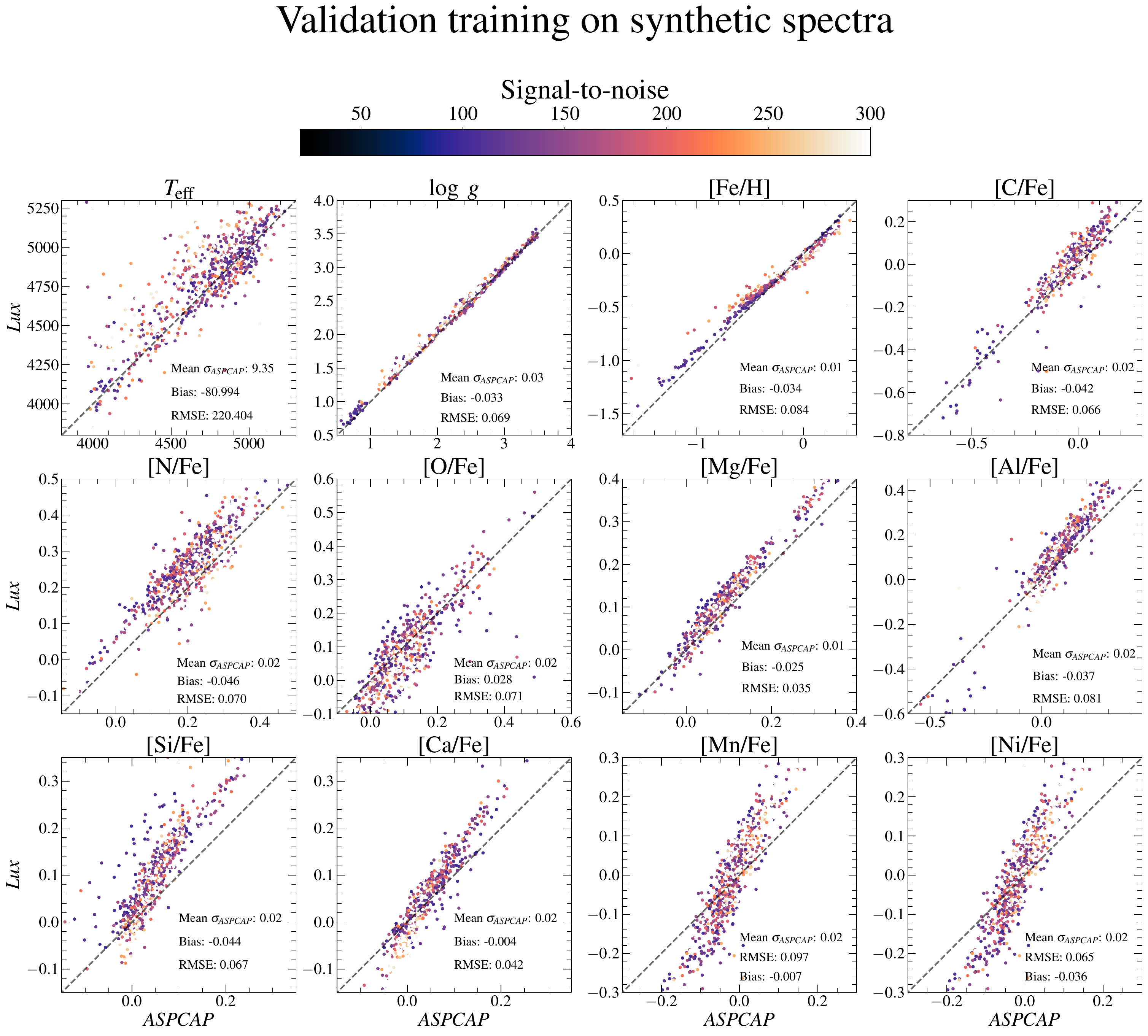}
    \caption{Validation results on 1,000 test stars with synthetic model \textsl{Korg} \citep[][]{Wheeler.2022} spectra. The \textsl{Lux} labels shown are determined by optimizing the latent representations for test stars using each star's synthetic (\textsl{Korg}) spectral fluxes. The model used to impute these validation results was trained on 4,000 stars with \textsl{APOGEE} stellar labels and \textsl{Korg} spectra. In each panel, we show the mean \textsl{ASPCAP} uncertainty, bias, and RMSE values. Overall, the RMSE value is reasonably low for all labels, and is extremely low for $\log~g$ and [Fe/H] (RMSE value below 0.1). However, there are some weak trends in some stellar labels (e.g., [Si/Fe], [Mn/Fe], and [Ni/Fe], for example), which could be due to limitations in the \textsl{Lux} model. 
    }
    \label{fig:cv-korg}
\end{figure*}

We also test how well our model operates by training \textsl{Lux} on synthetic model spectra determined using the \textsl{Korg} software \citep[][]{Wheeler.2022}, instead of using observational \textsl{APOGEE} spectra. To do so, we take the High-SNR field RGB-train sample of 5,000 stars, and compute synthetic model spectra using the \textsl{ASPCAP} outputted stellar parameters\footnote{We assume a $10\%$ per pixel uncertainty in the synethetic \textsl{Korg} spectra.}. We then divide this sample of 5,000 stars into a training set (4,000 stars) and a test set (1,000 stars), and train a new \textsl{Lux} model with the same set up (i.e., $P=4M$ and $\Omega=10^{3}$). We then compute predicted stellar labels using the same procedure for the 1,000 test stars. Overall, our model is able to infer reliable stellar labels using synthetic model spectra (see Figure~\ref{fig:cv-korg}). This result proves that \textsl{Lux} is able to train on both real observed and synthetic model spectra. For the resulting validation results when using the synthetic \textsl{Korg} spectra, see Figure~\ref{fig:cv-korg}.

\section{Dimensionality reduction of the latent vectors with t-SNE and principal component analysis}

\begin{figure*}
    \centering
    \includegraphics[width=1\textwidth]{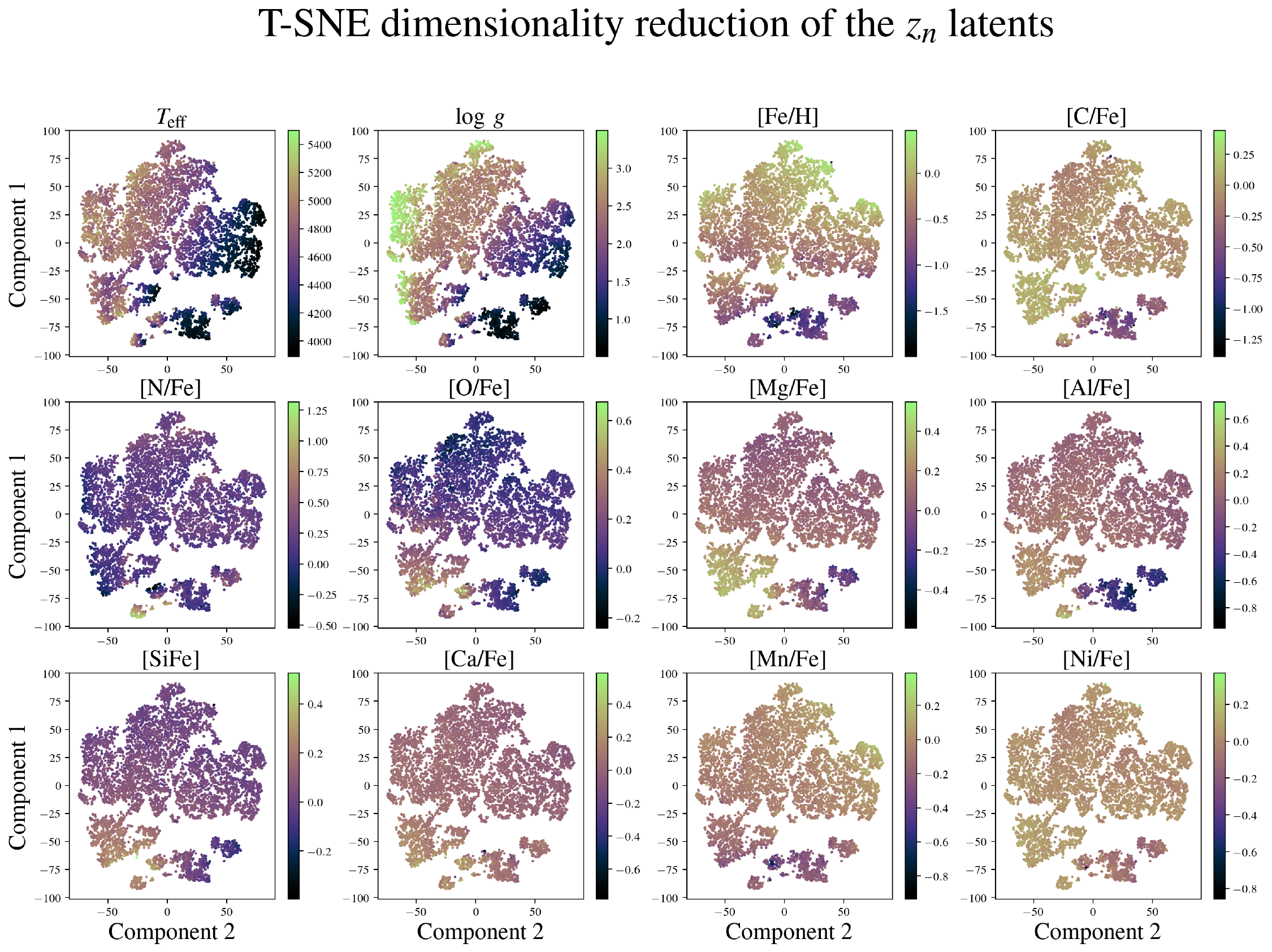}
    \caption{Components of a T-SNE dimensionality reduction (using two components) on the high-SNR field RGB-test set $\boldsymbol{z}$ latent parameters using a perplexity of 25. Each panel is color coded by one of the corresponding twelve stellar labels used to train the model for test set stars; these \textsl{Lux} stellar labels are computed by optimizing the test set latent representations using each star's spectral fluxes. Overall, there is structure in the latent space that correlates with the stellar labels.}
    \label{fig:tsne-latents}
\end{figure*}

Figure~\ref{fig:tsne-latents} shows the T-SNE dimensionality reduction results using two components for the high-SNR field RGB-test set $\boldsymbol{z}$ latent parameters. Here, we have chosen a perplexity of 25. Each panel is color-coded by a stellar label that was used to train and test the model. Overall, this result shows that \textsl{Lux} is mapping the stellar labels into the $\boldsymbol{z}$ latent parameters well. For example, low [Al/Fe] stars (typically associated with stellar halo debris) appear as a clear separated locci. Similarly, there are clear trends in this mapping with important labels ($T_{\mathrm{eff}}, \log~g$, and [Fe/H], for example). In summary, these results allow us to superficially interpret the mapping \textsl{Lux} is performing between the stellar labels and the $\boldsymbol{z}$ latent parameters, and highlights how the model is training effectively for the stellar labels. A similar exploration of the latent representations with spectral fluxes could also be performed.

\end{document}